\documentclass[aps,
               prx,
               superscriptaddress,
               twocolumn,
               floatfix,
               nofootinbib,
               amsfonts,
               amsmath,
               amssymb,
               10pt]{revtex4-2}

\usepackage[T1]{fontenc}
\usepackage{microtype}

\usepackage{braket}
\usepackage{bm}
\usepackage{bbold}
\usepackage{siunitx}
\usepackage[version=4]{mhchem}

\usepackage{graphicx}
\usepackage{xcolor}
\usepackage{booktabs}
\usepackage{comment}

\newcommand{\eff}{\mathrm{eff}}

\usepackage{hyperref}
\hypersetup{
    colorlinks,
    linkcolor={blue!80!black},
    citecolor={blue!80!black},
    urlcolor={blue!80!black},
    pdftitle={All-electrical dephasing-protected spin qubits in altermagnets},
    pdfauthor={José Carlos Abadillo-Uriel, Andrea Maiani, Alberto Cortijo, Ramón Aguado, Rubén Seoane Souto},
}

\usepackage[capitalize]{cleveref}
\crefformat{equation}{Eq.~(#2#1#3)}

\definecolor{revorange}{rgb}{0.80,0.40,0.0}

\begin{document}

\title{All-electrical dephasing-protected spin qubits in altermagnets}

\author{Jos\'e Carlos Abadillo-Uriel}
\thanks{These authors contributed equally to this work.}
\affiliation{Quantum Advanced Research Center (QuARC), Consejo Superior de Investigaciones Cient{\'i}ficas (CSIC), Sor Juana In{\'e}s de la Cruz 3, 28049 Madrid, Spain}
\affiliation{Instituto de Ciencia de Materiales de Madrid (ICMM), Consejo Superior de Investigaciones Cient{\'i}ficas (CSIC), Sor Juana In{\'e}s de la Cruz 3, 28049 Madrid, Spain}

\author{Andrea Maiani}
\thanks{These authors contributed equally to this work.}
\affiliation{Nordita, KTH Royal Institute of Technology, and Stockholm University, Hannes Alfv\'ens v\"ag 12, SE-10691 Stockholm, Sweden}

\author{Alberto Cortijo}
\affiliation{Instituto de Ciencia de Materiales de Madrid (ICMM), Consejo Superior de Investigaciones Cient{\'i}ficas (CSIC), Sor Juana In{\'e}s de la Cruz 3, 28049 Madrid, Spain}

\author{Ram{\'o}n Aguado}
\affiliation{Quantum Advanced Research Center (QuARC), Consejo Superior de Investigaciones Cient{\'i}ficas (CSIC), Sor Juana In{\'e}s de la Cruz 3, 28049 Madrid, Spain}
\affiliation{Instituto de Ciencia de Materiales de Madrid (ICMM), Consejo Superior de Investigaciones Cient{\'i}ficas (CSIC), Sor Juana In{\'e}s de la Cruz 3, 28049 Madrid, Spain}

\author{Rub\'en Seoane Souto}
\affiliation{Quantum Advanced Research Center (QuARC), Consejo Superior de Investigaciones Cient{\'i}ficas (CSIC), Sor Juana In{\'e}s de la Cruz 3, 28049 Madrid, Spain}
\affiliation{Instituto de Ciencia de Materiales de Madrid (ICMM), Consejo Superior de Investigaciones Cient{\'i}ficas (CSIC), Sor Juana In{\'e}s de la Cruz 3, 28049 Madrid, Spain}

\date{July 3, 2026}

\begin{abstract}
We propose altermagnetic semiconductors as a platform for field-free, all-electrically controlled spin qubits in gate-defined quantum dots. The momentum-dependent spin splitting of an altermagnet produces a Zeeman-like qubit splitting whose magnitude and sign are set by the dot ellipticity, enabling local frequency tunability without external magnetic fields or micromagnets. Because the splitting is tied to a fixed altermagnetic quantization axis, electric-field noise is longitudinally suppressed at leading order, while quantization-axis fluctuations couple transversely and therefore cause relaxation rather than pure dephasing. The compensated magnetic order also avoids stray fields, making the platform naturally compatible with superconducting resonators and dispersive circuit-QED readout through the qubit's spin-dependent electric dipole. Starting from an effective quantum-dot model, supported by a microscopic lattice model, we derive the single- and two-dot Hamiltonian models. We show that electric-dipole spin resonance enables single-qubit control, while tunable exchange and electrically addressable qubit frequencies realize fSim two-qubit gates. The same double-dot architecture also supports singlet--triplet qubits with electrical control of both exchange and splitting gradients, removing the need for micromagnets or nuclear-polarization gradients. These results establish altermagnetic quantum dots as a route to field-free spin qubits with intrinsic electrical tunability and enhanced dephasing protection.
\end{abstract}

\maketitle

\section{Introduction}

Among the leading candidates for scalable quantum processors, spin qubits encode quantum information in the spin of a single electron or hole confined in a gate-defined quantum dot (QD)~\cite{Loss_1998_Quantum,Kane_1998_Silicon,Leuenberger_2001_Quantum,Burkard_2023_Semiconductor}. Their long intrinsic coherence times~\cite{Yoneda_2018_Quantum, Kobayashi_2021_Engineering}, compact footprint, and compatibility with industrial semiconductor fabrication have made them a leading candidate for scalable quantum processors, with recent demonstrations steadily improving single- and two-qubit gate fidelities~\cite{Noiri_2022_Fast, Xue_2022_Quantum, Mills_2022_Two} and the size of coupled qubit arrays~\cite{Hendrickx_2021_Four, Philips_2022_Universal, Borsoi_2024_Shared, John_2025_Robust}. Importantly, spin qubits are not limited to nearest-neighbor exchange: coupling through shared microwave modes or coherent spin shuttling offers routes to reconfigurable long-range connectivity~\cite{Vandersypen_2017_Interfacing,Dijkema_2024_Cavity}. This flexibility is valuable both for scalable processor layouts and for error-correction schemes, including quantum low-density parity-check codes that benefit from sparse nonlocal interactions~\cite{Breuckmann_2021_Quantum,Bravyi_2024_High}.

\begin{figure}[htbp]
    \includegraphics[width=\linewidth]{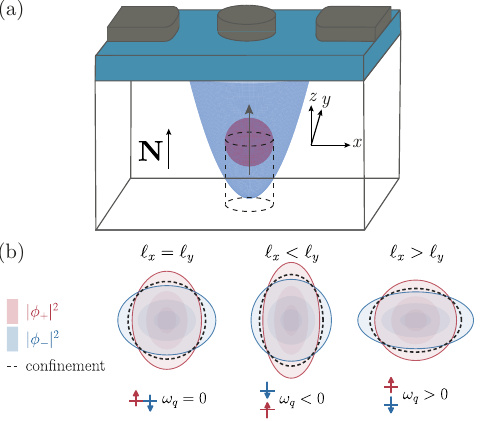}
    \caption{
    Altermagnetic semiconductor platform for field-free, gate-defined spin qubits.
    (a) Schematic device geometry. Metallic gates deposited above a dielectric layer electrostatically define a single-electron quantum dot in an altermagnetic semiconductor. The N\'eel vector is taken along the out-of-plane direction, providing a fixed spin-quantization axis, while lateral gate voltages tune the dot position and confinement anisotropy.
    (b) Shape-controlled altermagnetic spin splitting. The dashed contours indicate the electrostatic confinement, while the red and blue contours show the spin-dependent orbital densities $|\phi_+|^2$ and $|\phi_-|^2$ for up and down spins. For a circular dot, $\ell_x=\ell_y$, the two spin sectors have equal zero-point energy and the qubit splitting vanishes. An elliptic confinement samples the $d$-wave altermagnetic band splitting anisotropically, producing a finite $\omega_q$ whose sign is reversed by interchanging the long and short axes of the dot.
    }
    \label{fig:1}
\end{figure}

In practice, however, the Zeeman splitting that defines the qubit frequency is determined by an external magnetic field, which introduces several experimental constraints. Because the field acts globally, the splitting is fixed uniformly across the device, offering only limited per-qubit tunability---largely via voltage-controlled modulation of the $g$-factor in hole-spin qubits~\cite{Voisin_2015_Electrical,Crippa_2018_Electrical}. The accompanying stray fields are, moreover, difficult to reconcile with the superconducting circuitry used for readout, and achieving first-order insensitivity to charge noise requires tuning the device to dedicated operating points~\cite{Medford_2013_Quantum,Piot_2022_Single,Hendrickx_2024_Sweet,Bassi_2025_Optimal,Ungerer_2025_Dephasing}. Engineered field gradients from micromagnets or from dynamic nuclear polarization can restore local, qubit-resolved control~\cite{PioroLadriere_2008_Electrically, Wu_2014_Two, Foletti_2009_Universal}, but they reintroduce stray fields or fabrication overhead and offer little \emph{in situ} tunability. These limitations motivate the search for a material platform with intrinsic, gate-tunable spin splitting that would eliminate the need for external magnetic fields, thereby enhancing qubit performance and facilitating circuit-QED integration.

Altermagnets meet precisely these requirements~\cite{Mazin_2022_Editorial,Smejkal_2022_Emerging,Autieri_2024_New,Bai_2024_Altermagnetism,Song_2025_Altermagnets}. In these collinear compensated magnets, the net magnetization vanishes as in an antiferromagnet, while crystal symmetries relating the magnetic sublattices allow a momentum-dependent spin splitting of the electronic bands, reminiscent of exchange splitting in a ferromagnet but without a macroscopic magnetic moment. The symmetry principles, candidate materials, and microscopic and transport consequences of altermagnetism have now been developed across a broad range of systems~\cite{Ahn_2019_Antiferromagnetism,Hayami_2019_Momentum,Naka_2019_Spin,Naka_2020_Anomalous,Smejkal_2020_Crystal,Smejkal_2022_Giant,Chen_2024_Impurity,Fernandes_2024_Topological,Sukhachov_2024_Impurity,Zhu_2024_Field,Sheoran_2026_Tuning,Gomonay_2024_Structure,Fu_2025_All,Cortijo_2026_Quantum}, and spin-split bands have been observed directly in several candidate materials~\cite{Krempasky_2024_Altermagnetic,Lee_2024_Broken,Osumi_2024_Observation,Reichlova_2024_Observation,Jiang_2025_Metallic,Amin_2024_Nanoscale,Yamamoto_2025_Altermagnetic,Reimers_2024_Direct,Fedchenko_2024_Observation,Bai_2022_Observation}.

This combination is especially appealing for spin-based quantum devices. Because its magnetic moments compensate, the order carries no net magnetization and thus avoids the stray fields of micromagnet-based schemes, yet it still defines an intrinsic spin-quantization axis. Crucially, the altermagnetic spin splitting is set by exchange and crystal symmetry rather than by relativistic spin--orbit coupling, so it follows electronic energy scales and can far exceed the relativistic spin--orbit splittings that hole-spin qubits rely on for electrical control~\cite{Smejkal_2022_Emerging}. At the same time, its anisotropy in momentum space makes the splitting sensitive to confinement: in an anisotropic gate-defined dot, the confined carrier samples the altermagnetic band structure in a geometry-dependent way, producing a spin splitting that can be tuned electrically through the dot shape, see Fig.~\ref{fig:1}. The result is a field-free analogue of a Zeeman energy whose strength is controlled by gates rather than by an external magnetic field, opening a route to all-electrical spin manipulation. Thus altermagnets combine antiferromagnetic compensation with an accessible, gate-tunable spin splitting~\cite{Jungwirth_2016_Antiferromagnetic}, distinguishing them from conventional antiferromagnetic, micromagnet-based, and spin--orbit-based qubit platforms.

In this work, we show that altermagnets are promising candidates for all-electrical spin-based quantum processing. Building on this geometric control of the qubit splitting, electrostatically tunable from zero to tens of gigahertz at realistic confinement lengths, with no applied magnetic field, we find that an altermagnetic double dot supports two complementary encodings. The first is a Loss--DiVincenzo single-spin qubit at each dot, with individually addressable splittings and entangling two-qubit gates from the fermionic-simulation (fSim) family---including $\sqrt{\text{iSWAP}}$ and CZ---generated by the tunable interdot exchange. The second is a singlet--triplet qubit in the $(1,1)$ charge configuration, in which the exchange and the per-dot splitting difference, itself set by the relative orientation of the two dots, natively realize the two control axes that conventionally require micromagnets or nuclear-polarization gradients. We characterize both qubits as a function of the confinement potential, analyze the dominant error channels, propose protocols for single- and two-qubit gates in each encoding, and discuss readout via both conventional spin-to-charge conversion and dispersive circuit-QED. We finally distill these findings into concrete design criteria for an ideal intrinsic altermagnetic host and further argue that the platform is not even tied to a single such compound, since the requisite splitting can also be proximity-induced in a conventional host and thus leaving substantial freedom in the choice of material.

\section{Effective model}
\label{sec:effective-model}
The qubit physics follows from how an electrostatically confined carrier samples the spin-split bands of the altermagnetic host, which we capture here in an effective low-energy model. \cref{fig:1}(a) shows a representative conduction band: its momentum-dependent spin splitting reflects the $C_4\mathcal{T}$ symmetry of the $d$-wave altermagnetic order, and in a semiconducting host a band gap separates it from the valence states. For concreteness, we assume hereafter that the relevant conduction-band edge has a single minimum at $\Gamma$, which removes the valley degree of freedom present in, e.g., electron-based Si QDs~\cite{Burkard_2023_Semiconductor}. This is a simplification for clarity rather than a fundamental requirement since spin qubits can also operate in multi-valley hosts.

The presence of a band gap enables conventional fabrication and gating in semiconductor devices. Metallic gate electrodes deposited on a dielectric barrier electrostatically define a confinement potential that traps individual electrons in QDs, as sketched in \cref{fig:1}(b). Each dot is tunable via gate voltages: the number of confined electrons can be controlled down to the single-electron regime, where a single spin-$1/2$ degree of freedom constitutes a spin qubit following the Loss--DiVincenzo paradigm~\cite{Loss_1998_Quantum}. The altermagnetic host provides these qubits with a momentum-dependent spin splitting that enables qubit manipulation without external magnetic fields.

Near the $\Gamma$ point, one can derive an effective low-energy theory for the conduction-band electrons (see Appendix~\ref{app:micro}). The most general leading-order continuum Hamiltonian consistent with the $d$-wave altermagnetic symmetry and Rashba spin--orbit coupling (SOC) is
\begin{equation}
\label{eq:continuum-hamiltonian}
H_{\mathrm{eff}}
=
\frac{\bm{k}^2}{2m}\sigma_0
-
\eta\, g_d(\bm{k}) \sigma_z
+
\alpha_R
\left(
\sigma_y k_x-\sigma_x k_y
\right)
+
V(\bm r)
\,,
\end{equation} 
where $\bm{k} = (k_x, k_y)$ is the crystal momentum, $m$ is the effective mass of the active band, $\eta$ is the $d$-wave altermagnetic coefficient, $g_d(\bm{k}) = k_x^2-k_y^2$ is the altermagnetic form factor that introduces the $k$-dependent spin splitting, $\alpha_R$ is the effective Rashba coefficient, and $V(\bm r)$ is the electrostatic confinement potential (throughout we adopt units with $\hbar=1$). The term $\eta\, g_d(\bm{k})$ is the altermagnetic non-relativistic spin--orbit coupling and the central ingredient that provides a field-free, electrically tunable spin splitting, aligning the spin with the N\'eel vector $\bm{n}$. 

We choose the N\'eel vector to be aligned out of plane, $\bm{n}\parallel\hat z$ for convenience, since it renders the altermagnetic splitting longitudinal ($\propto\sigma_z$) and the Rashba coupling purely transverse, which is what makes the qubit physics below most transparent. An in-plane component of the N\'eel vector merely rotates the altermagnetic axis into the $\sigma_{x,y}$ plane, and the scheme carries over up to negligible corrections in the weak spin--orbit limit $O(\alpha_R^2)$, as discussed in Appendix~\ref{app:inplane}. We argue in \cref{sec:materials} that an out-of-plane easy axis is moreover realistic.

The relevant figure of merit is the dimensionless combination $m\eta$, which measures the fractional spin-dependence of the band curvature and is independent of the wavevector near the band edge. As we will show, it sets both the orbital squeezing and, through it, the qubit-frequency scale, so that the field-free splitting follows directly from the gate-defined confinement anisotropy without reference to the absolute band splitting at any particular momentum.

\section{Single-electron spin qubit}
\label{sec:single-electron-qubit}

\begin{figure}[ht]
    \includegraphics[width=\linewidth]{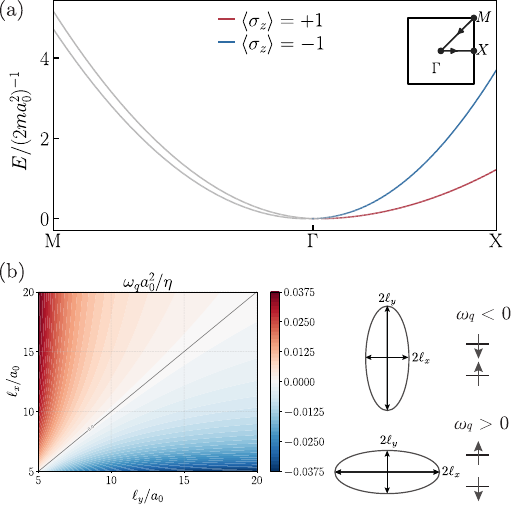}
    \caption{
        Band-structure origin and magnitude of the shape-controlled qubit splitting.
        (a) Representative lowest conduction band of a $d$-wave altermagnetic semiconductor along the high-symmetry path shown in the inset. The altermagnetic term produces opposite dispersions for the two spin sectors, with $\langle\sigma_z\rangle=\pm1$, while the splitting vanishes along the symmetry direction where the $d$-wave form factor is zero, up to spin--orbit corrections. Parameters used: $m=0.1$, $\eta=0.25$, and $\alpha_R=0.05$.
        (b) Qubit frequency generated by anisotropic harmonic confinement, plotted as the dimensionless quantity $\omega_q a_0^2/\eta$ versus the confinement lengths $\ell_x$ and $\ell_y$. The diagonal line marks the circular-dot condition $\ell_x=\ell_y$, where $\omega_q=0$. Away from this line, the sign and magnitude of $\omega_q$ are controlled electrically by the dot ellipticity. For $\eta>0$, elongation along $y$ gives $\omega_q<0$, while elongation along $x$ gives $\omega_q>0$, as illustrated by the schematics on the right.
}
    \label{fig:2}
\end{figure}

The in-plane confinement potential $V(\mathbf{r})$ in \cref{eq:continuum-hamiltonian} is generated by the metallic gates and, in the single-electron regime, can be approximated by a parabolic confining potential $V(\mathbf{r})\approx \tfrac{1}{2}m(\omega^2_x x^2+\omega^2_y y^2)$, where $\omega_x$ and $\omega_y$ are independently tunable via the gate stack and $x,y$ denote the dot's principal axes, taken aligned with the crystal axes. A generic parabola, with a cross term $\propto xy$, is recovered by a rigid rotation of these axes by an angle $\theta_\mathrm{d}$ relative to the crystal; this orientation rescales the splitting by $\cos2\theta_\mathrm{d}$ and leaves the results below unchanged for aligned axes (Appendix~\ref{app:higher-order-altermagnets}). It is convenient to parametrize the confinement by the harmonic lengths $\ell_{x,y}=\sqrt{1/(m\omega_{x,y})}$.

When the Rashba coupling $\alpha_R$ is small compared with the orbital excitation gap, we construct the qubit basis from the $\alpha_R=0$ Hamiltonian and treat the Rashba term perturbatively. In this limit, the unperturbed Hamiltonian commutes with $\sigma_z$ and decomposes into spin sectors $s_z=\pm1$, each described by an anisotropic harmonic oscillator with squeezed widths
\begin{equation}
\begin{split}
    &\ell_{x,s_z}=\ell_x(1-2s_zm\eta)^{1/4}, \\
    &\ell_{y,s_z}=\ell_y(1+2s_zm\eta)^{1/4}.
\end{split}
\label{eq:squeezedlengths}
\end{equation}

We obtain the effective spin-qubit Hamiltonian by projecting onto the two-dimensional subspace spanned by the harmonic ground states of the two spin sectors, $\ket{\phi_+}$ and $\ket{\phi_-}$. 

Throughout this work, the qubit basis states $\ket{\uparrow}$ and $\ket{\downarrow}$ denote these dressed spin-orbital eigenstates,
\begin{equation}
\ket{\uparrow}\equiv\ket{\phi_+}\otimes\ket{\zeta_\uparrow},
\qquad
\ket{\downarrow}\equiv\ket{\phi_-}\otimes\ket{\zeta_\downarrow},
\label{eq:qubit-basis-def}
\end{equation}
with $\zeta_{\uparrow,\downarrow}$ the bare spin factor; the arrow kets used in the remainder of the paper are built from this definition. 

Projecting onto this subspace gives
\begin{equation}
H_q=\frac{\omega_q}{2}\sigma_z,
\label{eq:prelarmor}
\end{equation}
where the splitting is given by the difference of the corresponding zero-point energies,
\begin{equation}
\begin{split}
\omega_q
&=\tfrac{1}{2}(\omega_x-\omega_y)
\bigl[\sqrt{1-2m\eta}-\sqrt{1+2m\eta}\bigr] \\
&\approx \eta\left(\frac{1}{\ell^2_y}-\frac{1}{\ell^2_x}\right),
\end{split}
\label{eq:larmor}
\end{equation}
with the approximation valid for $|m\eta|\ll 1$. The splitting therefore grows with the dot anisotropy and changes sign when $\ell_x$ and $\ell_y$ are interchanged: for $\ell_x<\ell_y$ (elongation along $y$) it is negative with a spin-up ground state, and for $\ell_x>\ell_y$ (elongation along $x$) it is positive with a spin-down ground state (taking $\eta>0$), as illustrated in \cref{fig:2}(b). Notably, the \emph{exact} expression in \cref{eq:larmor} describes a vanishing qubit frequency at the circular geometry $\ell_x=\ell_y$ \emph{non-perturbative}: the squeezed eigenstates are spin-polarized Gaussians whose widths (\emph{cf.}~\cref{eq:squeezedlengths}) simply interchange under $s_z\to-s_z$, so their zero-point energies cancel identically. The full derivation of \cref{eq:larmor}, together with the perturbative effect of the Rashba spin--orbit coupling, is given in Appendix~\ref{app:eigenstates}.

Taking a \SI{1}{\milli\electronvolt} ellipticity-induced confinement-energy difference as a practical lower target (gate-defined Si/SiGe dots routinely reach orbital energies of several \si{\milli\electronvolt}~\cite{Kunne_2024_SpinBus}), Eq.~\eqref{eq:larmor} gives $|\omega_q|/2\pi\simeq\SI{242}{\giga\hertz}\,|m\eta|$, spanning \SIrange{2.4}{60}{\giga\hertz} for $|m\eta|=0.01$--$0.25$. For $m^*\!\approx\!0.1\,m_0$, this confinement-energy difference corresponds, for example, to harmonic lengths of about \SI{15}{\nano\meter} and \SI{18}{\nano\meter}. The $\alpha$-MnTe benchmark $m\eta\sim0.05$--$0.25$ therefore gives $|\omega_q/2\pi|\sim\SIrange{12}{60}{\giga\hertz}$~\cite{Belashchenko_2025_Giant}

Importantly, the qubit splitting in \cref{eq:larmor} requires no external magnetic field: it is set entirely by the N\'eel order, through $\eta$, and the dot geometry $\ell_x\neq\ell_y$, so reshaping the dot ellipticity with gate voltages tunes the frequency of each qubit individually, as shown in \cref{fig:2}(a). This is in stark contrast to conventional semiconductor spin qubits, whose Zeeman splitting is fixed by an external field, increasing complexity and limiting their coupling to superconductors. This tunability is not a phenomenological assumption but microscopic in origin: the effective $d$-wave spin-dependent term is symmetry-allowed only in the presence of both N\'eel order and next-nearest-neighbor hopping anisotropy, so the geometric dependence of $\omega_q$ is inherited directly from the lattice symmetries. The ability to tune $\omega_q$ through zero parallels strongly squeezed hole spin qubits~\cite{Michal_2021_Longitudinal, Bosco_2021_Squeezed, AbadilloUriel_2023_Hole, Martinez_2026_Disorder} and the proposed five-electron quadrupole qubit~\cite{Caporaletti_2025_Proposed}, here with the crucial differences that the scheme is field-free, operates at the single-electron level, and places the zero-splitting naturally at the circular geometry.

The altermagnetism-induced shape-dependent frequency in \cref{eq:larmor} provides single-axis ($\sigma_z$) control via gate tuning of the dot. However, full single-qubit manipulation requires rotations about a second spin axis. This is achieved through electric-dipole spin resonance (EDSR)~\cite{Rashba_2003_Orbital, Golovach_2006_Electric}, exploiting the Rashba spin--orbit coupling already present in \cref{eq:continuum-hamiltonian}. For instance, an oscillating in-plane electric field along $x$, $-eF_x(t)\,x$, applied via an ac gate voltage, displaces the electron wavefunction within the dot and activates the Rashba term, generating a time-dependent transverse spin coupling
\begin{equation}
    H_{\text{EDSR}} = -e\frac{m\ell_x^4}{\ell_{\text{so}}}\,\partial_t F_x(t)\sigma_y,
    \label{eq:edsr}
\end{equation}
where $\ell_{\text{so}}=1/(m\alpha_R)$ is the spin--orbit length,  see Appendix~\ref{app:edsr} for the full derivation. By symmetry, an analogous drive along $y$ generates the complementary coupling $H^{(y)}_{\text{EDSR}}=-e(m\ell_y^4/\ell_{\text{so}})\,\partial_t F_y(t)\sigma_x$, so the two in-plane drives access both transverse Bloch-sphere axes. Combined with $H_q\propto\sigma_z$, this enables arbitrary single-qubit gates entirely through electric fields: a scheme well established in SOC-based spin qubits \cite{Nowack_2007_Coherent,NadjPerge_2010_Spin,vandenBerg_2013_Fast,PitaVidal_2023_Direct}. 

Another possible control mechanism is provided by strain. Since strain can rotate the N\'eel vector~\cite{LiebmanPelaez_2026_Strain}, local strain fields, for instance arising from thermal strain in the gate stack, could rotate the magnetic order and thereby provide an additional handle on the qubit. This would enable a complementary EDSR-mechanism to electrostatic deformation of the dot, in close analogy with strain-assisted manipulation protocols proposed and implemented for hole spin qubits~\cite{AbadilloUriel_2023_Hole, Wang_2024_Operating, Martinez_2026_Disorder}.

The mechanism above is not specific to the $d$-wave form factor. For any collinear altermagnet the splitting stays locked to the N\'eel axis, $\omega_q=-2\eta_\ell\langle g_\ell(\bm k)\rangle_\mathrm{dot}$, with $\langle\cdots\rangle_\mathrm{dot}$ the average over the momentum spread of the confined orbital; reshaping the dot then tunes the magnitude of $\omega_q$ but never tilts the quantization axis. What the form-factor order changes is how strongly the splitting couples to the dot shape: the $d$-wave case ($\ell=2$) responds linearly to the ellipticity, whereas the $g$- and $i$-wave ($\ell=4,6$) cases couple only to higher-order deformations and require a dot rotated relative to the crystal axes. The $d$-wave is moreover special in that its quadratic form factor is absorbed exactly into the orbital squeezing of \cref{eq:squeezedlengths}, while higher harmonics enter only perturbatively in $\eta_\ell$ (Appendix~\ref{app:higher-order-altermagnets}). Consequently, every result relying only on a fixed-axis splitting and Rashba-activated driving carries over directly, while the genuinely $d$-wave-specific quantities, such as the exchange anisotropy, the charge-noise form factor, and the cavity polarizability of the next sections, must be recomputed for a $g$- or $i$-wave host.

\section{Double-quantum-dot exchange Hamiltonian}
\label{sec:dqd}
\begin{figure}[htbp]
    \includegraphics[width=0.7\linewidth]{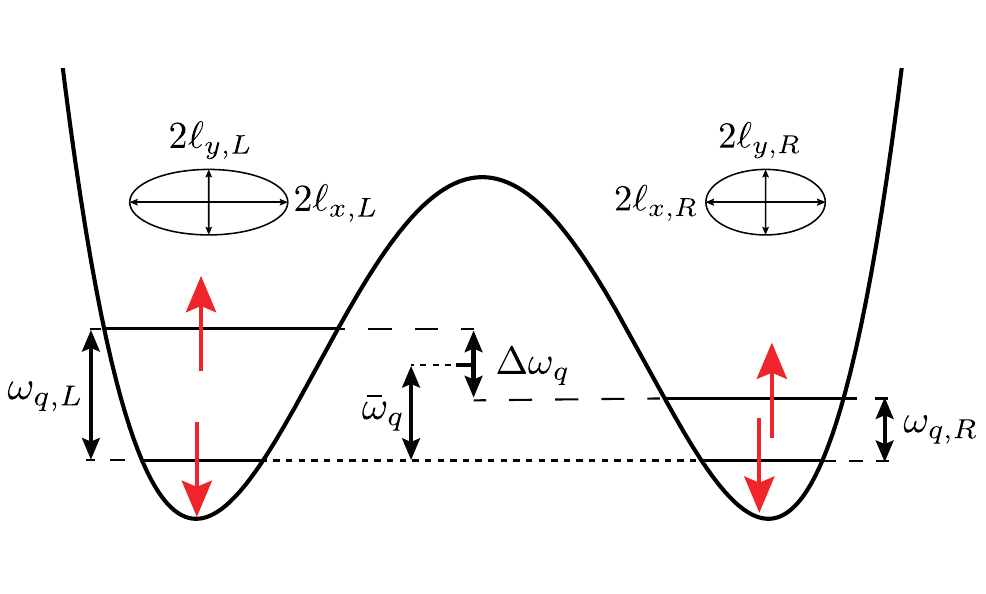}
    \caption{A double quantum dot provides independent electrical control of the average splitting, splitting gradient, and exchange. Two adjacent QDs of unequal ellipticity (left: $\ell_{x,L},\ell_{y,L}$; right: $\ell_{x,R},\ell_{y,R}$); the ellipticity sets $\omega_{q,L}$ and $\omega_{q,R}$, with average $\bar{\omega}_q=(\omega_{q,L}+\omega_{q,R})/2$; the inter-dot barrier sets the exchange $\mathcal{J}$. When the ellipticities are unequal, a gradient $\Delta\omega_q=\omega_{q,L}-\omega_{q,R}$ develops.}
    \label{fig:3}
\end{figure}

When two occupied neighboring QDs are tunnel-coupled, the wavefunction overlap gives rise to an exchange interaction controlled by the inter-dot barrier gate. The resulting double-dot device is sketched in \cref{fig:3}: two dots sit in a double-well confinement potential, each shaped by its own ellipticity so that each carries an independently tunable qubit splitting set by \cref{eq:larmor}, while the barrier between them controls their mutual exchange. 

In an altermagnetic dot, the overlaps between neighboring dots, and therefore the tunneling and Coulomb matrix elements entering exchange, are spin dependent. Consequently, the exchange exhibits an anisotropic structure. In addition, the Rashba spin--orbit coupling rotates the local spin frames between the two dots~\cite{Aleiner_2001_Spin,Geyer_2024_Anisotropic}. Both ingredients endow the exchange with a non-scalar tensor structure.

Projecting the two-electron problem onto these localized orbitals and eliminating the doubly occupied charge states gives, up to a spin-independent constant, the effective $(1,1)$ anisotropic-exchange Hamiltonian
\begin{equation}
H_\mathrm{DQD}^\mathrm{gen} = \frac{\omega_{q,L}}{2}\sigma_z^L + \frac{\omega_{q,R}}{2}\sigma_z^R + \frac{1}{4}\bm{\sigma}_L^T \mathcal{J}\bm{\sigma}_R,
\label{eq:HDQD-general}
\end{equation}
where $\omega_{q,L}$ and $\omega_{q,R}$ are the per-dot qubit splittings of the left and right dots, each fixed by the ellipticity of that dot through \cref{eq:larmor}, and $\mathcal{J}$ is a $3\times 3$ anisotropic exchange tensor. It is convenient to characterize the pair of splittings by their common-mode average and difference,
\begin{equation}
\bar\omega_q = \frac{\omega_{q,L}+\omega_{q,R}}{2},\qquad
\Delta\omega_q = \omega_{q,L}-\omega_{q,R},
\label{eq:omega-sumdiff}
\end{equation}
both indicated in \cref{fig:3}. The gradient $\Delta\omega_q$ is the differential frequency between the dots: it vanishes when the two are identically shaped and grows with their ellipticity difference. The average $\bar\omega_q$ is the common splitting of the pair and, as the two-electron spectrum below makes explicit, equals the gap that separates the spin-polarized triplets from the unpolarized states. Because each per-dot splitting is fixed by the shape of its own dot, $\bar\omega_q$ and $\Delta\omega_q$ are independent, purely electrical knobs.

As derived in Appendix~\ref{sec:anisotropic-exchange}, $\mathcal{J}$ receives two distinct contributions: the altermagnetic spin-dependence of the squeezed orbitals renders it diagonally anisotropic, such that even without SOC, the exchange is anisotropic $\mathcal{J}_0 = \mathrm{diag}(J_\perp,J_\perp,J_z)$, while the Rashba spin--orbit coupling rotates it as $\mathcal{J} = \mathcal{J}_0\mathcal{R}(2\theta_\mathrm{so},\hat{y})$ for dots separated by $d$ along $\hat{\mathbf{x}}$, where $\mathcal{R}(\theta,\hat{\mathbf{n}})$ denotes the $3\times3$ matrix that rotates a vector by angle $\theta$ about the axis $\hat{\mathbf{n}}$, with $\theta_\mathrm{so} = d/\ell_\mathrm{so}$.

Despite the naturally complicated exchange structure, we find that the deviations from a scalar exchange coupling are negligible in the $(1,1)$ configuration considered in this manuscript, as discussed in Appendix~\ref{sec:anisotropic-exchange}. The squeezing-induced anisotropy is parametrically small in the charging-energy-dominated regime characteristic of gate-defined dots, with $J_z-J_\perp \propto \tau_c^2(m\eta)^2/E_C$, where $\tau_c$ is the inter-dot tunnel coupling and $E_C$ the charging energy, so that $J_z\simeq J_\perp\equiv J$ (Appendix~\ref{sec:spin-dependent-exchange}). The SOC rotation $\mathcal{R}(2\theta_\mathrm{so},\hat{y})$ generates only couplings between two-spin states of different total spin projection along $\hat{z}$; these are off-resonant by the polarized-triplet gap $|\bar\omega_q|$ in both control schemes below and are dropped at leading order by the rotating-wave approximation (Appendix~\ref{app:soc-exchange}). Within these approximations we recover the scalar-exchange Hamiltonian
\begin{equation}
H_\mathrm{DQD} = \frac{J}{4}\!\left(\boldsymbol{\sigma}_L\cdot\boldsymbol{\sigma}_R-\sigma_L^0\sigma_R^0\right) + \frac{\omega_{q,L}}{2}\sigma_z^L + \frac{\omega_{q,R}}{2}\sigma_z^R,
\label{eq:HDQD}
\end{equation}
which we use throughout the rest of this section. Written out in the two-electron basis $\{\ket{\uparrow\uparrow},\ket{\uparrow\downarrow},\ket{\downarrow\uparrow},\ket{\downarrow\downarrow}\}$, and using the average and gradient of \cref{eq:omega-sumdiff}, \cref{eq:HDQD} reads
\begin{equation}
H_\mathrm{DQD} = \begin{pmatrix}
\bar\omega_q & 0 & 0 & 0 \\
0 & \dfrac{\Delta\omega_q}{2}-\dfrac{J}{2} & \dfrac{J}{2} & 0 \\[4pt]
0 & \dfrac{J}{2} & -\dfrac{\Delta\omega_q}{2}-\dfrac{J}{2} & 0 \\[4pt]
0 & 0 & 0 & -\bar\omega_q
\end{pmatrix}.
\label{eq:HDQD-matrix}
\end{equation}
In this form, the polarized triplets $\ket{\uparrow\uparrow}$ and $\ket{\downarrow\downarrow}$ are split off by $\pm\bar\omega_q$ and decouple, while the exchange $J$ and the gradient $\Delta\omega_q$ act entirely within the inner $\{\ket{\uparrow\downarrow},\ket{\downarrow\uparrow}\}$ doublet: the structure that the resonant exchange drive of \cref{subsec:fsim} exploits.

Three calibrated electrical knobs, $\bar\omega_q$, $\Delta\omega_q$, and the barrier-controlled exchange $J$, span the control space of the double dot in the isotropic-exchange limit. The exchange $J$ in particular is set by the inter-dot barrier gate through the tunnel coupling $\tau_c$ (Appendix~\ref{sec:spin-dependent-exchange}), and can therefore be pulsed and modulated in time, providing the dynamical control exploited by the gate protocols below. As we show, these knobs support both the implementation of fSim gates between Loss--DiVincenzo qubits and a singlet--triplet encoding within the $(1,1)$ charge configuration of a single double-dot pair.

\subsection{Two-qubit gates between Loss--DiVincenzo qubits}
\label{subsec:fsim}
Among these three knobs, the interplay between $J$ and $\Delta\omega_q$ generates two-qubit entanglement, while the third, $\bar\omega_q$, sets the local phase frame, as we now show. For resonant splittings, $\Delta\omega_q=0$, pulsing $J$ for a controlled duration implements the $\sqrt{\text{iSWAP}}$ gate, which can be compiled into a CNOT following the original Loss--DiVincenzo prescription~\cite{Loss_1998_Quantum}. More generally, the simultaneous presence of $J$ and $\Delta\omega_q$ natively enable the fSim gates~\cite{Kivlichan_2018_Quantum}, a two-parameter family $\text{fSim}(\theta,\phi)$ that combines iSWAP-like exchange rotations with a controlled-phase component, simplifying quantum simulation with linear circuit depth.

To go beyond $\sqrt{\text{iSWAP}}$ and access the full fSim family, we modulate the exchange resonantly with the splitting gradient, which selectively activates the inner doublet while leaving the polarized triplets idle~\cite{Yu_2026_Robust}. Concretely, we set
\begin{equation}
    J(t) = J_0 + 2j(t)\cos(|\Delta\omega_q|\,t),
    \label{eq:Jdrive}
\end{equation}
where $J_0$ is a static exchange working point and $j(t)$ is a slowly varying envelope. Moving to the rotating frame defined by $U_R(t)=\exp\bigl[-i\tfrac{\Delta\omega_q\,t}{4}(\sigma_z^L-\sigma_z^R)\bigr]$ and applying the rotating-wave approximation (RWA), we discard the off-resonant flip-flop terms generated by $J_0$, the counter-rotating terms generated by the resonant modulation, and the SOC-induced couplings to the polarized triplets. The two-qubit Hamiltonian in the basis $\{\ket{\uparrow\uparrow},\ket{\uparrow\downarrow},\ket{\downarrow\uparrow},\ket{\downarrow\downarrow}\}$ then takes the block-diagonal form (setting the $\ket{\uparrow\uparrow}$ energy to zero),
\begin{equation}
    \tilde{H} = \begin{pmatrix}
    0 & 0 & 0 & 0 \\
    0 & -J(t)/2 - \bar{\omega}_q & j(t)/2 & 0 \\
    0 & j(t)/2 & -J(t)/2 - \bar{\omega}_q & 0 \\
    0 & 0 & 0 & -2\bar{\omega}_q
    \end{pmatrix}.
    \label{eq:Hrwa}
\end{equation}
The resonant component of the exchange rotates the inner doublet $\{\ket{\uparrow\downarrow},\ket{\downarrow\uparrow}\}$ with amplitude $j(t)/2$, while the full exchange waveform contributes to its accumulated phase. The polarized triplets remain dynamically decoupled at this order.

The fSim gate is defined in the computational basis as
\begin{equation}
    \text{fSim}(\theta,\phi) = \begin{pmatrix}
    1 & 0 & 0 & 0 \\
    0 & \cos\theta & -i\sin\theta & 0 \\
    0 & -i\sin\theta & \cos\theta & 0 \\
    0 & 0 & 0 & e^{i\phi}
    \end{pmatrix},
    \label{eq:fsimdef}
\end{equation}
combining an iSWAP-like rotation of angle $\theta$ in the $\{\ket{\uparrow\downarrow},\ket{\downarrow\uparrow}\}$ subspace with a controlled phase $\phi$ on the $\ket{\downarrow\downarrow}$ state. Up to calibrated single-qubit $Z$ rotations, the evolution generated by \cref{eq:Hrwa} is an fSim gate with
\begin{subequations}\label{eq:fsimcond}
\begin{align}
    \theta &= \tfrac{1}{2}\int_0^T j(t)\,dt,
    \label{eq:fsim_swap}\\
    \phi &= -\int_0^T J(t)\,dt \pmod{2\pi}.
    \label{eq:fsim_cphase}
\end{align}
\end{subequations}
Here $T$ is the gate time. The first condition fixes the iSWAP-like angle through the area of the resonant exchange envelope, while the second fixes the nonlocal controlled phase through the total exchange area. The inner doublet acquires the common phase
\begin{equation}
\bar{\omega}_q T+\frac{1}{2}\int_0^T J(t)\,dt,
\end{equation}
whereas the $\ket{\downarrow\downarrow}$ state acquires the phase $2\bar{\omega}_qT$. These phases, together with those introduced by the rotating-frame transformation, correspond to calibrated single-qubit $Z$ rotations that can be performed locally. Thus $\Delta\omega_q$ sets the resonant modulation frequency, $j(t)$ controls the exchange rotation, the total exchange area controls the nonlocal phase, and $\bar\omega_q$ sets the local phase frame while keeping the polarized triplets spectrally separated, with all quantities controlled electrically through the dot shape and the barrier gate.

While the exchange interaction provides a natural mechanism for nearest-neighbour two-qubit gates, scalable spin-qubit architectures also require mechanisms to connect more distant sites. In gate-defined quantum-dot arrays, this can be pursued by physically moving spin states through the device using spin shuttling~\cite{Mills_2019_Shuttling,Zwerver_2023_Shuttling,vanRiggelenDoelman_2024_Coherent}, or by coupling distant qubits through shared bosonic modes~\cite{Dijkema_2024_Cavity}. In the present platform, the vanishing net magnetization of the altermagnetic host suppresses stray fields and makes this second route particularly appealing, since it is natively compatible with superconducting microwave resonators. This opens a complementary path to long-range qubit interactions via cavity-mediated coupling, in direct analogy with recent demonstrations in gate-defined spin systems~\cite{Dijkema_2024_Cavity}. We return to the dot--cavity coupling in \cref{subsec:cqed-readout} in the context of dispersive readout.

\subsection{Singlet--triplet encoding}
\label{subsec:st}

The double-dot Hamiltonian, \cref{eq:HDQD}, naturally supports an alternative encoding in which a single qubit is stored in the two-electron spin state of the $(1,1)$ charge configuration. In the singlet--triplet (ST) basis $\{\ket{T_+},\ket{T_0},\ket{S},\ket{T_-}\}$ with $\ket{T_+}=\ket{\uparrow\uparrow}$, $\ket{T_-}=\ket{\downarrow\downarrow}$, and $\ket{S},\ket{T_0}=(\ket{\uparrow\downarrow}\mp\ket{\downarrow\uparrow})/\sqrt{2}$, \cref{eq:HDQD} takes the form
\begin{equation}
\label{eq:HDQDST}
\begin{split}
H_\mathrm{DQD} = \mathrm{diag}\bigl(\bar\omega_q,\,0,\,-J,\,-\bar\omega_q\bigr) \\
+ \frac{\Delta\omega_q}{2}\bigl(\ket{T_0}\bra{S}+\mathrm{h.c.}\bigr).
\end{split}
\end{equation}
The polarized triplets are split off by $\bar\omega_q$, while the qubit lives in the $\{\ket{S},\ket{T_0}\}$ doublet, where it satisfies
\begin{equation}
\label{eq:HST}
H_\mathrm{ST} = -\frac{J}{2}\,\tilde\sigma_z + \frac{\Delta\omega_q}{2}\,\tilde\sigma_x ,
\end{equation}
with $\tilde\sigma_z = \ket{S}\bra{S}-\ket{T_0}\bra{T_0}$ and $\tilde\sigma_x = \ket{T_0}\bra{S}+\mathrm{h.c.}$. We note the effect of SOC has been neglected since the spin--orbit field lies in the plane perpendicular to the qubit-subspace quantization axis $\hat{z}$, and has no matrix element within $\{\ket{S},\ket{T_0}\}$. Thus, SOC only renormalizes the coefficients without changing the structure of $H_\mathrm{ST}$ (Appendix~\ref{app:soc-exchange}).

The two control axes of the ST Bloch sphere therefore correspond to the two electrically tunable knobs of the platform: $J$, set by the inter-dot barrier gate, and $\Delta\omega_q$, set by the \emph{difference} between the ellipticities of the two dots through \cref{eq:larmor}. This is in striking contrast with conventional ST implementations, where the second axis is supplied by an external Zeeman gradient $\Delta B_z$ engineered with on-chip micromagnets~\cite{PioroLadriere_2008_Electrically,Wu_2014_Two}, built up dynamically through nuclear polarization gradients~\cite{Foletti_2009_Universal}, or set by inter-dot $g$-factor differences~\cite{Jirovec_2021_Singlet}. In the altermagnetic platform, $\Delta\omega_q$ is generated by the same physics that produces the single-qubit splittings, $d$-wave spin splitting projected onto an anisotropic confinement, and is thus intrinsically electrical, locally addressable, and reversible on gate-pulse timescales.

By construction, $\Delta\omega_q$ vanishes at geometric symmetry between the two dots, $\ell_{x,L}=\ell_{x,R}$ and $\ell_{y,L}=\ell_{y,R}$, and grows linearly away from this point with a slope set by $\eta$. The exchange axis inherits the standard symmetric-operation-point sweet spot deep in the $(1,1)$ charge configuration~\cite{Reed_2016_Reduced,Martins_2016_Noise}, where $J$ is first-order insensitive to detuning fluctuations through a mechanism that is independent of the altermagnetic ingredients.

A distinguishing feature of the altermagnetic ST encoding is that the polarized-triplet gap $|\bar\omega_q|=|\omega_{q,L}+\omega_{q,R}|/2$ is itself a gate-tunable quantity, set by the symmetric (common-mode) ellipticity of the two dots. In conventional ST devices this gap is fixed by the external Zeeman field; here it can be raised in operation to push $\ket{T_\pm}$ off resonance with any drive used to manipulate $\{\ket{S},\ket{T_0}\}$, and adjusted to a chosen value during idle or readout (see \cref{sec:readout}).

In summary, the altermagnetic double dot realizes a ST qubit whose two control axes, $J$ and $\Delta\omega_q$, are both purely electrical and locally tunable per qubit, eliminating the principal hardware overhead of conventional ST devices (micromagnets, dynamical nuclear polarization) while inheriting standard charge-sensing readout and the symmetric-operation point. 

\section{Noise sensitivity}
\label{sec:noise}

\begin{figure*}[t]
    \includegraphics[width=\linewidth]{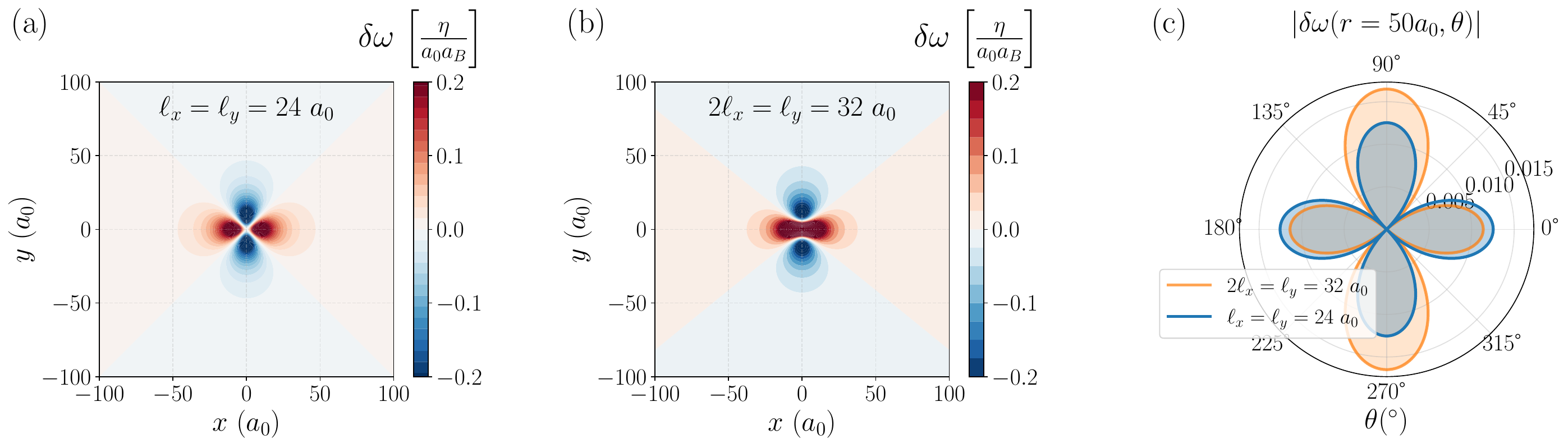}
    \caption{Anisotropic sensitivity to second-order charge noise. (a) Change in the qubit frequency $\delta\omega_q$ due to a Coulomb charge trap positioned at $\mathbf{r}'=(x',y',z')$ with respect to the QD center for an isotropic QD with $z'=10a_0$, $\ell_x=\ell_y=24a_0$. (b) Same as (a) but with $2\ell_x=\ell_y=32a_0$. (c) Polar plot of $|\delta\omega_q|$ as a function of the in-plane angle $\theta$, measured from the $y'$ axis so that $\theta=0^\circ$ corresponds to $x'=0$, with $r'=50\,a_0$ for the cases in (a) blue and (b) orange; values are reported in units of $\eta/(a_0a_B)$.}
    \label{fig:4}
\end{figure*}

As seen from \cref{eq:larmor}, the qubit splitting $\omega_q$ depends on the altermagnetic parameter $\eta$, set by the N\'eel order and hopping anisotropy, and on the confinement anisotropy $(1/\ell_y^2-1/\ell_x^2)$. Noise can therefore enter through fluctuations of $\eta$ via changes in the N\'eel vector, or of the confinement lengths via charge noise.

\subsection{N\'eel-vector fluctuations}

Writing the N\'eel vector as $\mathbf N(\mathbf r)=[N_0+\delta N(\mathbf r)]n(\mathbf r)$, fluctuations of the amplitude $\delta N$ are energetically costly in the ordered phase. The order parameter $N_0$ is robust against $\delta N(\mathbf r)$ fluctuations and its magnitude can, therefore, be treated as approximately fixed for small fluctuations. The softer low-energy degrees of freedom are rotations of the unit vector $\mathbf n(\mathbf r)$, which can be distorted into spatial textures by disorder, strain, gate-induced inhomogeneity, or substrate mismatch.

In the regime of a single-domain collinear background with $\bm n=\hat z+\delta\bm n_\perp$, $\delta\bm n_\perp=(\delta n_x,\delta n_y,0)$, and smooth in-plane fluctuations, we align the spin quantization axis with the instantaneous N\'eel direction through a local spin rotation. The gradients and time derivative of this rotation are what couple to the qubit, with two properties that control the response (see Appendix~\ref{app:neel} for the full derivation): the spatial part vanishes at leading order by parity of the orbital ground state, while the temporal part is purely transverse on the qubit subspace. Projecting onto the qubit subspace then yields
\begin{equation}
\label{eq:main-text-neel-noise}
\delta H_q
=
\frac{\Lambda}{2}
\Bigl[
\dot{n}_y\,\sigma_x
-
\dot{n}_x\,\sigma_y
\Bigr]
+
\mathcal{O}(\ell^2_i/L_m^2,\delta n^2),
\end{equation}
where $\Lambda\equiv\braket{\phi_+|\phi_-}$ is the orbital overlap and $L_m$ the magnetic-texture length scale. Two features of \cref{eq:main-text-neel-noise} are immediately useful. 
First, as direct consequence of the rigidity of the N\'eel amplitude $N$ combined with the transverse character of the temporal coupling, no $\sigma_z$ term arises so smooth N\'eel fluctuations cannot generate pure dephasing.
Second, the qubit couples to the time derivative $\dot{\bm n}_\perp$ rather than $\bm n_\perp$ itself, so low-frequency $1/f$ magnetic noise is further suppressed by the frequency weighting $\omega^2 S_n(\omega)$. The dominant channel is therefore relaxation, scaling as
\begin{equation}
\label{eq:main-text-t1-scaling}
T_1^{-1}
=
\frac{\Lambda^2\omega_q^2}{4}
\Bigl[
\mathcal S_{n_x n_x}(|\omega_q|)
+
\mathcal S_{n_y n_y}(|\omega_q|)
\Bigr],
\end{equation}
with $\mathcal S_{AB}(\omega)$ the noise spectral density of the dot-averaged N\'eel fluctuation. To connect this scaling to material properties, we parametrize the long-wavelength transverse N\'eel dynamics by damped antiferromagnetic-resonance modes. 

This is the natural low-energy description of a collinear compensated magnet and is consistent with recent measurements in altermagnetic $\alpha$-MnTe, where split altermagnetic magnons and a THz antiferromagnetic-resonance mode have been reported~\cite{Liu_2024_Chiral, Dzian_2025_Antiferromagnetic}. For the out-of-plane easy-axis configuration assumed here, the two transverse Néel modes are expected to be gapped by the anisotropy that pins $\mathbf n \parallel \hat z$. The size of these gaps is nevertheless material- and device-dependent, and must exceed the intended qubit-frequency window to suppress real magnon emission. 

Weighting these modes by the dot form factor and applying the fluctuation-dissipation theorem expresses $1/T_1$ in terms of material parameters namely the transverse gaps $\Delta_a$, spin-wave velocities, damping, and spectral weights (Appendix~\ref{app:neel}). This material-level estimate separates the symmetry protection from the material constraint. If the qubit frequency lies below the relevant transverse gaps, $|\omega_q|<\Delta_a$, relaxation is controlled only by the damping tail of the magnon susceptibility. If a nearly gapless easy-plane mode remains, real low-energy magnon emission can contribute, although the time-derivative coupling still suppresses the rate at low frequency. The most favorable regime is therefore a stiff, anisotropy-pinned or strain-pinned altermagnet whose transverse modes coupled to the dot lie above the electrically tunable qubit-frequency window.

The ST encoding introduced in \cref{subsec:st} is inherently insensitive to global magnetic fluctuations since the ST splitting is set by $\Delta\omega_q$ rather than the average $\bar\omega_q$. A N\'eel fluctuation $\delta\bm n_\perp(\bm r,t)$ would only introduce effective corrections to the anticrossings with the polarized triplets, leaving the qubit subspace $\{\ket{S},\ket{T_0}\}$ untouched at linear order.

The effect of N\'eel fluctuations on altermagnetic spin qubits contrasts sharply with that in conventional spin qubits, where strain also modifies the quantization axis and, importantly, the \emph{amplitude} of the Zeeman splitting (e.g.\ through $g$-factor modulation~\cite{AbadilloUriel_2023_Hole}), opening a pure dephasing channel~\cite{Piot_2022_Single, Bassi_2025_Optimal}. Processes beyond the smooth-texture regime, such as sharp textures and domain-wall motion, are discussed alongside mitigation strategies in \cref{sec:materials}.

\subsection{Charge noise}

For charge noise~\cite{Paladino_2014_1} at the single-quantum-dot level, any uniform electric-field fluctuation $\delta F_i$ rigidly displaces the harmonic wavefunction, $x_i\rightarrow x_i+eF_i/(m\omega_i^2)$, without modifying the harmonic lengths. The qubit frequency $\omega_q$ therefore resides at a first-order charge-noise sweet spot by construction. In conventional spin qubits such sweet spots require careful tuning~\cite{Medford_2013_Quantum, Piot_2022_Single, Hendrickx_2024_Sweet, Bassi_2025_Optimal, Ungerer_2025_Dephasing}; here it is intrinsic to the $d$-wave splitting. Neither the Rashba interaction nor N\'eel-vector rotations compromise the sweet spot, since both couple transversely and leave the qubit frequency unchanged to first order.

The leading charge-noise sensitivity therefore arises at second order. Modeling individual charge-noise fluctuators as Coulomb impurities~\cite{Connors_2022_Charge}
\begin{equation}
    V^{(i)}_C(x,y)=\frac{e^2}{4\pi\epsilon\sqrt{(x-x')^2+(y-y')^2+z'^2}},
    \label{eq:coulomb}
\end{equation}
with $(x',y',z')$ the impurity position relative to the dot center and $\epsilon$ the host dielectric constant, and expanding to second order for $z'$ comparable to or larger than the dot size, only the quadratic terms renormalize the confinement anisotropy. With $r'=\sqrt{x'^2+y'^2+z'^2}$, the correction reads
\begin{equation}
    \delta\omega_q \simeq -\frac{\eta}{a_0a_B}\,\mathcal{F}(x',y',z',\ell_x,\ell_y),
    \label{eq:deltaomega}
\end{equation}
\begin{equation}
    \mathcal{F}=a_0\frac{\ell_x^2(3x'^2-r'^2)-\ell_y^2(3y'^2-r'^2)}{r'^5},
    \label{eq:formfactor}
\end{equation}
where $a_B=4\pi\epsilon/(m e^2)$ is the effective Bohr radius. The result factorizes into a material prefactor $\eta/(a_0 a_B)$ and a geometric form factor $\mathcal{F}$ encoding the anisotropic noise sensitivity. The complete derivation is given in Appendix~\ref{app:coulomb}.

\Cref{fig:4} shows $\delta\omega_q$ as a function of the in-plane fluctuator position at fixed $z'=10 a_0$. A circular dot (\cref{fig:4}a) displays the $d$-wave angular dependence with nodal directions along the diagonals, where the qubit is insensitive to charge noise even to second order. Breaking the dot symmetry (\cref{fig:4}b) reshapes the lobes and shifts the nodes, as made explicit in the polar plot of \cref{fig:4}c. This tunability allows the dephasing rate to be minimized by choosing the dot ellipticity to suppress the dominant noise sources.

In contrast, for the ST qubit encoding, the exchange interaction is first-order sensitive to electric fields. Therefore, the linear corrections that are ineffective for the altermagnetic Loss--DiVincenzo spin qubit become relevant for the ST qubit. It is always possible, however, to work at the symmetric-operation point~\cite{Reed_2016_Reduced,Martins_2016_Noise}, which is a sweet spot for the $J$ axis, so that the altermagnetic ST qubit naturally resides at a sweet spot. Away from this $J$ operating point, however, gate-voltage fluctuations on the inter-dot barrier couple directly to the ST splitting and become the dominant residual charge-noise channel.

\subsection{Hyperfine noise}

Finally, hyperfine coupling to host nuclear spins is a dominant dephasing mechanism in GaAs and natural-Si/Ge spin qubits~\cite{Reilly_2008_Suppressing, Bluhm_2010_Dephasing, RojasArias_2025_Origins}, as it introduces a noisy effective splitting via the Overhauser field. The ferromagnetic ordering within each sublattice tends to lock the nuclear spins along the local magnetization via the hyperfine interaction, potentially reducing the fluctuations of the Overhauser field. Even if this mechanism is not fully suppressed, several predicted $d$-wave altermagnets, including \ce{NdRuO3}, \ce{Ca3Cr2O7}, \ce{ZrCrO3}, and \ce{RuO2}~\cite{Gao_2025_AI}, are composed of elements whose dominant natural isotopes carry zero nuclear spin, suggesting that hyperfine noise may be strongly suppressed in these hosts. In terms of this noise source, the ST encoding would be a better choice than the single-spin Loss--DiVincenzo qubit since it is encoded in a zero-spin-projection subspace. 

\begin{figure*}[t]
    \includegraphics[width=1\linewidth]{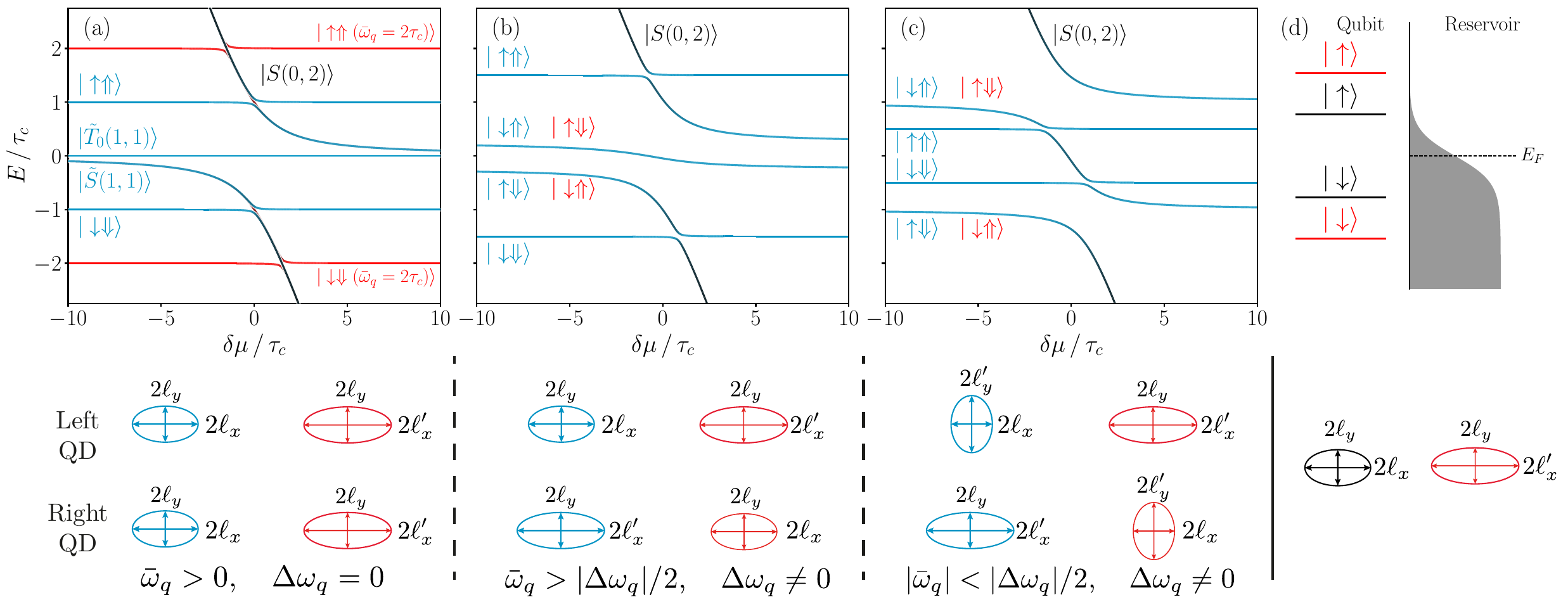}
    \caption{%
        Readout mechanisms for altermagnetic spin qubits.
        (a)-(c) Five-level spectra near the $(1,1)$--$(0,2)$ transition [\cref{eq:Hpsb}] versus detuning $\delta\mu$, for $\theta_\mathrm{so}=5^\circ$, $\beta=5^\circ$; energies and detuning are in units of $\tau_c$.
        Blue denotes $(1,1)$ charge character; increasing black weight denotes growing $\ket{S(0,2)}$ occupation.
        The narrow first arrow labels the data spin in the left dot; the wide second arrow labels the ancilla in the right dot.
        (a) For $\Delta\omega_q=0$, $\ket{\tilde S(1,1)}$ transfers to $\ket{S(0,2)}$ while the triplets are blockaded, apart from weak SOC anticrossings with $\ket{T_\pm}$. Red levels show the larger splitting $\bar\omega_q=2\tau_c$, which moves these anticrossings away from the transfer region.
        (b) For $\bar\omega_q>|\Delta\omega_q|/2$, a finite gradient produces product-state eigenstates and a finite single-spin readout window. Blue and red labels correspond to $\Delta\omega_q<0$ and $\Delta\omega_q>0$, respectively; reversing the gradient exchanges the branch labels without changing the eigenenergies.
        (c) For $|\Delta\omega_q|/2>\bar\omega_q$, the unpolarized product states become the outermost $(1,1)$ levels. With the ancilla chosen consistently with the gradient sign [\eqref{eq:Hpsb}], one data state ends in $(0,2)$ and the other in $(1,1)$, robustly against the weak SOC anticrossings.
        (d) For Elzerman readout the Fermi level $E_F$ is aligned so that $E_{\downarrow}<E_F<E_{\uparrow}$; the splitting $\omega_q$ (enhanced via dot ellipticity, red)
        sets the energy-selectivity window $\omega_q/k_{\mathrm{B}}T$ before the
        measurement event.
    }
    \label{fig:readout}
\end{figure*}

\section{Readout}
\label{sec:readout}

\subsection{Spin-to-charge readout: Pauli blockade and Elzerman protocols}
Standard spin-qubit readout techniques carry over directly to the altermagnetic platform. The two encodings introduced in \cref{sec:dqd} naturally pair with complementary readout mechanisms, the Pauli spin blockade for the ST encoding, and energy-selective Elzerman readout for the Loss--DiVincenzo encoding, but, as we discuss below, both mechanisms can be adapted to either encoding with minor modifications. Both reduce the spin measurement to a charge measurement, which is then performed with established rf charge sensors.

We begin with Pauli spin blockade, which turns a spin measurement into a charge measurement near the $(1,1)$--$(0,2)$ transition. The mechanism relies on a single exclusion rule: two electrons can crowd into the same dot, the doubly occupied $(0,2)$ configuration, only if they have opposite spins, since the shared orbital ground state is spatially symmetric. During readout the interdot detuning $\delta\mu$ (the relative chemical potential of the two dots) is swept from the $(1,1)$ sector toward positive detuning, where $\ket{S(0,2)}$ is the only accessible doubly occupied state. The singlet $(1,1)$ state can then move one electron and adiabatically become $(0,2)$ singlet state, whereas the triplet $(1,1)$ state is forbidden from doing so and remains in $(1,1)$; an rf charge sensor or reflectometry on the plunger-gate tank circuit distinguishes the two charge configurations. We track this conversion by color in \cref{fig:readout}(a)-(c): blue marks $(1,1)$ charge character and increasing black weight marks growing $\ket{S(0,2)}$ occupation. Following a labeled eigenstate from the negative-detuning side as $\delta\mu$ is swept, a branch that turns black is unblocked [charge transfers to $(0,2)$], while one that stays blue is blockaded in $(1,1)$. Throughout, we use the shorthand $\ket{\uparrow\Downarrow}$: the narrow arrow ($\uparrow$, $\downarrow$) is the data spin in the left dot, the readout target, and the wide arrow ($\Uparrow$, $\Downarrow$) the ancilla spin in the right dot, which we assume initialized to $\Downarrow$ before readout.

Close to the readout transition, the natural exchange anisotropy of the altermagnetic system introduces a correction to the spin-conserving tunneling by making it spin-dependent, as discussed in Appendix~\ref{sec:spin-dependent-exchange}. We introduce $\tau_\uparrow$ and $\tau_\downarrow$ as the tunnel couplings for each spin state, and the average and difference tunnel couplings as 
\begin{equation}
\label{eq:tunnel-symmetric-asymmetric-main}
\tau_c
=
\frac{\tau_\uparrow+\tau_\downarrow}{2},
\qquad
\delta\tau
=
\frac{\tau_\uparrow-\tau_\downarrow}{2}.
\end{equation}
This altermagnetic correction to the tunnel coupling can weakly couple $\ket{S(0,2)}$ to $\ket{T_0(1,1)}$. This effect is most simply absorbed into a rotation of the unpolarized $(1,1)$ ST states,
\begin{equation}
    \begin{aligned}
        \ket{\tilde S(1,1)}&=\cos\beta\,\ket{S(1,1)}+\sin\beta\,\ket{T_0(1,1)},\\
\ket{\tilde T_0(1,1)}&=-\sin\beta\,\ket{S(1,1)}+\cos\beta\,\ket{T_0(1,1)},
    \end{aligned}
\end{equation}
where, microscopically,  $\tan\beta\simeq\delta\tau/\tau_c$.
The five-level Hamiltonian in the basis $\{\ket{S(0,2)},\ket{\tilde S(1,1)},\ket{T_+},\ket{\tilde T_0},\ket{T_-}\}$ is
\begin{equation}
H_{\rm PSB}=
\begin{pmatrix}
-\delta\mu & \tau_{\rm sc} & \tau_{\rm sf} & 0 & \tau_{\rm sf}\\
\tau_{\rm sc} & \tfrac{\Delta\omega_q}{2}\sin 2\beta & 0 & \tfrac{\Delta\omega_q}{2}\cos 2\beta & 0\\
\tau_{\rm sf} & 0 & \bar\omega_q & 0 & 0\\
0 & \tfrac{\Delta\omega_q}{2}\cos 2\beta & 0 & -\tfrac{\Delta\omega_q}{2}\sin 2\beta & 0\\
\tau_{\rm sf} & 0 & 0 & 0 & -\bar\omega_q
\end{pmatrix}.
\label{eq:Hpsb}
\end{equation}
Each diagonal entry corresponds to the energy of the states for large $\delta \mu$ values, while the off-diagonal terms correspond to the avoided crossings between the states in \cref{fig:readout}(a)-(c). 

Increasing $\delta\mu$ lowers the energy of $\ket{S(0,2)}$, producing a descending black branch, Fig.~\ref{fig:readout}. The spin-conserving tunneling, $\tau_\mathrm{sc}=\tau_c\cos\theta_\mathrm{so}$, couples $\ket{\tilde S(1,1)}$ to $\ket{S(0,2)}$ and opens the large anticrossing shown in Fig.~\ref{fig:readout}(a). The smaller SOC-induced amplitude, $\tau_\mathrm{sf}=\tau_c\sin\theta_\mathrm{so}$, couples $\ket{S(0,2)}$ to $\ket{T_\pm}$ but not to $\ket{\tilde T_0}$~\cite{Sen_2023_Classification}, opening the weak anticrossings that are responsible for leakage. The gradient $\Delta\omega_q$ mixes $\ket{\tilde S(1,1)}$ and $\ket{\tilde T_0}$, while $\bar\omega_q$ sets the energies $\pm\bar\omega_q$ of the polarized triplets $\ket{T_\pm}$. The readout pulse can be engineered to exploit this separation of energy scales: sufficiently slow to adiabatically traverse the $\tau_\mathrm{sc}$ anticrossing, but sufficiently fast to cross the gap due to $\tau_\mathrm{sf}$, so population is not lost to the triplets.

Consider first two equally elongated dots, so the two Larmor splittings are equal and the gradient vanishes, $\Delta\omega_q=0$ [\cref{fig:readout}(a)]. We relax below this condition. The relevant unpolarized states are then the bright/dark pair $\ket{\tilde S}$ and $\ket{\tilde T_0}$, which are adiabatically connected to the usual $\ket{S}$ and $\ket{T_0}$ states as the spin-dependent tunneling is turned off in the (1,1) configuration. As $\delta\mu$ increases, only $\ket{\tilde S(1,1)}$ connects to $\ket{S(0,2)}$ through the strong $\tau_{\rm sc}$ anticrossing (its branch turns black) while $\ket{\tilde T_0}$ and the polarized triplets stay blockaded [blue color in \cref{fig:readout}(a)]. This singlet-versus-triplet contrast is precisely what reads out the ST qubit. The one imperfection is leakage: on its way down, the singlet branch meets the weak SOC anticrossings with $\ket{T_\pm}$, which can bleed population out of the transferred state and reduce the contrast~\cite{Jirovec_2021_Singlet, Hendrickx_2024_Sweet}. Since $\bar\omega_q$ sets the energies $\pm\bar\omega_q$ of $\ket{T_\pm}$, elongating both dots to raise $\bar\omega_q$ slides these anticrossings (red levels, $\bar\omega_q=2\tau_c$) away from the singlet-transfer region, cleanly separating the leakage channels from the readout interval. This is a knob freely available in the altermagnetic platform through dot ellipticity (lower part of the panel), whereas in conventional devices the same gap is fixed by the external field.

Making the QDs unequally elongated turns on a finite gradient, $\Delta\omega_q\neq0$, which splits the two unpolarized $(1,1)$ states into the product states $\ket{\uparrow\Downarrow}$ and $\ket{\downarrow\Uparrow}$. With the ancilla fixed, the data spin can now be read out on its own [\cref{fig:readout}(b)]. Take $\Delta\omega_q<0$ (blue labels) and a $\Downarrow$ ancilla. The data-$\uparrow$ state $\ket{\uparrow\Downarrow}$ is the lower of the two unpolarized branches; it acquires $\ket{S(0,2)}$ character and turns black (unblocked, $\to(0,2)$). The data-$\downarrow$ state is $\ket{\downarrow\Downarrow}=\ket{T_-}$, which stays blue and blockaded ($\to(1,1)$). The two spin outcomes thus map onto two charge states. The complication is that the unblocked branch still passes through the weak SOC anticrossing, which opens an error channel: the readout window is finite, since the sweep must be slow enough to follow the strong $\tau_\mathrm{sc}$ anticrossing yet fast enough to cross the SOC gap diabatically. Reversing the gradient to $\Delta\omega_q>0$ (red labels) leaves the eigenenergies untouched but swaps the branch identities, so that $\ket{\uparrow\Downarrow}$ becomes the second excited branch, which connects adiabatically back to a $(1,1)$ configuration and yields low contrast unless the ancilla is re-prepared to $\ket{\Uparrow}$ instead.

The cleanest operating point is reached when the gradient dominates the average splitting, $|\Delta\omega_q|/2>\bar\omega_q$ [\cref{fig:readout}(c)]. The two unpolarized product states $\ket{\uparrow\Downarrow}$ and $\ket{\downarrow\Uparrow}$ are then pushed all the way out to become the \emph{outermost} $(1,1)$ levels, with the polarized triplets sandwiched between them: an inverted Pauli-blockade ordering of the kind observed in hole-spin qubits~\cite{Hendrickx_2021_Four}. For $\Delta\omega_q<0$, $\ket{\uparrow\Downarrow}$ is now the lowest $(1,1)$ level and connects directly to the descending $\ket{S(0,2)}$ branch: it turns black and stays black out to large positive detuning, while $\ket{\downarrow\Downarrow}=\ket{T_-}$ remains blue. The decisive advantage is that the weak SOC anticrossings are encountered \emph{earlier}, i.e., for smaller $\delta \mu$ values, before the final strong anticrossing; once the electron has transferred, no lower $(1,1)$ branch is left to pull the $\ket{S(0,2)}$ character back out. The charge assignment is therefore robust against SOC, and the finite-window constraint of panel (b) disappears. For $\Delta\omega_q>0$ one simply chooses the $\ket{\Uparrow}$ ancilla to recover the same high contrast. Reaching the inverted ordering requires the two dot splittings to have opposite signs, realized by elongating the dots along orthogonal axes (lower part of the panel).

Alternatively, energy-selective Elzerman readout~\cite{Elzerman_2004_Single} converts the spin into a transient charge signal by aligning the dot's chemical potential between the two qubit levels. With the Fermi level $E_F$ of a nearby reservoir set so that $E_\downarrow<E_F<E_\uparrow$ [\cref{fig:readout}(d)], only the $\ket{\uparrow}$ electron is energetically allowed to tunnel out of the dot, after which a $\ket{\downarrow}$ electron tunnels back in to refill it. A charge sensor monitoring the dot occupation therefore detects a brief excursion to $N{-}1$ electrons if and only if the qubit was in $\ket{\uparrow}$. The energy-selectivity window is set by the ratio $\omega_q/k_B T$, with $T$ the electron temperature of the reservoir; high-fidelity readout requires $\omega_q$ to be substantially larger than $k_B T$. This requirement can be naturally controlled in the altermagnetic platform: $\omega_q$ is set in operation by the QD ellipticity through \cref{eq:larmor}, so it can be pulsed to a large value before measurement to maximize the energy selectivity (red doublet, \cref{fig:readout}(d)), held there during the tunneling event, and returned to the operating point afterwards. This mechanism is not as sensitive to spin--orbit coupling. The same trick, applied to ST qubits through $\bar\omega_q$, raises the polarized-triplet gap during readout in the two-qubit case. 

\subsection{Circuit-QED dispersive readout}
\label{subsec:cqed-readout}
Beyond the spin-to-charge mechanisms above, the altermagnetic  compatibility with superconducting microwave resonators due to the zero-net magnetization enables a complementary readout channel for the single-spin (Loss--DiVincenzo) qubit via circuit-QED dispersive detection~\cite{Blais_2004_Cavity,Blais_2021_Circuit}. We consider a resonator mode of frequency $\omega_c$ with photon ladder operators $a$, $a^\dagger$, whose zero-point electric field at the dot location is polarized along $\hat x$ and couples to the confined electron through the dipole interaction $H_d=-eE_{\mathrm{zpf}}\,x(a+a^\dagger)$, where $E_{\mathrm{zpf}}$ is the zero-point field amplitude. Combined with the dot Hamiltonian of \cref{eq:continuum-hamiltonian}, a second-order projection onto the qubit doublet yields the effective spin--photon Hamiltonian
\begin{equation}
\label{eq:Heff-main}
H_{\mathrm{eff}}
=
\frac{\omega_q}{2}\sigma_z
+
\tilde\omega_c a^\dagger a
+
g_X X\sigma_x
+
g_P P\sigma_y
+
\frac{\chi_{\mathrm{pol}}}{2}X^2\sigma_z,
\end{equation}
with $X=a+a^\dagger$, $P=i(a^\dagger-a)$, and $\tilde\omega_c$ the cavity frequency renormalized by the spin-independent dot polarizability. To leading order in $\eta$, the transverse couplings and the polarizability read
\begin{align}
\label{eq:gX-gY-chipol-main}
g_X&\simeq\frac{\alpha_R\,eE_{\mathrm{zpf}}}{2}\frac{\omega_q}{\omega_x^2-\omega_q^2}, \\
g_P&\simeq-\frac{\alpha_R\,eE_{\mathrm{zpf}}}{2}\frac{\omega_c}{\omega_x^2-\omega_c^2},\\
\chi_{\mathrm{pol}}&\simeq-2(eE_{\mathrm{zpf}})^2\eta\,\frac{\omega_c^2}{(\omega_x^2-\omega_c^2)^2}.
\end{align}
The full derivation is given in Appendix~\ref{app:cqed}. The polarizability $\chi_{\mathrm{pol}}\propto\eta$ is a purely altermagnetic term: it is generated by the $d$-wave $\eta$ term, which makes the QD's $x$-polarizability spin-dependent. It grows with the vacuum field $E_{\mathrm{zpf}}$, the altermagnetic coefficient $\eta$, and the cavity-to-orbital frequency ratio $\omega_c/\omega_x$. The transverse coupling $g_X$, on the other hand, inherits the gate-tunability of the qubit splitting $\omega_q$ through \cref{eq:larmor}: reshaping the dot ellipticity tunes $g_X$ continuously between a leading-order zero and its maximum, allowing the spin--photon coupling to be switched on and off with the same gates that define the qubit, without any external Zeeman field.

\begin{figure}[t]
    \includegraphics[width=1\linewidth]{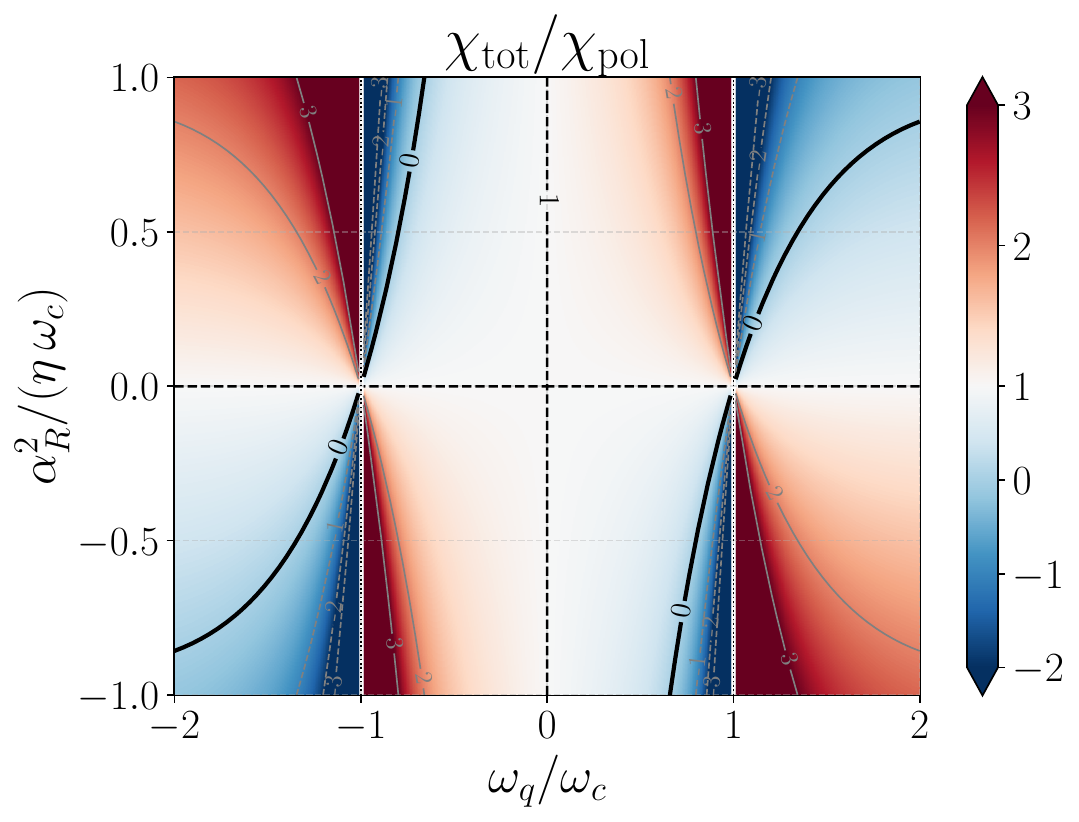}
    \caption{%
        Interplay of the dispersive readout mechanisms.
        Ratio between the total dispersive coupling $\chi_\text{tot}=\chi_\text{pol}+\chi_\text{lin}$ and the polarizability dispersive coupling $\chi_\text{pol}$ as a function of the ratio between qubit frequency and the cavity frequency $\omega_q/\omega_c$ as well as the normalized Rashba coefficient $\alpha_R^2/(\eta\omega_c)$. We mark the cases where the dispersive coupling cancels out, $\chi_\text{tot}=0$ (solid), and where it is contributed by the polarizability term only, $\chi_\text{tot}=\chi_\text{pol}$ (dashed). The plot shows how to choose $\omega_q$ to enhance the dispersive readout mechanism depending on the Rashba and $\eta$ parameters. At $\omega_q\approx\pm\omega_c$, the dispersive approximation breaks down and is masked in the plot.
    }
    \label{fig:dispersive}
\end{figure}

The dispersive regime $|\omega_q\pm\omega_c|\gg g_X,g_P$ is the natural operating point for measurement: a Schrieffer--Wolff transformation eliminates the transverse couplings of \cref{eq:Heff-main} in favor of a dispersive correction $\chi_{\mathrm{lin}}$, and the rotating-wave approximation reduces the polarizability piece $\frac{\chi_{\mathrm{pol}}}{2}X^2\sigma_z$ to $\chi_{\mathrm{pol}}a^\dagger a\sigma_z$ up to a Lamb shift. Absorbing the corresponding qubit Lamb shift into the calibrated qubit frequency, the qubit--cavity dispersive Hamiltonian becomes
\begin{equation}
\label{eq:Hdisp-main}
H_{\mathrm{disp}}
=
\frac{\omega_q}{2}\sigma_z
+
\tilde\omega_c a^\dagger a
+
\chi_{\mathrm{tot}}a^\dagger a\sigma_z,
\end{equation}
with $\chi_{\mathrm{tot}}=\chi_{\mathrm{pol}}+\chi_{\mathrm{lin}}$. The Rashba contribution to the dispersive shift, obtained by eliminating $g_X,g_P$, is (for $\omega_x\gg\omega_c,\omega_q$)
\begin{equation}
\label{eq:chilin-main}
\chi_{\mathrm{lin}}\simeq\frac{(\alpha_R)^2(eE_{\mathrm{zpf}})^2}{2\omega_x^4}\,\frac{\omega_q(\omega_q^2+3\omega_c^2)}{\omega_q^2-\omega_c^2},
\end{equation}
the full expression being given in Appendix~\ref{app:cqed}. Like $\chi_{\mathrm{pol}}$ it scales as $(eE_{\mathrm{zpf}})^2$, but it is controlled by the Rashba coupling $(\alpha_R)^2$ rather than $\eta$ and, crucially, changes sign across $\omega_q=0$ and across the dispersive poles $|\omega_q|=\omega_c$.

Because both shifts carry the same $(eE_{\mathrm{zpf}})^2$ prefactor, their interplay is governed by only two dimensionless ratios: the qubit--cavity detuning $\omega_q/\omega_c$ and the normalized Rashba strength $(\alpha_R)^2/(\eta\omega_c)$:
\begin{equation}
\label{eq:chiratio-main}
\frac{\chi_{\mathrm{tot}}}{\chi_{\mathrm{pol}}}
=1-\frac{\alpha_R^2}{4\eta\omega_c}\,
\frac{(\omega_q/\omega_c)\bigl[(\omega_q/\omega_c)^2+3\bigr]}{(\omega_q/\omega_c)^2-1},
\end{equation}
plotted in \cref{fig:dispersive}; cast through these ratios the comparison is material-agnostic. The polarizability and Rashba mechanisms cooperate ($\chi_{\mathrm{tot}}/\chi_{\mathrm{pol}}>1$, red) or compete ($<1$, blue), and interfere destructively along a cancellation locus $\chi_{\mathrm{tot}}=0$ (solid curve) set purely by these ratios and distinct from the spin-inversion locus $\omega_q=0$. Since the dot ellipticity tunes $\omega_q$ electrically, including a sign reversal across the circular geometry [\cref{eq:larmor}], the operating point can be steered into a cooperative region by placing $\omega_q$ above or below $\omega_c$, or by reversing its sign according to the material's $\alpha_R$ and $\eta$, so as to maximize $|\chi_{\mathrm{tot}}|$ while staying away from the cancellation locus: an optimization unavailable to spin qubits whose splitting is fixed by an external field. Distinguishing the two spin states then requires their state-dependent cavity pull to exceed the resonator linewidth, $2|\chi_{\mathrm{tot}}|\gtrsim\kappa$~\cite{Blais_2021_Circuit}.

As in standard circuit-QED, this dispersive readout is quantum-nondemolition: the readout operator is $\sigma_z$ and the leading Hamiltonian \eqref{eq:Hdisp-main} commutes with the qubit, so the cavity probes the spin without electron tunneling or a spin-to-charge conversion pulse. What the altermagnetic platform contributes is the polarizability channel $\chi_{\mathrm{pol}}$ itself, a longitudinal photon-number coupling supplied by the $d$-wave term, absent in a spin-degenerate dot, that combines with the Rashba shift as above, together with a field-free, gate-tunable splitting that lets the cavity pull be optimized electrically. Should the residual transverse couplings, Purcell relaxation, or charge noise dominate, or $|\chi_{\mathrm{tot}}|$ prove too small, the Pauli-blockade and Elzerman channels remain available. The dispersive scheme above targets the single-spin qubit; the ST encoding of \cref{subsec:st} instead lends itself to a \emph{longitudinal} resonator coupling, since its quantization axis is set by the electrically tunable exchange $J$ [\cref{eq:HST}], so that modulating $J$ at the cavity frequency yields a parametric $\sigma_z$ photon coupling, as demonstrated for ST spin qubits~\cite{Bottcher_2022_Parametric}.

\section{Materials requirements}
\label{sec:materials}

After discussing the main features, we now suggest a set of material design criteria for altermagnetic spin qubits. 

The essential requirement to implement our proposal is an altermagnetic semiconductor with a well-defined bulk gap, so that gate-defined QDs can be formed in the single-electron regime; for example, $\alpha$-MnTe establishes this ingredient experimentally~\cite{Krempasky_2024_Altermagnetic, Lee_2024_Broken}. Beyond the gap, the ideal host should exhibit a simple, isolated band edge formed by a single valley carrying both spin species, so that the qubit is encoded in the two spin states of one confined orbital, as assumed in our model. The position of this valley in the Brillouin zone is not essential: a nondegenerate edge at $\Gamma$ or another high-symmetry point realizes the same squeezed-orbital mechanism after projection onto the dot, the predicted $\Gamma$-centered valence edge of \ce{MnGeP2} is one good example material~\cite{Turan_2025_Electronic}. What matters is the absence of symmetry-related copies: the logical two-level system must not require changing valley. By contrast, a symmetry-related $X/Y$ pair with spin--valley locking, whose low-energy states are predominantly of the form $\{\ket{X,\uparrow},\ket{Y,\downarrow}\}$, realizes a distinct spin--valley qubit. In that case, a qubit flip requires both a spin flip and an intervalley momentum transfer, so a smooth EDSR drive cannot directly couple the two states in the clean limit; any effective coupling would rely on atomically sharp disorder, interface-induced intervalley mixing, or additional engineered mechanisms, and the splitting would inherit the usual sensitivity of valley physics to short-range disorder~\cite{Burkard_2023_Semiconductor}. We therefore regard $X/Y$ spin--valley-locked materials as promising extensions beyond the minimal single-valley squeezed-orbital encoding analyzed here, rather than as direct realizations of it.

The relevant figure of merit is not the maximum spin splitting at generic momenta but the leading \emph{near-edge} coefficient of the collinear altermagnetic form factor that survives projection onto the confined orbital; large absolute splittings away from the band edge are correspondingly less informative than a sizable dot-projected near-edge form factor. For the $d$-wave case treated explicitly, this reduces to the dimensionless band-edge curvature $m\eta$ of \cref{eq:larmor}, $\omega_q\propto m\eta\,(\omega_y-\omega_x)$; for $g$- or $i$-wave hosts the analogous coefficient must be combined with the appropriate fourth- or sixth-order dot moment derived in Appendix~\ref{app:higher-order-altermagnets}. A high ordering temperature is likewise desirable, as it makes the ordered phase more robust against thermal fluctuations and device processing; the predicted high $T_N$ of \ce{V2Se2O} shows that this scale is plausible~\cite{Singh_2025_V2Se2O}.

The ordered phase should remain collinear, so that the spin quantization axis stays fixed along the N\'eel vector and reshaping the QD tunes only the qubit frequency, and robust on device scales, with domain sizes exceeding the dot size, sufficient isotropy to stabilize a well-defined N\'eel axis, and ideally a gapped and stiff magnetic sector to suppress soft long-wavelength fluctuations. 
In the notation of Eq.~\eqref{eq:neel-material-t1}, this means that the relevant transverse gaps $\Delta_a$ should exceed the intended qubit-frequency window, while the low-frequency damping tail and dot-filtered spectral weight should be small; the observed gapped transverse mode in $\alpha$-MnTe illustrates this favorable regime~\cite{Dzian_2025_Antiferromagnetic}.

The out-of-plane orientation assumed throughout (\cref{sec:effective-model}) is set by magnetocrystalline anisotropy, which determines whether the N\'eel vector prefers to align along the out-of-plane direction or to lie within the plane. Both situations occur generically among collinear (alter)magnets, and the out-of-plane case is realized within the relevant $d$-wave symmetry class by the rutile fluorides $\alpha$-\ce{MnF2}, \ce{FeF2}, and \ce{CoF2}, whose N\'eel vector points along the $c$-axis~\cite{Smejkal_2022_Emerging}. The anisotropy scale is, moreover, small, of order $0.1$,meV per magnetic ion, so even in hosts with a bulk easy-plane orientation the N\'eel axis can be reoriented out of plane by epitaxial strain or interfacial engineering at modest cost~\cite{LiebmanPelaez_2026_Strain, Polewczyk_2025_Control}, with the electron spin remaining locked to $\mathbf{N}$.

A practical prerequisite is that the active device region sits within a single altermagnetic domain that is static on qubit timescales. With the N\'eel axis fixed out of plane the only residual domain freedom is the \emph{sign} of the order: time-reversed ($\mathbf{N}\!\to\!-\mathbf{N}$) and $C_4$-related domains both flip $\eta$ and therefore send $\omega_q\!\to\!-\omega_q$, which for a single dot merely relabels the two spin states and is harmless. The genuine requirement is thus mild, i.e., no domain wall crossing the dot, and a static axis, so the domain need only exceed the dot footprint ($\sim 10$--$50$\,nm), far below the micrometer-scale single-domain antiferromagnetic films routinely achieved in spintronics. Several established handles ensure this: epitaxial strain that breaks the in-plane $C_4$ symmetry pins the orientation of the altermagnetic anisotropy through magnetoelastic coupling, without changing the magnitude of the order and hence leaving the charge-noise sweet spot of \cref{sec:noise} intact~\cite{LiebmanPelaez_2026_Strain}; field-cooling through $T_N$, biased by small uncompensated interfacial moments or an exchange-bias layer, selects a single sign domain at the $\sim\SI{10}{\milli\tesla}$ level~\cite{Jungwirth_2016_Antiferromagnetic}; and below a critical device size domain-wall formation becomes energetically unfavorable, enforcing single-domain behavior geometrically~\cite{Rickhaus_2024_Antiferromagnetic}. Combined, these make the single-domain assumption well justified in realistic device geometries.

Moreover, several further properties are highly desirable from the qubit perspective. A material with a sufficiently large spin--orbit coupling would be optimal for EDSR-based manipulation. However, if a candidate exhibits low spin--orbit coupling, the ST encoding is still a field-free fully gate-tunable alternative. The material should be, preferably, composed of elements with a low abundance of non-zero nuclear-spin isotopes that would suppress hyperfine-induced dephasing; on this criterion $\alpha$-MnTe falls short, since its only stable isotope $^{55}$Mn carries nuclear spin $I=5/2$, motivating hosts assembled predominantly from zero-nuclear-spin elements, such as the predicted chromium oxychalcogenides \ce{Cr2Se2O} and \ce{Cr2SeO}~\cite{Khan_2025_Altermagnetism,Guo_2026_Ultrafast}. Clean interfaces and compatibility with electrostatic gating are important to minimize disorder and charge noise; the van-der-Waals geometry of \ce{V2Se2O} is favorable in this respect~\cite{Singh_2025_V2Se2O}. Moderate effective masses are advantageous for lithographic confinement.

A route that sidesteps these materials trade-offs altogether is the recently proposed altermagnetic proximity effect~\cite{Zhu_2026_Altermagnetic}: when a nonmagnetic layer is placed in contact with an altermagnet, the momentum-alternating spin splitting is imprinted across the interface onto the otherwise spin-degenerate bands, transferring the altermagnetic spin texture, with its momentum structure preserved, without net magnetization or stray fields. First-principles calculations predict this ``altermagnetization'' for, e.g., a monolayer of the conventional semiconductor PbO proximitized by \ce{V2Se2O}, with the induced splitting tunable through the interlayer spacing. If realized experimentally, this mechanism decouples the host requirements from the source of the splitting: the dot could be defined in a conventional, well-optimized semiconductor, selected for clean gating, favorable isotopes, and a single simple band edge, while the field-free altermagnetic splitting is supplied by an adjacent altermagnet, including the metallic or insulating altermagnets (such as \ce{CrSb}) that cannot themselves host gate-defined dots. Such a proximitized geometry would broaden the accessible materials space enormously, at the cost of requiring the confined channel to lie within the short proximity range of the interface.

\section{Conclusion}
\label{sec:conclusion}
We have proposed altermagnetic semiconductors as a host for spin qubits that operate entirely without external magnetic fields. Starting from a microscopic lattice model, we derived an effective low-energy Hamiltonian combining a $d$-wave spin splitting with Rashba spin--orbit coupling, and showed that electrostatically confining single electrons in QDs yields a qubit whose frequency is set by the intrinsic altermagnetic order and tuned continuously (through zero, with a sign reversal across the circular geometry) by the dot ellipticity alone. This geometric splitting replaces the globally fixed Zeeman energy of conventional spin qubits with an individually addressable, all-electrical knob, while the compensated magnetic order leaves the device free of stray fields and inter-dot cross-talk. The same Hamiltonian furnishes a complete gate set: Rashba-mediated electric-dipole spin resonance drives single-qubit rotations, while the combination of tunable interdot exchange with individually addressable qubit splittings natively generates the fSim family of two-qubit gates, including $\sqrt{\text{iSWAP}}$ and CZ. Within the same double dot, the $(1,1)$ charge configuration realizes a singlet--triplet qubit whose two control axes, the exchange and the per-dot splitting difference, are both purely electrical, removing the micromagnets or nuclear-polarization gradients that conventional implementations require.

A defining feature of the platform is its built-in noise protection. The altermagnetic splitting resides at a first-order charge-noise sweet spot by construction, since a uniform electric field merely displaces the confining potential, and the residual second-order sensitivity inherits a $d$-wave angular structure that can itself be minimized by shaping the dot. Magnetically, smooth N\'eel-vector fluctuations couple to the qubit only transversely, so that they drive relaxation rather than pure dephasing and are further suppressed by their time-derivative coupling. This protection is matched by flexible readout: alongside conventional Pauli-blockade and Elzerman spin-to-charge conversion, the vanishing net magnetization makes the qubit natively compatible with superconducting microwave resonators, enabling a quantum-nondemolition dispersive readout in which a spin-dependent, longitudinal polarizability (a hallmark of the altermagnetic $d$-wave term, absent in spin-degenerate dots) supplements the Rashba dispersive shift and can be optimized electrically. This compatibility is further strengthened by the extensive study of the interplay between altermagnets and superconductivity~\cite{Li_2023_Majorana,Ghorashi_2024_Altermagnetic,Fu_2025_Light,Fukaya_2026_Crossed,Fukaya_2025_Superconducting,Hong_2025_Unconventional,Kazmin_2025_Andreev,Lu_2026_Engineering,Maeda_2025_Classification,Maiani_2025_Impurity,Vosoughinia_2025_Altermon,Fu_2026_Floquet}. Together with the materials criteria laid out in \cref{sec:materials}, these results establish altermagnetic QDs as a scalable, all-electrical route to spin-based quantum information processing, in either the Loss--DiVincenzo or singlet--triplet encoding.

Crucially, realizing this architecture need not await the discovery of an ideal intrinsic altermagnetic semiconductor. Through the altermagnetic proximity effect, the same field-free, momentum-dependent splitting can be imprinted onto a conventional, well-optimized host placed in contact with an altermagnet, decoupling the qubit's host requirements from the source of the splitting. This separation substantially broadens the accessible materials space---the dot may be defined in a clean, gate-friendly semiconductor while the splitting is supplied by an adjacent altermagnet---and offers a flexible, potentially near-term route to these devices.

Looking further ahead, the underlying mechanism need not be restricted to collinear altermagnets: spin-space-group analyses predict momentum-even, spin--orbit-free spin splittings in a broader class of compensated magnets, such as the plaid-like texture observed in the noncoplanar antiferromagnet \ce{MnTe2}~\cite{Zhu_2024_Observation}, where a geometry-dependent qubit axis could even enable all-electric geometric gates, at the cost of the protections that rely on collinearity, which we leave for future work.

\paragraph*{Note added.} During the final preparation of this manuscript, a related preprint appeared proposing gate-controlled spin qubits in confined altermagnetic quantum dots, with quadrupolar single-qubit control and double-dot entangling dynamics in an atomistic model~\cite{Vakili_2026_Gate}. Our work is complementary: we develop a magnetic-field-free architecture based on Rashba-assisted electric-dipole spin resonance, derive an analytic exchange Hamiltonian supporting native fSim gates, introduce a fully electrical singlet--triplet encoding, and analyze noise, readout, cavity coupling, and materials constraints for scalable devices.

\begin{acknowledgments}
Work supported by the Spanish Ministry of Science, Innovation, and Universities through Grants RYC2022-037527-I, PID2023-148257NA-I00, and PID2024-161156NB-I00 funded by MICIU/AEI/10.13039/501100011033 and by FSE+ and FEDER, UE, ``ERDF A way of making Europe'' and European Union Next Generation EU/PRTR. JCAU, RSS, and RA acknowledge the support of the CSIC's Quantum Technologies Platform (QTEP) and the Severo Ochoa Centers of Excellence program through Grant CEX2024-001445-S. RSS acknowledges funding from the Spanish Comunidad de Madrid (CM) ``Talento Program'' (Project No. 2022-T1/IND-24070). AM acknowledges funding from the Wallenberg Initiative on Networks and Quantum Information (WINQ).
\end{acknowledgments}

\appendix

\section{Microscopic lattice model and Schrieffer--Wolff derivation}
\label{app:micro}

\subsection{Derivation of the long-wavelength model}
\label{app:lattice1}

\begin{figure}[htp]
    \centering
    \includegraphics[width=5cm]{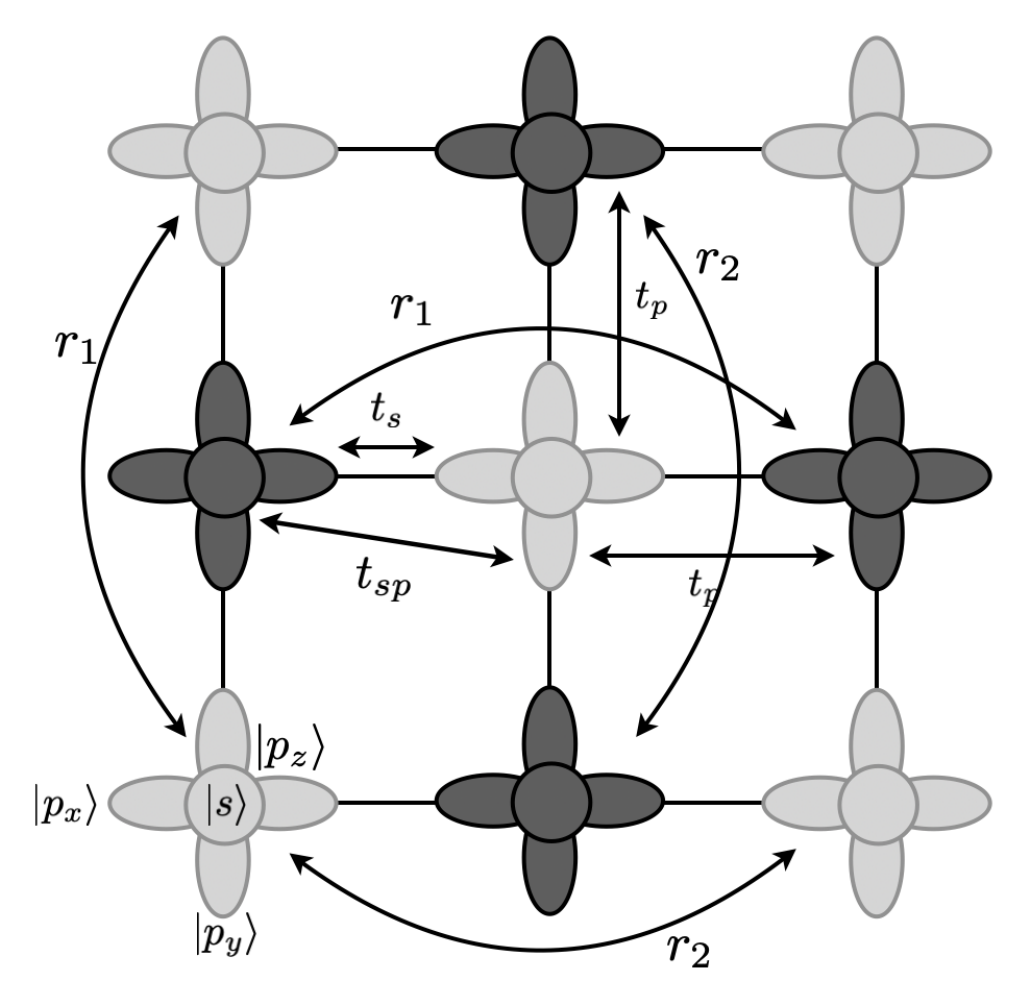}
    \caption{Scheme of hopping parameters of the microscopic BHZ-inspired model.}
    \label{fig:lattice}
\end{figure}

We start from a variant of the lattice version of the Bernevig--Hughes--Zhang (BHZ) model on a square lattice. At each site, we have 8 angular momentum-adapted orbitals: $(\ket{s},\ket{p_{1}},\ket{p_{-1}},\ket{p_z})\otimes(\uparrow,\downarrow)$, as shown in \cref{fig:lattice}.

The Hamiltonian comprises the kinetic terms in momentum space:
\begin{equation}
\begin{split}
H_\mathrm{kin}=&\sum_{\bm{k},\alpha,\sigma}\varepsilon^{s}_{\bm{k}}s^{\dagger}_{\sigma}s_{\sigma}+\varepsilon^{p}_{\bm{k}}p^{\dagger}_{\alpha,\sigma}p_{\alpha,\sigma}\\
+&t_{sp}\sum_{\bm{k}\alpha\sigma}f_{\alpha}(\bm{k})s^{\dagger}_{\sigma}p_{\alpha\sigma}+\text{h.c.},
\end{split}
\end{equation}
where $f_{\pm}(\bm{k})=\tfrac{1}{\sqrt{2}}(\mp e^{-ia_0k_x}-ie^{-ia_0k_y})$, the index $\alpha$ relates to the $m=\pm1$ angular momentum projection, the local spin--orbit coupling 
\begin{equation}
H_\mathrm{SOC} =\sum_{\bm{k}}\frac{\lambda}{2}\left(\underbrace{L_z \sigma_z}_{\text{intrinsic}}+\underbrace{L_{+}\sigma_{-}+L_{-}\sigma_{+}}_{\text{spin-flip}}\right),  
\end{equation}
where the operators $(L_z, L_{\pm})$ operate in the $(l=1)$ subspace. The operator $L_z$ is $L_z=\text{Diag}(0,+1,-1)$ in the basis $(\ket{p_z},\ket{p_+},\ket{p_-})$,
and finally structural inversion asymmetry (SIA) Stark terms, $\Delta_\mathrm{SIA}=\bra{s}eE_z z\ket{p_z}$, read
\begin{equation}
H_\mathrm{SIA} = \sum_{\bm{k},\sigma}\Delta_\mathrm{SIA} (s^{\dagger}_{\sigma}p_{z,\sigma}+p^{\dagger}_{z,\sigma}s_{\sigma}).
\end{equation}
In the kinetic term, we have neglected hopping terms from $s$ and $p_z$ orbitals as they just renormalize $\Delta_\mathrm{SIA}$.
The SIA term couples the $\ket{s,\sigma}$ and $\ket{p_z,\sigma}$ orbitals, while $\ket{p_z,\sigma}$ is coupled to $\ket{p_{\pm},\sigma'}$ by the spin-flip part of the SOC Hamiltonian. 

The procedure to obtain the Rashba spin--orbit coupling terms then consists of two steps. First, project out the orbitals $\ket{p_z,\sigma}$ by a Schrieffer--Wolff (SW) transformation. This transformation maps the original Hamiltonian into a new one where the subspace of $(\ket{s},\ket{p_+},\ket{p_-})$ decouples from the $\ket{p_z}$ subspace. The part of the kinetic Hamiltonian for the $\ket{s}$ states reads
\begin{equation}
H'_s=\sum_{\bm{k},\sigma}\left(\varepsilon^{s}_{\bm{k}}-\frac{\Delta_\mathrm{SIA}^2}{\underbrace{\varepsilon^{p}_{\bm{k}}-\varepsilon^{s}_{\bm{k}}}_{\Delta_{ps}(\bm{k})}}\right)s^{\dagger}_{\sigma}s_{\sigma},
\end{equation}
the kinetic term for the $p_{\alpha}$ orbitals is
\begin{equation}
H'_p=\sum_{\bm{k},\alpha=\pm,\sigma}\left(\varepsilon^{p}_{\bm{k}}-\frac{\lambda^2}{2\Delta_{ps}(\bm{k})}\right)p^{\dagger}_{\alpha,\sigma}p_{\alpha,\sigma},
\end{equation}
and a new coupling between $s$ and $p_{\pm}$ orbitals with opposite spin projection appears, that adds to the hopping term between $s$ and $p_\pm$ states:
\begin{align}\label{eq:Hsp1}
H'_{sp}={}&t_{sp}\sum_{\bm{k},\sigma\alpha}f_{\alpha}(\bm{k})s^{\dagger}_{\sigma}\sigma_0p_{\alpha,\sigma}\nonumber\\
&-\sum_{\bm{k},\sigma\sigma'}\frac{\lambda\Delta_\mathrm{SIA}}{\sqrt{2}\Delta_{ps}(\bm{k})}\left(s^{\dagger}_{\sigma}\sigma_{+}p_{+\sigma'}+s^{\dagger}_{\sigma}\sigma_{-}p_{-\sigma'}\right)\nonumber\\
&+\text{h.c.}
\end{align}
In the Hamiltonian $H'_p$, the difference $\Delta_{ps}$ appears in the denominator, as we have expanded around the reference energy $\varepsilon^{s}_{\bm{k}}$ when computing the SW transformation. This comes after assuming that there is a global gap between $s$ and $p$ states in the whole Brillouin Zone, $\Delta_{ps}(\bm{k})>0,\forall \bm{k}\in\mathrm{BZ}$. This first SW transformation is local, that is, projecting out the orbitals $p_z$ leads to on-site spin-flip terms that couple $s$ and $p_\pm$ orbitals at each site, although this is the seed of the Rashba spin--orbit terms.

The second step is to perform a second SW transformation to decouple the $\ket{p_{\pm},\sigma}$ and $\ket{s,\sigma}$ subspaces, keeping only the $\ket{s,\sigma}$ part:
\begin{equation}
\begin{aligned}
H''_s &= \sum_{\bm{k}}\hat{\varepsilon}^{s}_{\bm{k}}s^{\dagger}_{\sigma}s_{\sigma}\\
&\quad -\frac{t_{sp}\lambda\Delta_\mathrm{SIA}}{\tilde{\Delta}_{ps}(\bm{k})\Delta_{ps}(\bm{k})}\bigl(f_{+}(\bm{k})s^{\dagger}_{\sigma}\sigma_{+}s_{\sigma'}\\
&\qquad +f_{-}(\bm{k})s^{\dagger}_{\sigma}\sigma_{-}s_{\sigma'}\bigr)\\
&= \sum_{\bm{k}}\hat{\varepsilon}^{s}_{\bm{k}}s^{\dagger}_{\sigma}s_{\sigma}\\
&\quad -\sum_{\bm{k}}\alpha_R(\bm{k})s^{\dagger}_{\sigma}\bigl(\sigma_x \sin(a_0k_y)-\sigma_y\sin(a_0k_x)\bigr)s_{\sigma'},
\end{aligned}
\end{equation}
with
\begin{equation}
\label{eq:rashba-coefficients}
\begin{aligned}
\alpha_R(\bm{k})&=\frac{t_{sp}\lambda\Delta_\mathrm{SIA}}{\tilde{\Delta}_{ps}(\bm{k})\Delta_{ps}(\bm{k})},\\
\tilde{\varepsilon}^{s}_{\bm{k}}&=\varepsilon^{s}_{\bm{k}}-\frac{\Delta_\mathrm{SIA}^2}{\Delta_{ps}(\bm{k})},\\
\tilde{\varepsilon}^{p}_{\bm{k}}&=\varepsilon^{p}_{\bm{k}}-\frac{\lambda^2}{2\Delta_{ps}},\\
\tilde{\Delta}_{ps}&=\tilde{\varepsilon}^{p}_{\bm{k}}-\tilde{\varepsilon}^{s}_{\bm{k}},
\end{aligned}
\end{equation}
and $\hat{\varepsilon}^s_{\bm{k}}=\tilde{\varepsilon}^s_{\bm{k}}+\mathcal{O}(\lambda^2\Delta_\mathrm{SIA}^2, t^2_{sp})$. This is how a Rashba spin--orbit coupling appears in the lattice model. This term arises because in \cref{eq:Hsp1}, $H'_{sp}$ comprises two contributions, one proportional to $\lambda\Delta_\mathrm{SIA}$ that produces spin precession, and one proportional to $t_{sp}$ that introduces the $\bm{k}$ dependence. Terms proportional to $t^2_{sp}$ and $\lambda^2\Delta_\mathrm{SIA}^2$ are also generated, but they only renormalize the dispersion relation $\tilde{\varepsilon}^{s}_{\bm{k}}$.

\subsection{Including anisotropic and antiferromagnetic order}
\label{app:lattice2}

Let us add the ingredients that lead to altermagnetic order in the model: the anisotropic next-to-nearest-neighbor hopping terms $r_{1,2}$ and the antiferromagnetic (AFM) or N\'eel order. The main effect of these two elements is to create a new unit cell in the lattice comprised of two atomic sites, each one with 8 states (orbitals times spin). This new sublattice content will be incorporated through a new pseudospin structure, denoted by $\tau$ Pauli matrices. In this way, nearest-neighbor hoppings will be now proportional to $\tau_x\sigma_0$, while the $r_{1,2}$ hoppings will enter through the combination $(r_1+r_2)\tau_0\sigma_0\equiv r\tau_0\sigma_0$, and $(r_1-r_2)\tau_z\sigma_0\equiv\delta\tau_z\sigma_0$. The AFM order, that will be considered to point along the third direction, will be $N\tau_z\sigma_z$, with $N=|\mathbf{N}|$ the AFM order parameter. The presence of the AFM order breaks the spin degeneracy of the on-site energies, $\varepsilon^{s,p}\sim\varepsilon^{s,p}+N\sigma_z\tau_z$. This will affect the structure of the denominators in the SW transformations.

As before, the SOC and SIA terms are local, so they couple orbitals of the same sublattice. We can thus perform the same first SW transformation to remove the $p_z$ orbitals, obtaining
\begin{equation}
\begin{aligned}
H'_{sp} &= -\sum_{\bm{k}}\frac{\lambda\Delta_\mathrm{SIA}}{\sqrt{2}\Delta_{ps}}\bigl(s^{\dagger}_{\sigma\tau}\sigma_{+}\tau_0 p_{+\sigma'\tau}\\
&\qquad +s^{\dagger}_{\sigma\tau}\sigma_{-}\tau_0 p_{-\sigma'\tau}\bigr)+\text{h.c.}
\end{aligned}
\end{equation}
The other element we need is the hopping term between $s$ and $p_\pm$ states. This term is proportional now to $\tau_x$, as we are connecting $s$ and $p_{\pm}$ states of different sublattices:
\begin{equation}
H^{(0)}_{sp}=t_{sp}\sum_{\bm{k}\alpha\sigma,\tau\tau'}f_{\alpha}(\bm{k})s^{\dagger}_{\sigma\tau}\tau_x\sigma_0p_{\alpha\sigma\tau'}.
\end{equation}
The functions $f_{\alpha}(\bm{k})$ are the same as in the previous section. 

The kinetic terms read
\begin{align}
H'_s
&=
\sum_{\bm{k}}
s^{\dagger}_{\sigma\tau}
\bigl[
    (\varepsilon_s-r g_0(\bm{k}))\tau_0\sigma_0
    +
    \delta g_3(\bm{k})\tau_z\sigma_0
    \notag
    \\
&\qquad\qquad
    +
    t_s h_s(\bm{k})\tau_x\sigma_0
    +
    N\tau_z\sigma_z
\bigr]
s_{\sigma'\tau'} ,
\end{align}
Here $r$ and $\delta$ are the symmetric and anisotropic next-nearest-neighbor hopping amplitudes introduced above, $t_s$ is the nearest-neighbor $s$-orbital (inter-sublattice) hopping amplitude---the subscript $s$ labeling the orbital, not the spin---and $g_0(\bm{k})$, $g_3(\bm{k})$, and $h_s(\bm{k})$ are the corresponding dimensionless lattice form factors of these three hoppings, while $f_\alpha(\bm{k})$ is the $s$--$p$ form factor of the previous section. Their long-wavelength expansions, used in the projection below, are given in \cref{eq:bonding-g0-hs-expansion,eq:bonding-g3-r-expansion}; in particular $g_3(\bm{k})\propto k_x^2-k_y^2$ carries the $d$-wave anisotropy generated by $\delta$.

The analysis simplifies considerably if we assume that the bandwidth of these bands is much smaller than the energy difference, so in the intermediate steps we use
\begin{align}
H'_s&\simeq\sum_{\bm{k}}s^{\dagger}_{\sigma\tau}\left[\varepsilon_s\tau_0\sigma_0+N\tau_z\sigma_z\right]s_{\sigma'\tau'},\\
H'_p&\simeq\sum_{\bm{k}\alpha}p^{\dagger}_{\alpha\sigma\tau}\left[\varepsilon_p\tau_0\sigma_0+N\tau_z\sigma_z\right]p_{\alpha\sigma'\tau'}.
\end{align}
Considering the whole perturbation $H'_{sp}+H^{(0)}_{sp}$, the $s$ state subspace formed by the states $\ket{s,\sigma\tau}$ ($\sigma=\pm1$ is the spin projection and $\tau=\pm$) splits into two subspaces, $(\sigma\cdot\tau=1)$ with energies $\varepsilon_s+N$, and $(\sigma\cdot\tau=-1)$ with energies $\varepsilon_s-N$. The term $H'_{sp}$ flips the spin but not the sublattice, so the first subspace is connected to the $p_{\alpha}$ subspace with $(\sigma\cdot\tau=-1)$, and because $H^{(0)}_{sp}$ flips sublattice but not spin, we connect back to the $s$ subspace with $(\sigma\cdot\tau=1)$, but flipping both spin and sublattice. Since the energies of the states are corrected by $\sigma\tau N$, in all cases we are connecting $s$ and $p$ states with opposite value of the quantity $(\sigma\cdot\tau)$. We can thus write the part of the transformed Hamiltonian that contains spin (and sublattice) flips as
\begin{align}\label{eq:formalH}
\delta H''_s={}&H^{0}_{sp}\left(\frac{1}{\Delta_{ps}-2\sigma\tau N}\right)H'_{sp}\nonumber\\
&+H'_{sp}\left(\frac{1}{\Delta_{ps}-2\sigma\tau N}\right)H^{(0)}_{sp},
\end{align}
keeping in mind that the $p$ subspace contains two angular projections $\alpha=\pm1$. This formal expression is valid when working in the corresponding subspaces $(\ket{s,\sigma\tau=\pm1})$ connected to $(\ket{p_\alpha,\sigma\tau=\mp1})$. 

Since the N\'eel field shifts the intermediate energies in a spin- and sublattice-dependent way, the relevant denominator has the structure
\begin{equation}
    \frac{1}{\Delta_{ps}-2\sigma\tau N}
    =
    \frac{1}{\Delta_{ps}}
    \left[
        \frac{1}{1-x^2}
        +
        \frac{x}{1-x^2}\sigma_z\tau_z
    \right]\,,
\label{eq:denominator_expansion}
\end{equation}
with $x = \tfrac{2N}{\Delta_{ps}}$. This separates the induced Rashba coupling into an even and an odd component. We define
\begin{equation}
    \alpha_R^e
    =
    \frac{t_{sp}\lambda\Delta_\mathrm{SIA}}{\Delta_{ps}^2}
    \frac{1}{1-x^2} \, ,
    \qquad
    \alpha_R^o
    =
    \frac{t_{sp}\lambda\Delta_\mathrm{SIA}}{\Delta_{ps}^2}
    \frac{x}{1-x^2} \, .
\label{eq:alpha_even_odd}
\end{equation}
The reduced $s$-orbital Hamiltonian then takes the form
\begin{equation}
\begin{aligned}
    H_s^\mathrm{red}
    &=
    \sum_{\bm{k}}
    s_{\bm{k}}^\dagger
    \bigg[
        -r g_0(\bm{k})\tau_0\sigma_0
        + t_s h_s(\bm{k})\tau_x\sigma_0\\
        &+ \delta g_3(\bm{k})\tau_z\sigma_0
        + N\tau_z\sigma_z
        \\
        &-
        \alpha_R^e
        \tau_x
        \left(
            \sigma_x\sin(a_0k_y)
            -
            \sigma_y\sin(a_0k_x)
        \right)
        \\
        &+
        \alpha_R^o
        \tau_y
        \left(
            \sigma_x\sin(a_0k_x)
            +
            \sigma_y\sin(a_0k_y)
        \right)
    \bigg]
    s_{\bm{k}} \, .
\end{aligned}
\label{eq:Hs_red_sublattice}
\end{equation}
The coefficient $\alpha_R^e$ is the even Rashba coupling, present also in the absence of altermagnetism. The coefficient $\alpha_R^o$ is the odd Rashba coupling generated by the staggered N\'eel splitting in the virtual denominators, and therefore vanishes when $N=0$.

\subsection{Projection onto the bonding subspace}
\label{app:bonding-sw-projection}

A simple unitary transformation allows us to rewrite the operator $H''_s$ in the bonding-antibonding basis, while leaving the spin structure unchanged:
\begin{equation}
\label{eq:bonding-full-hamiltonian}
\begin{split}
H''_s
&=
-r g_0(\bm{k}) \tau_0 \sigma_0
-
t_s h_s(\bm{k}) \tau_z \sigma_0
\\
&\quad
-
\alpha_R^{e}\tau_z\bigl(\sigma_y \sin(a_0 k_x)-\sigma_x \sin(a_0 k_y)\bigr)
\\
&\quad
+
N \tau_x \sigma_z
+
\delta g_3(\bm{k}) \tau_x \sigma_0
\\
&\quad
+
\alpha_R^{o}\tau_y\bigl(\sigma_x \sin(a_0 k_x)+\sigma_y \sin(a_0 k_y)\bigr).
\end{split}
\end{equation}
The first three terms are diagonal in the bonding-antibonding basis, whereas the N\'eel term, the anisotropic hopping term proportional to $\delta$, and the $\alpha_R^o$ Rashba term are off-diagonal.

To make the structure transparent, define
\begin{equation}
\label{eq:bonding-def-rq}
\begin{aligned}
R(\bm{k}) &\equiv \sigma_y \sin(a_0 k_x)-\sigma_x \sin(a_0 k_y),\\
Q(\bm{k}) &\equiv \sigma_x \sin(a_0 k_x)+\sigma_y \sin(a_0 k_y),
\end{aligned}
\end{equation}
and
\begin{equation}
\label{eq:bonding-def-e0-t}
\varepsilon_0(\bm{k}) \equiv -r g_0(\bm{k}),
\qquad
t(\bm{k}) \equiv t_s h_s(\bm{k}).
\end{equation}
Then \cref{eq:bonding-full-hamiltonian} can be rewritten as
\begin{equation}
\label{eq:bonding-compact-hamiltonian}
\begin{split}
H''_s
=
&\varepsilon_0(\bm{k}) \tau_0 \sigma_0
-
\tau_z\Bigl[t(\bm{k}) \sigma_0 + \alpha_R^e R(\bm{k})\Bigr]\\
&+
\tau_x B(\bm{k})
+
\tau_y C(\bm{k}),
\end{split}
\end{equation}
with
\begin{equation}
\label{eq:bonding-def-bc}
B(\bm{k}) \equiv N \sigma_z + \delta g_3(\bm{k}) \sigma_0,
\qquad
C(\bm{k}) \equiv \alpha_R^o Q(\bm{k}).
\end{equation}

In the $\tau_z$ eigenbasis, this takes the block form
\begin{equation}
\label{eq:bonding-block-form}
H''_s =
\begin{pmatrix}
\varepsilon_0 - t - \alpha_R^e R & B - i C \\[4pt]
B + i C & \varepsilon_0 + t + \alpha_R^e R
\end{pmatrix}.
\end{equation}
Near $\Gamma$, one has $h_s(\bm{0})=2$, so that $t(\bm{k})>0$ for $t_s>0$. Therefore, the lower-energy bonding block is the $\tau_z=+1$ block with unperturbed Hamiltonian
\begin{equation}
\label{eq:bonding-low-energy-block}
H_b^{(0)}(\bm{k})=\varepsilon_0(\bm{k})\sigma_0-t(\bm{k})\sigma_0-\alpha_R^e R(\bm{k}).
\end{equation}

We now perform a second Schrieffer--Wolff transformation, projecting onto this bonding subspace. Assuming $\alpha_R^e \ll 2 t_s$, we neglect $\alpha_R^e$ in the energy denominator. To second order in the off-diagonal terms, the effective Hamiltonian is
\begin{equation}
\label{eq:bonding-sw-effective}
\begin{split}
H_{\mathrm{eff}}
&\simeq
\varepsilon_0(\bm{k}) \sigma_0
-
t(\bm{k}) \sigma_0
-
\alpha_R^e R(\bm{k})
\\
&\quad-
\frac{1}{2 t(\bm{k})}\,(B-iC)(B+iC),
\end{split}
\end{equation}
where the product $(B-iC)(B+iC)=B^2+C^2+i[B,C]$. We evaluate each contribution separately. First,
\begin{equation}
\label{eq:bonding-b-squared}
\begin{split}
B^2
&=
\bigl(N \sigma_z + \delta g_3 \sigma_0\bigr)^2
=
N^2 \sigma_0
+
\delta^2 g_3^2 \sigma_0
+
2 N \delta g_3 \sigma_z.
\end{split}
\end{equation}
Second,
\begin{equation}
\label{eq:bonding-c-squared}
\begin{split}
C^2
&=
\alpha_R^{o\,2} Q^2
=
\alpha_R^{o\,2}\bigl[\sin^2(a_0 k_x)+\sin^2(a_0 k_y)\bigr]\sigma_0,
\end{split}
\end{equation}
since the mixed term vanishes by the anticommutation of $\sigma_x$ and $\sigma_y$. Finally,
\begin{equation}
\label{eq:bonding-commutator-cb}
[B,C]
=
\alpha_R^o N\,[\sigma_z,Q],
\end{equation}
because the term proportional to $\delta g_3 \sigma_0$ commutes with $Q$. Using Pauli algebra,
\begin{equation}
\label{eq:bonding-q-commutator}
\begin{split}
[\sigma_z,Q]
&=
[\sigma_z,\sigma_x]\sin(a_0 k_x)+[\sigma_z,\sigma_y]\sin(a_0 k_y)\\
&=
2 i\,\bigl(\sigma_y \sin(a_0 k_x)-\sigma_x \sin(a_0 k_y)\bigr)\\
&=
2 i\,R(\bm{k}).
\end{split}
\end{equation}
Therefore,
\begin{equation}
\label{eq:bonding-i-bc}
i[B,C]
=
-2 \alpha_R^o N\,R(\bm{k}).
\end{equation}

Collecting the terms,
\begin{equation}
\label{eq:bonding-product-expanded}
\begin{split}
(B-iC)(B+iC)
&=
\Bigl[
N^2+\delta^2 g_3^2
\\
&\quad
+\alpha_R^{o\,2}\bigl(\sin^2(a_0 k_x)+\sin^2(a_0 k_y)\bigr)
\Bigr]\sigma_0
\\[4pt]
&\quad
+
2 N \delta g_3(\bm{k}) \sigma_z
-
2 \alpha_R^o N\,R(\bm{k}).
\end{split}
\end{equation}
The terms proportional to $N^2$, $\delta^2$, and $\alpha_R^{o\,2}$ only renormalize the spin-independent dispersion and will be omitted. Keeping only the leading spin-dependent contributions, we obtain
\begin{equation}
\label{eq:bonding-delta-heff-general}
\delta H_{\mathrm{eff}}
\simeq
-\frac{N \delta}{t(\bm{k})} g_3(\bm{k}) \sigma_z
+
\frac{\alpha_R^o N}{t(\bm{k})} R(\bm{k}).
\end{equation}
Near $\Gamma$, one has $h_s(0)=2$, hence $t(\bm{k}) \simeq 2 t_s$ in the denominator to this order, and therefore
\begin{equation}
\label{eq:bonding-delta-heff-gamma}
\delta H_{\mathrm{eff}}
\simeq
-\frac{N \delta}{2 t_s} g_3(\bm{k}) \sigma_z
+
\frac{\alpha_R^o N}{2 t_s} R(\bm{k}).
\end{equation}

The effective low-energy Hamiltonian in the bonding subspace is thus
\begin{equation}
\label{eq:bonding-heff-lattice}
\begin{split}
H_{\mathrm{eff}}
&=
-r g_0(\bm{k}) \sigma_0
-
t_s h_s(\bm{k}) \sigma_0
-
\frac{N \delta}{2 t_s} g_3(\bm{k}) \sigma_z
\\[4pt]
&\quad
+
\left(
-\alpha_R^e
+
\frac{\alpha_R^o N}{2 t_s}
\right)
\bigl(\sigma_y \sin(a_0 k_x)-\sigma_x \sin(a_0 k_y)\bigr),
\end{split}
\end{equation}
up to terms of order $\mathcal{O}(N^2/t_s^2,\delta^2/t_s^2,\alpha_R^{o\,2}/t_s^2)$ that only renormalize the scalar dispersion at this level.

We finally arrive at the effective $2\times 2$ Hamiltonian for the low-energy $s$ orbitals. Taylor expanding the form factors around $\Gamma$,
\begin{equation}
\label{eq:bonding-g0-hs-expansion}
\begin{aligned}
g_0(\bm{k}) &= 2 - 2 a_0^2 (k_x^2+k_y^2)+\cdots,\\
h_s(\bm{k}) &= 2 - \frac{a_0^2}{2}(k_x^2+k_y^2)+\cdots,
\end{aligned}
\end{equation}
\begin{equation}
\label{eq:bonding-g3-r-expansion}
\begin{aligned}
g_3(\bm{k}) &= -2 a_0^2 (k_x^2-k_y^2)+\cdots,\\
R(\bm{k}) &= a_0\bigl(\sigma_y k_x-\sigma_x k_y\bigr)+\cdots,
\end{aligned}
\end{equation}
then subtracting the constant energy offset and defining the continuum coefficients
\begin{equation}
\label{eq:bonding-continuum-coefficients}
\begin{aligned}
\frac{1}{2m}
&\equiv
a_0^2\left(2r+\frac{t_s}{2}\right),
\qquad
\eta
\equiv
-\frac{N\delta a_0^2}{t_s},\\
\alpha_R
&\equiv
a_0\left(-\alpha_R^e+\frac{\alpha_R^o N}{2 t_s}\right),
\end{aligned}
\end{equation}
the continuum Hamiltonian becomes \cref{eq:continuum-hamiltonian} in the main text:
\begin{equation}
\label{eq:bonding-continuum-final}
H_{\mathrm{eff}}
=
\frac{\bm{k}^2}{2m}\sigma_0
-
\eta (k_x^2-k_y^2)\sigma_z
+
\alpha_R(\sigma_y k_x-\sigma_x k_y).
\end{equation}
The last term has the standard Rashba structure, while the spin-dependent anisotropic term is proportional to the product of the hopping anisotropy $\delta$ and the N\'eel order $N$, as expected.

\section{Effect of a canted N\'eel vector}
\label{app:inplane}
The main text assumes a uniform out-of-plane N\'eel vector, $\mathbf n=\hat{\mathbf z}$. Here we summarize how the results change for a uniform canted direction. The single-particle Hamiltonian may be written as
\begin{equation}
H=
\frac{\mathbf k^2}{2m}+V(\mathbf r)
-\eta g_d(\mathbf k)\,\mathbf n\cdot\boldsymbol{\sigma}
+\alpha_R(\sigma_y k_x-\sigma_x k_y).
\label{eq:canted-H}
\end{equation}
At $\alpha_R=0$, a global spin rotation that maps $\mathbf n$ onto $\hat{\mathbf z}$ makes \cref{eq:canted-H} identical to the Hamiltonian analyzed in the main text. The spin-dependent squeezing and the resulting splitting are therefore unchanged, with $\sigma_z$ understood as the Pauli operator along $\mathbf n$.

To make the effect of Rashba coupling explicit, consider a canting in the $xz$ plane,
\begin{equation}
\mathbf n=(\sin\vartheta,0,\cos\vartheta),
\label{eq:canted-n-xz}
\end{equation}
and define the rotated spin axes 
\begin{equation}
\begin{aligned}
\sigma_{z'}&=\sin\vartheta\,\sigma_x+\cos\vartheta\,\sigma_z,\\
\sigma_{x'}&=\cos\vartheta\,\sigma_x-\sin\vartheta\,\sigma_z,\\
\sigma_{y'}&=\sigma_y,
\end{aligned}
\end{equation}
which aligns the N\'eel vector with $\sigma_z'$ at the expense of rotating the Rashba
\begin{equation}
H_R=\alpha_R\left(
k_x\sigma_{y'}
-k_y\cos\vartheta\,\sigma_{x'}
-k_y\sin\vartheta\,\sigma_{z'}
\right).
\label{eq:canted-rashba}
\end{equation}
The last term is longitudinal with respect to the altermagnetic qubit axis, while the first two are transverse. In particular, the $x$-directed electric drive used in the main text activates the $k_x\sigma_{y'}$ channel and remains purely transverse for every $\vartheta$. More generally, for any uniform $\mathbf n$ one can choose an in-plane electric-field direction whose Rashba drive is perpendicular to $\mathbf n$, so canting does not obstruct EDSR.

Because $H_R$ is odd in momentum, its direct static expectation value in a centered ground orbital vanishes. The leading orientation-dependent static correction is second order in $\alpha_R$. At the circular-dot point, a Schrieffer--Wolff projection to first order in $\eta$ and second order in $\alpha_R$ gives
\begin{equation}
\delta H_{q}^{(2)}=\frac{\eta}{\ell_\mathrm{so}^2}
\left(n_x\sigma_x-n_y\sigma_y\right),
\label{eq:canted-static}
\end{equation}
For the representative $xz$-plane tilt this becomes
\begin{equation}
\delta H_{q}^{(2)}
=
\frac{\eta}{\ell_\mathrm{so}^2}
\left(
\sin^2\vartheta\,\sigma_{z'}
+\sin\vartheta\cos\vartheta\,\sigma_{x'}
\right).
\label{eq:canted-static-xz}
\end{equation}
The first term shifts the longitudinal splitting, whereas the second tilts the qubit axis. Near the out-of-plane configuration the axis-tilting correction scales as $O(\eta m^2\alpha_R^2\vartheta)$, while the longitudinal shift starts at $O(\vartheta^2)$. Consequently, for $\mathbf n\parallel\hat{\mathbf z}$ the circular-dot crossing remains unchanged; for $\mathbf n$ along an in-plane principal crystal axis the correction is collinear and shifts the zero-splitting point to a weakly elliptical dot; and for a generic partially tilted direction it produces a negligible avoided crossing $O(\eta m^2\alpha_R^2\vartheta)$. Eq.~\eqref{eq:canted-static} is the new correction at the circular point; away from that point it contributes to the leading ellipticity-dependent field and to the Rashba renormalization already derived in Appendix~\ref{app:eigenstates}.

The first-order charge-noise sweet spot is unaffected by a uniform canting because a uniform electric field still only translates the harmonic confinement. Likewise, at $\alpha_R=0$ the coupling of smooth fixed-length N\'eel fluctuations remains purely transverse after the same global spin rotation. For a generic canted equilibrium direction, however, the small correction in \cref{eq:canted-static} makes the true qubit axis differ slightly from $\mathbf n$ and permits a parametrically negligible longitudinal noise component of order $O[(\eta/\ell_\mathrm{so}^2)\sin\vartheta\,\delta n]$.

For the $xz$-plane canting of \cref{eq:canted-n-xz}, the double-dot geometry used in the main text remains particularly simple. Choosing the inter-dot axis along $\hat{\mathbf x}$ gives a Rashba tunneling axis along $\hat{\mathbf y}$, which is perpendicular to $\mathbf n$ for every $\vartheta$. Within the Rashba model considered here, the spin--orbit tunneling is therefore purely transverse, and the exchange and Pauli-blockade selection rules used in the main text remain unchanged. For a more general canting, the same result is obtained by aligning the double-dot axis with the in-plane projection of $\mathbf n$; other orientations can introduce a longitudinal spin--orbit tunneling component.

\section{Confined eigenstates}
\label{app:eigenstates}

We consider the confined continuum Hamiltonian
\begin{equation}
\label{eq:confined-continuum-hamiltonian}
H
=
H_{\eff}+
\frac{m}{2}\left(\omega_x^2 x^2+\omega_y^2 y^2\right)\sigma_0.
\end{equation}
In the regime where $\alpha_R$ is small compared to the orbital level spacing, the static qubit basis is obtained by first setting $\alpha_R=0$.

At $\alpha_R=0$, the Hamiltonian commutes with $\sigma_z$, so it decomposes into spin sectors $s_z=\pm 1$,
\begin{equation}
\label{eq:spin-sector-hamiltonian}
\begin{split}
H_s
&=
\left(\frac{1}{2m}-s_z\eta\right)k_x^2+\left(
\frac{1}{2m}+s_z\eta\right)k_y^2
\\
&\quad
+
\frac{m}{2}\omega_x^2 x^2
+
\frac{m}{2}\omega_y^2 y^2.
\end{split}
\end{equation}
Each spin sector is therefore an anisotropic harmonic oscillator with frequencies
\begin{equation}
\label{eq:spin-sector-frequencies}
\Omega_{x,s_z}
=
\omega_x\sqrt{1-2s_zm\eta},
\qquad
\Omega_{y,s_z}
=
\omega_y\sqrt{1+2s_zm\eta},
\end{equation}
and oscillator lengths
\begin{equation}
\label{eq:spin-sector-lengths}
\begin{aligned}
\ell_{x,s_z}
&=
\ell_x\left(1-2s_zm\eta\right)^{1/4},
\\
\ell_{y,s_z}
&=
\ell_y\left(1+2s_zm\eta\right)^{1/4},
\end{aligned}
\end{equation}
with
\begin{equation}
\label{eq:bare-lengths}
\ell_x=\sqrt{\frac{1}{m\omega_x}},
\qquad
\ell_y=\sqrt{\frac{1}{m\omega_y}}.
\end{equation}
The ground state in sector $s$ is
\begin{equation}
\label{eq:spin-sector-ground-state}
\phi_{s_z}(x,y)
=
\frac{1}{\sqrt{\pi\,\ell_{x,s}\ell_{y,s}}}
\exp\!\left[
-\frac{x^2}{2\ell_{x,s_z}^2}
-\frac{y^2}{2\ell_{y,s_z}^2}
\right],
\end{equation}
with energy
\begin{equation}
\label{eq:spin-sector-ground-state-energy}
E_{s_z}
=
\frac{1}{2}\Omega_{x,s_z}
+
\frac{1}{2}\Omega_{y,s_z}.
\end{equation}

The exact qubit energy splitting at $\alpha_R=0$ is, writing $E_\uparrow\equiv E_{s_z=+1}$ and $E_\downarrow\equiv E_{s_z=-1}$ for the energies of $\ket{\uparrow}$ and $\ket{\downarrow}$,
\begin{equation}
\label{eq:delta-q-exact}
\omega_q
\equiv
E_\uparrow-E_\downarrow
=
\frac{1}{2}\left(\omega_x-\omega_y\right)
\left[
\sqrt{1-2m\eta}
-
\sqrt{1+2m\eta}
\right].
\end{equation}
For $|m\eta|\ll 1$, the widths reduce to
\begin{equation}
\label{eq:small-beta-width-expansion}
\begin{aligned}
\ell_{x,\pm}
&=
\ell_x\left(1\mp \frac{m\eta}{2}\right)
+
\mathcal{O}(m^2\eta^2),\\
\ell_{y,\pm}
&=
\ell_y\left(1\pm \frac{m\eta}{2}\right)
+
\mathcal{O}(m^2\eta^2),
\end{aligned}
\end{equation}
and the qubit splitting becomes
\begin{equation}
\label{eq:delta-q-small-beta}
\omega_q
=
\eta
\left(
\frac{1}{\ell_y^2}
-
\frac{1}{\ell_x^2}
\right)
+
\mathcal{O}(\eta^3).
\end{equation}
Thus the static qubit eigenstates are squeezed Gaussians with opposite spin polarization. The splitting is generated by the combined action of the $d$-wave altermagnetic term and the ellipticity of the confinement. In particular, for a circular dot, $\ell_x=\ell_y$, one has $\omega_q=0$ at this order.

\subsection{Perturbative effect of Rashba spin--orbit coupling}
\label{app:perturbative-effect-rashba}

We now treat
\begin{equation}
\label{eq:rashba-perturbation}
V_R
=
\alpha_R\left(\sigma_y k_x-\sigma_x k_y\right)
\end{equation}
perturbatively in the static basis, with qubit eigenstates $\ket{\uparrow},\ket{\downarrow}$ [\cref{eq:qubit-basis-def}]. Since $V_R$ is odd in momentum and flips spin, its matrix elements within the qubit doublet vanish,
\begin{equation}
\label{eq:rashba-vanishing-matrix-elements}
\bra{a}V_R\ket{b}=0,
\qquad
a,b\in\{\uparrow,\downarrow\}.
\end{equation}
Thus Rashba spin--orbit coupling does not split the qubit at first order.

Its leading static effect appears at second order. In the exact oscillator basis of the spin-sector Hamiltonians, the second-order correction to the energy of the qubit state with spin $s_z=\pm1$ is
\begin{equation}
\label{eq:rashba-second-order-energy-exact}
\delta E_{s_z}^{(2)}
=
\sum_{n_x,n_y}
\frac{
\left|
\bra{n_x,n_y,-s_z}V_R\ket{0,0,s_z}
\right|^2
}{
E_{0,0,s_z}-E_{n_x,n_y,-s_z}
}.
\end{equation}
This expression is the complete second-order Rashba correction within the confined continuum model. Only opposite-spin excited orbitals with odd parity in $x$ or $y$ contribute.

For the purpose of estimating the correction analytically, it is sufficient to keep the leading order in $\eta$ inside the matrix elements and use the bare oscillator basis. Since
\begin{equation}
\label{eq:momentum-action-ground-state}
k_x\ket{0,0}
=
i\frac{1}{\sqrt{2}\ell_x}\ket{1,0},
\qquad
k_y\ket{0,0}
=
i\frac{1}{\sqrt{2}\ell_y}\ket{0,1},
\end{equation}
one obtains
\begin{equation}
\label{eq:rashba-second-order-energy-leading}
\begin{split}
\delta E_{s_z}^{(2)}
&=
-\frac{\left(\alpha_R\right)^2}{2}
\biggl[
\frac{1}{\ell_x^2\left(\omega_x-s_z\omega_q\right)}
\\
&\qquad
+
\frac{1}{\ell_y^2\left(\omega_y-s_z\omega_q\right)}
\biggr]
+
\mathcal{O}\left(\left(\alpha_R\right)^2\eta^2\right).
\end{split}
\end{equation}
Therefore the Rashba-renormalized qubit splitting is, evaluating \cref{eq:rashba-second-order-energy-leading} at $s_z=+1$ and $s_z=-1$,
\begin{equation}
\label{eq:rashba-renormalized-qubit-splitting}
\widetilde{\omega}_q
=
\omega_q+\delta\omega_R^{(2)},
\qquad
\delta\omega_R^{(2)}
=
\delta E_{\uparrow}^{(2)}-\delta E_{\downarrow}^{(2)}.
\end{equation}
Equivalently,
\begin{equation}
\label{eq:rashba-second-order-splitting-leading}
\begin{split}
\delta\omega_R^{(2)}
&=
-\left(\alpha_R\right)^2\omega_q
\biggl[
\frac{1}{\ell_x^2\left(\omega_x^2-\omega_q^2\right)}
\\
&\qquad
+
\frac{1}{\ell_y^2\left(\omega_y^2-\omega_q^2\right)}
\biggr]
+
\mathcal{O}\left(\left(\alpha_R\right)^2\eta^2\right).
\end{split}
\end{equation}
Thus Rashba coupling does not generate the static qubit splitting by itself. Instead, it gives a perturbative renormalization of the splitting already produced by the altermagnetic anisotropy and the elliptic confinement. In particular, when $\omega_q=0$, the two states remain degenerate at this order.

\section{Electric-dipole spin resonance}
\label{app:edsr}

In this appendix, we derive the Rabi frequency for the qubit under an in-plane time-dependent electric field by adapting the reasoning given in Refs.~\cite{Rashba_2003_Orbital,Bosco_2021_Squeezed}. We start from the confined single-particle Hamiltonian before projection onto the qubit subspace,
\begin{equation}
\label{eq:edsr-starting-hamiltonian}
H
=
H_{\text{stat}}+H_R+H_F,
\end{equation}
with
\begin{equation}
\label{eq:edsr-static-part}
H_{\text{stat}}
=
\frac{\bm{k}^2}{2m}
-
\eta\left(k_x^2-k_y^2\right)\sigma_z
+
\frac{1}{2}m\left(\omega_x^2x^2+\omega_y^2y^2\right),
\end{equation}
\begin{equation}
\label{eq:edsr-rashba-part}
H_R
=
\alpha_R\left(\sigma_yk_x-\sigma_xk_y\right),
\end{equation}
and the time-dependent electric-field perturbation
\begin{equation}
\label{eq:edsr-electric-perturbation}
H_F
=
-e\mathbf{F}(t)\cdot\mathbf{r}
=
-eF_x(t)x-eF_y(t)y.
\end{equation}
We assume that the orbital confinement energy is much larger than the qubit splitting. For a harmonic potential, a spatially uniform electric field produces a rigid displacement of the parabola minimum:
\begin{equation}
\label{eq:edsr-displacement-components}
\delta x(t)=\frac{eF_x(t)}{m\omega_x^2},
\qquad
\delta y(t)=\frac{eF_y(t)}{m\omega_y^2}.
\end{equation}

To treat this exactly at the orbital level, we move to a frame co-moving with the displaced dot by applying the unitary transformation
\begin{equation}
\label{eq:edsr-translation-unitary}
\mathcal{V}(t)
=
\exp\left(-i\,\bm{k}\cdot\delta\bm{x}(t)\right),
\end{equation}
which translates position space by $\delta\bm{x}(t)$. Under this transformation the electric-field term $H_F$ is canceled, but since $\mathcal{V}$ is time-dependent, an inertial term arises from the time derivative. Up to an irrelevant scalar shift, the transformed Hamiltonian is
\begin{equation}
\label{eq:edsr-transformed-hamiltonian}
\tilde{H}
=
\mathcal{V}^\dagger H \mathcal{V}
-
i\,\mathcal{V}^\dagger\partial_t\mathcal{V}
=
H_{\text{stat}}+H_R-\bm{k}\cdot\partial_t\delta\bm{x}(t).
\end{equation}
Thus the electric drive appears as the inertial coupling
\begin{equation}
\label{eq:edsr-inertial-coupling}
H_{\mathrm{in}}(t)
=
-\bm{k}\cdot\partial_t\delta\bm{x}(t).
\end{equation}

We now project the inertial coupling onto the low-energy qubit subspace. This projection can be performed directly, without using the static Rashba-dressed wavefunctions. To leading order in $\alpha_R$, the spin-electric coupling arises from a second-order virtual process involving one insertion of $H_R$ and one insertion of $H_{\mathrm{in}}(t)$. Denoting by $P$ the projector onto the two orbital ground states and by $Q=1-P$ the projector onto excited orbitals, the leading projected coupling is
\begin{equation}
\label{eq:edsr-second-order-projection}
\begin{split}
H_{\mathrm{EDSR}}(t)
&=
P\biggl[
H_R Q\frac{1}{E_0-QH_{\mathrm{harm}}Q}QH_{\mathrm{in}}(t)
\\
&\qquad
+
H_{\mathrm{in}}(t)Q\frac{1}{E_0-QH_{\mathrm{harm}}Q}QH_R
\biggr]P.
\end{split}
\end{equation}
Here $H_{\mathrm{harm}}$ denotes the spin-independent harmonic oscillator Hamiltonian and $E_0$ its ground-state energy. The relevant virtual transitions are the first excited orbitals,
\begin{equation}
\label{eq:edsr-virtual-orbital-elements}
k_x\ket{0,0}
=
i\frac{1}{\sqrt{2}\ell_x}\ket{1,0},
\qquad
k_y\ket{0,0}
=
i\frac{1}{\sqrt{2}\ell_y}\ket{0,1},
\end{equation}
with excitation energies $\omega_x$ and $\omega_y$. Evaluating the matrix elements gives
\begin{equation}
\label{eq:edsr-effective-spin-drive-alpha}
H_{\mathrm{EDSR}}(t)
=
-e\frac{\alpha_R}{\omega_y^2}\partial_tF_y(t)\,\sigma_x
-e\frac{\alpha_R}{\omega_x^2}\partial_tF_x(t)\,\sigma_y.
\end{equation}
Equivalently, introducing the Rashba spin--orbit length
\begin{equation}
\label{eq:edsr-spin--orbit-length}
\ell_{\mathrm{so}}
=
\frac{1}{m\alpha_R},
\end{equation}
we can write
\begin{equation}
\label{eq:edsr-dipole-definitions}
\begin{aligned}
d_x(t)
&=
-\frac{m\ell_x^4}{\ell_{\mathrm{so}}}\,\partial_tF_x(t),
\\
d_y(t)
&=
-\frac{m\ell_y^4}{\ell_{\mathrm{so}}}\,\partial_tF_y(t).
\end{aligned}
\end{equation}
The projected driving Hamiltonian is therefore
\begin{equation}
\label{eq:edsr-projected-driving-hamiltonian}
H_{\mathrm{EDSR}}(t)
=
ed_y(t)\,\sigma_x
+
ed_x(t)\,\sigma_y.
\end{equation}
The $x$ component of the electric field drives $\sigma_y$ rotations, while the $y$ component drives $\sigma_x$ rotations. This follows directly from the Rashba structure $\sigma_y k_x-\sigma_x k_y$.

The full single-qubit Hamiltonian in the co-moving frame is
\begin{equation}
\label{eq:edsr-full-qubit-hamiltonian}
H_q(t)
=
\frac{\widetilde{\omega}_q}{2}\sigma_z
+
ed_y(t)\,\sigma_x
+
ed_x(t)\,\sigma_y.
\end{equation}
For a resonant electric field, $F_j(t)=F_j^{\mathrm{ac}}\cos(\widetilde{\omega}_q t)$, the time derivative of the field provides a transverse drive at the qubit frequency. With the convention of \cref{eq:edsr-full-qubit-hamiltonian}, the corresponding Rabi scale is
\begin{equation}
\label{eq:edsr-rabi-frequency}
\Omega_R
=
e\alpha_R\widetilde{\omega}_q
\sqrt{
\frac{\left(F_x^{\mathrm{ac}}\right)^2}{\omega_x^4}
+
\frac{\left(F_y^{\mathrm{ac}}\right)^2}{\omega_y^4}
}.
\end{equation}

\section{Higher-order altermagnets}
\label{app:higher-order-altermagnets}

This appendix derives the geometric selection rule quoted in \cref{sec:single-electron-qubit}: which altermagnetic form factors a gate-defined dot can convert into a qubit splitting, and how that conversion depends on the dot shape and orientation.

For a general altermagnet characterized by $\ell=d,g,i$-wave symmetry, to leading order in the altermagnetic coupling the splitting is just the dot average of the collinear form factor,
\begin{equation}
\label{eq:app-ff-omega}
\omega_q=-2\eta_\ell\,\langle g_\ell(\bm k)\rangle_\mathrm{dot},
\end{equation}
so the confined orbital acts as a momentum-space filter that samples $g_\ell(\bm k)$ over the spread of momenta it occupies. For a harmonic dot this spread is Gaussian,
\begin{equation}
\label{eq:general-gaussian-momentum-distribution}
|\phi_0(\bm k)|^2\propto\exp\!\left[-\tfrac12 k_i(M^{-1})_{ij}k_j\right],
\quad M_{ij}=\langle k_i k_j\rangle,
\end{equation}
and all of the shape information sits in the second-moment tensor $M$. Its isotropic part cannot produce a splitting; the only anisotropy a Gaussian dot offers is its \emph{quadrupole}: the ellipticity $M_a-M_b$, set by the principal inverse-widths $M_{a,b}=1/(2\ell_{a,b}^2)$, and the orientation $\theta_\mathrm{d}$ of the dot axes relative to the crystal axes,
\begin{align}
\label{eq:rotated-moment-tensor}
M_{xx}-M_{yy}&=(M_a-M_b)\cos2\theta_\mathrm{d},
\\
2M_{xy}&=(M_a-M_b)\sin2\theta_\mathrm{d}.
\end{align}

This already fixes which form factors couple. An altermagnetic harmonic of angular order $2n$, $d$-, $g$-, or $i$-wave for $2n=2,4,6$, is a degree-$2n$ polynomial in $\bm k$, so averaging it over the Gaussian assembles it from products of $M$, each of which can supply at most one unit of the dot's quadrupolar anisotropy. A harmonic of order $2n$ can therefore survive only at order $n$ in the ellipticity. Making this explicit with the complex momentum $z=k_x+ik_y$, we have $z^2=k_x^2-k_y^2+2ik_xk_y$, which averages to $\langle z^2\rangle=(M_a-M_b)e^{2i\theta_\mathrm{d}}$ by \cref{eq:rotated-moment-tensor}. Because $z^{2n}$ contains only $z$ and no $\bar z$, its Gaussian average is built entirely from this one pairing, $\langle z^{2n}\rangle=(2n-1)!!\,\langle z^2\rangle^n$, where $(2n-1)!!$ counts the ways of pairing the $2n$ factors. This gives the compact selection rule
\begin{equation}
\label{eq:general-harmonic-average}
\big\langle \mathrm{Re}\!\left[e^{-i\varphi}z^{2n}\right]\big\rangle
=(2n-1)!!\,(M_a-M_b)^n\cos(2n\theta_\mathrm{d}-\varphi).
\end{equation}

In physical terms, a circular dot ($M_a=M_b$) averages every anisotropic harmonic to zero and gives no splitting at any order, while an elliptic dot couples to a $d$-wave edge linearly in the ellipticity, to a $g$-wave edge quadratically, and to an $i$-wave edge cubically. The orientation $\theta_\mathrm{d}$, through the phase $\varphi$ that the crystal symmetry of the band edge fixes, selects which member of the angular multiplet is sampled and, for the higher harmonics, whether an axis-aligned ellipse couples at all.

For $n=1$, taking $g_d(\bm k)=k_x^2-k_y^2=\mathrm{Re}\,z^2$ gives
\begin{equation}
\label{eq:appendix-d-wave-average}
\langle g_d(\bm k)\rangle
=M_{xx}-M_{yy}
=(M_a-M_b)\cos2\theta_\mathrm{d} .
\end{equation}
For the axis-aligned dots used in the main text, this reduces to $\langle g_d\rangle=1/(2\ell_x^2)-1/(2\ell_y^2)$. With the sign convention of \cref{eq:continuum-hamiltonian}, $H_\mathrm{AM}=-\eta g_d\sigma_z$, the projected field is $h_z=-\eta\langle g_d\rangle$ and therefore $\omega_q=2h_z\simeq\eta(1/\ell_y^2-1/\ell_x^2)$.

For $n=2$, the two real $g$-wave basis functions are proportional to $\mathrm{Re}\,z^4$ and $\mathrm{Im}\,z^4$. The component
\begin{equation}
\label{eq:appendix-g-wave-basis}
g_g(\bm k)=k_xk_y(k_x^2-k_y^2)=\frac{1}{4}\mathrm{Im}\,z^4
\end{equation}
has the dot average
\begin{equation}
\label{eq:appendix-g-wave-average}
\langle g_g(\bm k)\rangle
=\frac{3}{4}(M_a-M_b)^2\sin4\theta_\mathrm{d}
=3M_{xy}(M_{xx}-M_{yy}).
\end{equation}
This component is invisible to an axis-aligned elliptic Gaussian but becomes finite once the dot is rotated. The orthogonal component, proportional to $\mathrm{Re}\,z^4$, is instead proportional to $(M_a-M_b)^2\cos4\theta_\mathrm{d}$ and can be accessed by an axis-aligned ellipse. Which linear combination appears is fixed by the crystal symmetry of the specific altermagnetic band edge.

The $i$-wave case ($n=3$) follows identically, now cubic in the ellipticity, $\langle z^6\rangle\propto(M_a-M_b)^3$, so the splitting is correspondingly weaker and more orientation-sensitive.

We finally comment on Rashba spin--orbit coupling and EDSR in this notation. The altermagnetic term considered in this appendix is always collinear, $H_\mathrm{AM}=-\eta_\ell g_\ell(\bm k)\sigma_z$, whereas Rashba coupling is a separate perturbation $H_R=\alpha_R(\sigma_y k_x-\sigma_x k_y)$. Since $H_R$ is odd in momentum, its direct ground-orbital expectation value vanishes for a centered harmonic dot. Its leading role for control is instead activated by the time-dependent displacement of the orbital under an ac electric field.

Let the harmonic confinement be written as
\begin{equation}
\label{eq:rotated-dot-curvature}
V(\bm r)=\frac{1}{2}r_iK_{ij}r_j,
\end{equation}
where $K$ is the curvature matrix. An electric drive $-e\bm F(t)\cdot\bm r$ shifts the dot center by
\begin{equation}
\label{eq:rotated-dot-displacement}
\delta\bm r(t)=eK^{-1}\bm F(t).
\end{equation}
Repeating the co-moving-frame projection used in Appendix~\ref{app:edsr} gives, in the main-text sign convention,
\begin{equation}
\label{eq:appendix-general-edsr}
H_\mathrm{EDSR}
=-\frac{e}{\ell_\mathrm{so}}
\left[
(K^{-1}\partial_t\bm F)_x\sigma_y
+(K^{-1}\partial_t\bm F)_y\sigma_x
\right].
\end{equation}
For an axis-aligned dot, $K^{-1}_{xx}=1/(m\omega_x^2)=m\ell_x^4$ and $K^{-1}_{yy}=1/(m\omega_y^2)=m\ell_y^4$, recovering \cref{eq:edsr} and the corresponding $y$-drive expression. For a rotated dot, a drive along a laboratory axis generally excites both principal directions and therefore produces a linear combination of the two transverse Pauli operators. The form-factor symmetry enters EDSR primarily through the resonance condition, namely the value of $\omega_q=-2\eta_\ell\langle g_\ell\rangle_\mathrm{dot}$ selected by the dot geometry.

\section{Anisotropic exchange}
\label{sec:anisotropic-exchange}
\subsection{Spin-dependent exchange from altermagnetic orbital squeezing}
\label{sec:spin-dependent-exchange}

In the main text we mainly use the scalar-exchange Hamiltonian \cref{eq:HDQD}, appropriate when the altermagnetic squeezing only weakly modifies the exchange matrix elements. Here we show how this limit emerges from a microscopic projection and identify the leading corrections. The essential point is that the $d$-wave kinetic term makes the localized orbital associated with spin $s_z=\pm1$ spin dependent. Consequently, the inter-dot overlap, tunneling amplitudes, and Coulomb integrals acquire spin-symmetric and spin-asymmetric components.

Consider two dots $i=L,R$ centered at $\mathbf R_i=(X_i,Y_i)$. We first work in the spin basis aligned with the local N\'eel vector and neglect Rashba spin--orbit coupling in the orbital-envelope problem. The localized orbital in dot $i$ and spin sector $s_z=\pm1$ is then
\begin{equation}
\label{eq:spin-dependent-local-orbital}
\begin{aligned}
\phi_{i s_z}(\mathbf r)
&=
\frac{1}{\sqrt{\pi\ell_{x,i,s_z}\ell_{y,i,s_z}}}
\\
&\quad\times
\exp\left[
-\frac{(x-X_i)^2}{2\ell_{x,i,s_z}^2}
-\frac{(y-Y_i)^2}{2\ell_{y,i,s_z}^2}
\right],
\end{aligned}
\end{equation}
with
\begin{equation}
\label{eq:spin-dependent-local-lengths}
\begin{aligned}
\ell_{x,i,s_z}
&=
\ell_{x,i}(1-2s_z m\eta)^{1/4},
\\
\ell_{y,i,s_z}
&=
\ell_{y,i}(1+2s_z m\eta)^{1/4}.
\end{aligned}
\end{equation}
The same-spin overlap between the left and right localized orbitals is
\begin{equation}
\label{eq:spin-dependent-overlap-general}
\begin{split}
\Lambda_{s_z}
&=
\braket{\phi_{L s_z}|\phi_{R s_z}}
\\
&=
\prod_{\mu=x,y}
\left[
\frac{
2\ell_{\mu,L,s_z}\ell_{\mu,R,s_z}
}{
\ell_{\mu,L,s_z}^2+\ell_{\mu,R,s_z}^2
}
\right]^{1/2}
\\
&\quad\times
\exp\left[
-\frac{
(R_{L\mu}-R_{R\mu})^2
}{
2(\ell_{\mu,L,s_z}^2+\ell_{\mu,R,s_z}^2)
}
\right].
\end{split}
\end{equation}
For identical dots separated by $\mathbf d=\mathbf R_R-\mathbf R_L$, this reduces to
\begin{equation}
\label{eq:spin-dependent-overlap-identical}
\Lambda_{s_z}
=
\exp\left[
-\frac{d_x^2}{4\ell_{x,s_z}^2}
-\frac{d_y^2}{4\ell_{y,s_z}^2}
\right].
\end{equation}
Expanding for $|m\eta|\ll 1$ gives
\begin{equation}
\label{eq:spin-dependent-overlap-expansion}
\begin{aligned}
\Lambda_{s_z}
&\simeq
\Lambda_0
\left[
1
-
s_z m\eta
\left(
\frac{d_x^2}{4\ell_x^2}
-
\frac{d_y^2}{4\ell_y^2}
\right)
\right],\\
\Lambda_0
&=
\exp\left[
-\frac{d_x^2}{4\ell_x^2}
-\frac{d_y^2}{4\ell_y^2}
\right].
\end{aligned}
\end{equation}
This expression isolates the geometric origin of the spin dependence: for a double dot displaced mainly along $x$, the spin sector with the larger $x$-width has the larger orbital overlap, whereas for a displacement mainly along $y$ the sign of the spin-asymmetric overlap is reversed. This spin dependence of the overlap is the microscopic source of the spin-dependent tunneling amplitudes introduced below.

The one-body Hamiltonian in spin sector $s_z$ gives matrix elements
\begin{equation}
\label{eq:spin-dependent-one-body-elements}
h_{ij}^{s_z}
=
\bra{\phi_{i s_z}}h^{(s_z)}\ket{\phi_{j s_z}},
\end{equation}
where $h^{(s_z)}$ denotes the single-particle Hamiltonian projected to fixed spin sector $s_z=\pm1$ (the $\sigma_z$ eigenvalue). Orthogonalizing the two non-orthogonal localized orbitals to leading order gives the spin-dependent tunnel coupling
\begin{equation}
\label{eq:spin-dependent-tunnel}
\tau_{s_z}
\simeq
-\frac{
h_{LR}^{s_z}
-
S_{s_z}(h_{LL}^{s_z}+h_{RR}^{s_z})/2
}{
1-S_{s_z}^2
}.
\end{equation}
It is useful to decompose it into spin-symmetric and spin-asymmetric parts, writing $\tau_\uparrow\equiv\tau_{s_z=+1}$ and $\tau_\downarrow\equiv\tau_{s_z=-1}$ for the two evaluated sectors,
\begin{equation}
\label{eq:tunnel-symmetric-asymmetric}
\tau_c
=
\frac{\tau_\uparrow+\tau_\downarrow}{2},
\qquad
\delta\tau
=
\frac{\tau_\uparrow-\tau_\downarrow}{2}.
\end{equation}

The interaction part is obtained from Coulomb matrix elements between the same localized orbitals. For compactness, define the transition density
\begin{equation}
\label{eq:transition-density-definition}
\rho_{ij}^{s_z}(\mathbf q)
=
\int d^2r\,
\phi_{i s_z}^*(\mathbf r)
e^{i\mathbf q\cdot\mathbf r}
\phi_{j s_z}(\mathbf r).
\end{equation}
Then the Coulomb integrals are
\begin{equation}
\label{eq:coulomb-integral-four-index}
V_{ij;kl}^{s_z s_z'}
=
\int
\frac{d^2q}{(2\pi)^2}
V(q)\,
\rho_{ik}^{s_z}(\mathbf q)
\rho_{jl}^{s_z'}(-\mathbf q),
\end{equation}
where $V(q)$ is the screened two-dimensional Coulomb interaction. The relevant combinations are the on-site repulsion, inter-dot density interaction, and direct exchange,
\begin{equation}
\label{eq:hubbard-coulomb-definitions}
\begin{aligned}
U_i^{s_z s_z'}
&=
V_{ii;ii}^{s_z s_z'},
\\
V_{s_z s_z'}
&=
V_{LR;LR}^{s_z s_z'},
\\
K_{s_z s_z'}
&=
V_{LR;RL}^{s_z s_z'}.
\end{aligned}
\end{equation}
For the Gaussian orbitals in \cref{eq:spin-dependent-local-orbital}, the transition density factorizes as
\begin{equation}
\label{eq:transition-density-factorization}
\rho_{ij}^{s_z}(\mathbf q)
=
\rho_{ij,x}^{s_z}(q_x)
\rho_{ij,y}^{s_z}(q_y),
\end{equation}
with
\begin{equation}
\label{eq:transition-density-one-direction}
\begin{split}
\rho_{ij,\mu}^{s_z}(q_\mu)
&=
\left[
\frac{
2\ell_{\mu,i,s_z}\ell_{\mu,j,s_z}
}{
\ell_{\mu,i,s_z}^2+\ell_{\mu,j,s_z}^2
}
\right]^{1/2}
\\
&\quad\times
\exp\left[
-\frac{
(R_{i\mu}-R_{j\mu})^2
}{
2(\ell_{\mu,i,s_z}^2+\ell_{\mu,j,s_z}^2)
}
\right]
\\
&\quad\times
\exp\Bigl[
-\frac{
q_\mu^2
\ell_{\mu,i,s_z}^2
\ell_{\mu,j,s_z}^2
}{
2(\ell_{\mu,i,s_z}^2+\ell_{\mu,j,s_z}^2)
}
\\
&\qquad\qquad
+
iq_\mu
\frac{
R_{i\mu}\ell_{\mu,j,s_z}^2
+
R_{j\mu}\ell_{\mu,i,s_z}^2
}{
\ell_{\mu,i,s_z}^2+\ell_{\mu,j,s_z}^2
}
\Bigr].
\end{split}
\end{equation}

Keeping the dominant density, tunneling, and direct-exchange terms, the projected extended Hubbard Hamiltonian is
\begin{equation}
\label{eq:spin-dependent-hubbard}
\begin{split}
H_{\mathrm{Hub}}
&=
\sum_{i=L,R}\sum_{s_z}
\epsilon_{i s_z}n_{i s_z}
\\
&\quad
-
\sum_{s_z}
\tau_{s_z}
\left(
c_{L s_z}^\dagger c_{R s_z}
+
c_{R s_z}^\dagger c_{L s_z}
\right)
\\
&\quad
+
\sum_i
U_i^{\uparrow\downarrow}
n_{i\uparrow}n_{i\downarrow}
+
\sum_{s_z,s_z'}
V_{s_z s_z'}n_{L s_z}n_{R s_z}
\\
&\quad
-
\sum_{s_z,s_z'}
K_{s_z s_z'}
\\
&\qquad\times
c_{L s_z}^\dagger c_{R s_z'}^\dagger c_{L s_z'}c_{R s_z}.
\end{split}
\end{equation}
We now eliminate the doubly occupied charge states perturbatively. The energy of a $(1,1)$ configuration is
\begin{equation}
\label{eq:energy-one-one}
E_{LR}^{s_z s_z'}
=
\epsilon_{L s_z}
+
\epsilon_{R s_z'}
+
V_{s_z s_z'}.
\end{equation}
The energies of the two virtual doubly occupied states are
\begin{equation}
\label{eq:energy-double-occupancy}
\begin{aligned}
E_L^{s_z s_z'}
&=
\epsilon_{L s_z}
+
\epsilon_{L s_z'}
+
U_L^{s_z s_z'},
\\
E_R^{s_z s_z'}
&=
\epsilon_{R s_z}
+
\epsilon_{R s_z'}
+
U_R^{s_z s_z'}.
\end{aligned}
\end{equation}
The corresponding charge excitation energies are
\begin{equation}
\label{eq:charge-excitation-energies}
\Delta_L^{s_z s_z'}
=
E_L^{s_z s_z'}-E_{LR}^{s_z s_z'},
\qquad
\Delta_R^{s_z s_z'}
=
E_R^{s_z s_z'}-E_{LR}^{s_z s_z'}.
\end{equation}

The transverse exchange matrix element between $\ket{\uparrow_L\downarrow_R}$ and $\ket{\downarrow_L\uparrow_R}$ receives a superexchange contribution proportional to $\tau_\uparrow \tau_\downarrow$ and a direct-exchange contribution $K_{\uparrow\downarrow}$. The resulting coefficient of $\sigma_x^L\sigma_x^R+\sigma_y^L\sigma_y^R$ is
\begin{equation}
\label{eq:Jperp-spin-dependent}
J_\perp
=
\tau_\uparrow \tau_\downarrow
\left[
\frac{1}{\Delta_L^{\uparrow\downarrow}}
+
\frac{1}{\Delta_L^{\downarrow\uparrow}}
+
\frac{1}{\Delta_R^{\uparrow\downarrow}}
+
\frac{1}{\Delta_R^{\downarrow\uparrow}}
\right]
-
2K_{\uparrow\downarrow}.
\end{equation}
Using \cref{eq:tunnel-symmetric-asymmetric}, the superexchange part contains the product
\begin{equation}
\label{eq:tunnel-product-symmetric-asymmetric}
\tau_\uparrow \tau_\downarrow
=
\tau_c^2-\delta\tau^2.
\end{equation}
Thus, in a left-right symmetric device, the spin-asymmetric tunneling correction to the transverse exchange enters quadratically, while left-right asymmetry can generate linear corrections through the denominators and Coulomb integrals.

The diagonal energies in the $(1,1)$ manifold are
\begin{equation}
\label{eq:diagonal-energy-up-up}
E_{\uparrow\uparrow}
=
\epsilon_{L\uparrow}
+
\epsilon_{R\uparrow}
+
V_{\uparrow\uparrow}
-
K_{\uparrow\uparrow},
\end{equation}
\begin{equation}
\label{eq:diagonal-energy-down-down}
E_{\downarrow\downarrow}
=
\epsilon_{L\downarrow}
+
\epsilon_{R\downarrow}
+
V_{\downarrow\downarrow}
-
K_{\downarrow\downarrow},
\end{equation}
\begin{equation}
\label{eq:diagonal-energy-up-down}
E_{\uparrow\downarrow}
=
\epsilon_{L\uparrow}
+
\epsilon_{R\downarrow}
+
V_{\uparrow\downarrow}
-
\frac{\tau_\downarrow^2}{\Delta_L^{\uparrow\downarrow}}
-
\frac{\tau_\uparrow^2}{\Delta_R^{\uparrow\downarrow}},
\end{equation}
\begin{equation}
\label{eq:diagonal-energy-down-up}
E_{\downarrow\uparrow}
=
\epsilon_{L\downarrow}
+
\epsilon_{R\uparrow}
+
V_{\downarrow\uparrow}
-
\frac{\tau_\uparrow^2}{\Delta_L^{\downarrow\uparrow}}
-
\frac{\tau_\downarrow^2}{\Delta_R^{\downarrow\uparrow}}.
\end{equation}
The longitudinal exchange is therefore
\begin{equation}
\label{eq:Jz-from-diagonal-energies}
J_z
=
E_{\uparrow\uparrow}
+
E_{\downarrow\downarrow}
-
E_{\uparrow\downarrow}
-
E_{\downarrow\uparrow}.
\end{equation}
Equivalently,
\begin{equation}
\label{eq:Jz-spin-dependent-expanded}
\begin{split}
J_z
&=
V_{\uparrow\uparrow}
+
V_{\downarrow\downarrow}
-
V_{\uparrow\downarrow}
-
V_{\downarrow\uparrow}
-
K_{\uparrow\uparrow}
-
K_{\downarrow\downarrow}
\\
&\quad
+
\frac{\tau_\downarrow^2}{\Delta_L^{\uparrow\downarrow}}
+
\frac{\tau_\uparrow^2}{\Delta_R^{\uparrow\downarrow}}
+
\frac{\tau_\uparrow^2}{\Delta_L^{\downarrow\uparrow}}
+
\frac{\tau_\downarrow^2}{\Delta_R^{\downarrow\uparrow}}.
\end{split}
\end{equation}
The same diagonal energies also generate exchange-induced shifts of the local spin splittings,
\begin{equation}
\label{eq:exchange-induced-local-fields}
\begin{aligned}
\delta B_L
&=
\frac{
E_{\uparrow\uparrow}
+
E_{\uparrow\downarrow}
-
E_{\downarrow\uparrow}
-
E_{\downarrow\downarrow}
}{2},\\
\delta B_R
&=
\frac{
E_{\uparrow\uparrow}
-
E_{\uparrow\downarrow}
+
E_{\downarrow\uparrow}
-
E_{\downarrow\downarrow}
}{2}.
\end{aligned}
\end{equation}
The effective Hamiltonian in the $(1,1)$ charge sector is then
\begin{equation}
\label{eq:spin-dependent-exchange-final}
\begin{split}
H_{\mathrm{eff}}^{(1,1)}
&=
C
+
\frac{\widetilde\omega_{q,L}}{2}\sigma_z^L
+
\frac{\widetilde\omega_{q,R}}{2}\sigma_z^R
+
\frac{J_\perp}{4}
\left(
\sigma_x^L\sigma_x^R
+
\sigma_y^L\sigma_y^R
\right)
\\
&\quad
+
\frac{J_z}{4}
\left(
\sigma_z^L\sigma_z^R-\mathbb 1
\right),
\end{split}
\end{equation}
with
\begin{equation}
\label{eq:renormalized-local-splittings}
\widetilde\omega_{q,L}
=
\omega_{q,L}
+
\delta B_L,
\qquad
\widetilde\omega_{q,R}
=
\omega_{q,R}
+
\delta B_R.
\end{equation}
In the spin-independent limit,
\begin{equation}
\label{eq:isotropic-exchange-limit}
\tau_\uparrow=\tau_\downarrow=\tau,
\qquad
\Delta_L^{s_z s_z'}=\Delta_R^{s_z s_z'}=E_C,
\qquad
K_{s_z s_z'}=K,
\end{equation}
one obtains
\begin{equation}
\label{eq:isotropic-exchange-value}
J_\perp
=
J_z
=
\frac{4\tau^2}{E_C}
-
2K
\equiv
J.
\end{equation}
Therefore \cref{eq:spin-dependent-exchange-final} reduces, up to an irrelevant constant and calibrated local splittings, to the scalar exchange without spin-dependent tunneling.

To make explicit the size of the corrections to the scalar-exchange
limit, let us expand the spin-dependent tunnel couplings in the small
dimensionless squeezing parameter $m\eta$ and defining $\tau^\text{(n)}=\frac{1}{n!}\partial^{(n)}_{m\eta}\tau_{s_z}$, we get
\begin{equation}
\begin{aligned}
        \tau_c &= \tau^{(0)}+(m\eta)^2 \tau^{(2)}
        +\mathcal{O}(m^4\eta^4),\\
        \delta\tau &= m\eta \tau^{(1)}
        +\mathcal{O}(m^3\eta^3).
\end{aligned}
\end{equation}
The absence of a term linear in $m\eta$ in $\tau_c$ follows from
the fact that $\tau_c$ is the spin-symmetric part of the tunneling, whereas
$\delta\tau$ is the spin-antisymmetric part generated by the
altermagnetic orbital squeezing.

Another interesting limit, typical in quantum dots, is the charging-energy-dominated regime. In this scenario, the spin dependence of the charging energies may be neglected to leading order,
\begin{equation}
        \Delta^{s_z s_z'}_L \simeq \Delta^{s_z s_z'}_R
        \simeq E_C ,
\end{equation}
and the direct Coulomb exchange may similarly be replaced by its spin-independent value, $K_{s_z s_z'}\simeq K$. In this scenario, the exchange becomes
\begin{align}
        J_\perp &\simeq \frac{4\tau_\uparrow \tau_\downarrow}{E_C}-2K = \frac{4}{E_C}\left(\tau_c^2-\delta\tau^2\right)-2K ,
        \label{eq:Jperp_eta_scaling}
        \\
        J_z &\simeq \frac{2(\tau_\uparrow^2+\tau_\downarrow^2)}{E_C}-2K = \frac{4}{E_C}\left(\tau_c^2+\delta\tau^2\right)-2K .
        \label{eq:Jz_eta_scaling}
\end{align}
Because $\delta\tau=\mathcal{O}(m\eta)$, the exchange anisotropy is quadratic in the altermagnetic parameter,
\begin{equation}
        J_z-J_\perp \propto\tau_c^2\mathcal{O}\left(\frac{(m\eta)^2}{E_C}\right),
        \label{eq:exchange_anisotropy_eta2}
\end{equation}
Thus, although the altermagnetic squeezing produces a spin-dependent tunnel correction that is linear in $\eta$, the exchange anisotropy is proportional to $m^2\eta^2/E_C$ with a dominating charging energy. In this regime the scalar-exchange Hamiltonian used in the main text is therefore well justified.

Close to a $(1,1)$--$(0,2)$ readout transition, however, the accessible doubly occupied state is important and the anisotropic exchange correction is no longer suppressed by the charging energy. The linear spin-dependent tunneling component $\delta\tau$ can then appear directly in the coupling between $\ket{S(0,2)}$ and the unpolarized triplet $\ket{T(1,1)}$ state. Equivalently, by a proper state rotation the spin-conserving tunnel coupling selects a bright state $\ket{\tilde S(1,1)}$ and decouples from a dark state $\ket{\tilde T(1,1)}$
\begin{equation}
    \begin{aligned}
        \ket{\tilde S(1,1)}&=\cos\beta\,\ket{S(1,1)}+\sin\beta\,\ket{T_0(1,1)},\\
\ket{\tilde T_0(1,1)}&=-\sin\beta\,\ket{S(1,1)}+\cos\beta\,\ket{T_0(1,1)}.
    \end{aligned}
\end{equation}
Thus the leading readout effect of the spin-dependent tunneling is a rotation of the Pauli-blockade measurement basis for ST qubits, as discussed in section~\ref{sec:readout}.

\subsection{Inclusion of spin--orbit coupling in the exchange tensor}
\label{app:soc-exchange}
The derivation given above gauged the Rashba spin--orbit term at the level of the localized orbitals, which simplifies the previous analysis at the expense of rotating the local-site quantization axes of each dot with the Rashba field. It is possible to realign the local quantization axes with the N\'eel vector by a proper gauge choice~\cite{Aleiner_2001_Spin, Geyer_2024_Anisotropic}. Reinstating it endows the exchange tensor $\boldsymbol{\mathcal{J}}$ with the rotational structure shown in the main text. We take the dots to be separated by a distance $d$ along $\hat{\mathbf{x}}$ and derive the rotational structure of $\boldsymbol{\mathcal{J}}$ exactly; the resulting non-scalar contributions are then shown to be harmless to both control schemes of Sec.~\ref{sec:dqd}.

For pure Rashba SOC, a non-Abelian gauge transformation~\cite{Aleiner_2001_Spin,Geyer_2024_Anisotropic}
\begin{equation}
  U(\mathbf{r}) = \exp\!\left[
    i\bigl(\sigma_y\, x - \sigma_x\, y\bigr)/\ell_\mathrm{so}
  \right]
  \label{eq:AF_gauge}
\end{equation}
removes the quantization axes rotation, aligning the local axes with the N\'eel vector along $\hat{\mathbf{z}}$, and restoring the local-dot terms to the form $\omega_{q,i}\sigma_z^i/2$ at the expense of rotating the exchange tensor:
\begin{equation}
  \mathcal{J}
    = \mathcal{J}_0\mathcal{R}(2\theta_\mathrm{so},\hat{y}),
  \label{eq:J_rotation_app}
\end{equation}
with $\mathcal{J}_0=\mathrm{diag}(J_\perp,J_\perp,J_z)$ the SOC-free tensor of Sec.~\ref{sec:spin-dependent-exchange} and $\mathcal{R}(2\theta_\mathrm{so},\hat{\mathbf{y}})$ a rotation by angle $2\theta_\mathrm{so}$ about $\hat{y}$, with $\theta_\mathrm{so}=-d/\ell_\mathrm{so}$, since the dots are assumed to be separated along $\hat{x}$. The modified exchange tensor reads then as
\begin{equation}
    \mathcal{J}=\begin{pmatrix}
        J_\perp\cos2\theta_\text{so} & 0 & J_\perp\sin2\theta_\text{so} \\
        0 & J_\perp & 0 \\
        -J_z\sin2\theta_\text{so} & 0 & J_z\cos2\theta_\text{so}
    \end{pmatrix}
\end{equation}
In this frame, up to a spin-independent constant, chosen such that $\langle\uparrow\uparrow|H|\uparrow\uparrow\rangle=\bar{\omega}_q$ (i.e., subtracting $\tfrac{1}{4}J_z\cos 2\theta_\mathrm{so}$), the full (1,1) double-QD Hamiltonian reads in the $\{\ket{\uparrow\uparrow},\ket{\uparrow\downarrow},\ket{\downarrow\uparrow},\ket{\downarrow\downarrow}\}$ basis set as
\begin{widetext}
\begin{equation}
H_\mathrm{DQD}^\mathrm{gen} =
\begin{pmatrix}
\bar\omega_q
& -\dfrac{J_z\sin2\theta_\text{so}}{4}
& \dfrac{J_\perp\sin2\theta_\text{so}}{4}
& -\dfrac{J_\perp(1-\cos2\theta_\text{so})}{4} \\
-\dfrac{J_z\sin2\theta_\text{so}}{4}
& \dfrac{\Delta\omega_q}{2}-\dfrac{J_z\cos2\theta_\text{so}}{2}
& \dfrac{J_\perp(1+\cos2\theta_\text{so})}{4}
& -\dfrac{J_\perp\sin2\theta_\text{so}}{4} \\
\dfrac{J_\perp\sin2\theta_\text{so}}{4}
& \dfrac{J_\perp(1+\cos2\theta_\text{so})}{4}
& -\dfrac{\Delta\omega_q}{2}-\dfrac{J_z\cos2\theta_\text{so}}{2}
& \dfrac{J_z\sin2\theta_\text{so}}{4} \\
-\dfrac{J_\perp(1-\cos2\theta_\text{so})}{4}
& -\dfrac{J_\perp\sin2\theta_\text{so}}{4}
& \dfrac{J_z\sin2\theta_\text{so}}{4}
& -\bar\omega_q
\end{pmatrix}.
\end{equation}
\end{widetext}

The matrix elements of $H_\mathrm{DQD}^\mathrm{gen}$ separate into three groups. Those acting within the inner doublet with $S_z=0$ $\{\ket{\uparrow\downarrow},\ket{\downarrow\uparrow}\}$ -- the diagonal entries $-J_z\cos2\theta_\text{so}/2$ and the off-diagonal $J_\perp(1+\cos2\theta_\text{so})/4$ -- renormalize, respectively, the longitudinal and transverse exchange within that block. Those proportional to $\sin2\theta_\text{so}$ couple the $S_z=0$  block to the polarized states $\{\ket{\uparrow\uparrow},\ket{\downarrow\downarrow}\}$ across an energy separation $\bar\omega_q$. The remaining entry $-J_\perp(1-\cos2\theta_\text{so})/4$ couples the two polarized states across $2\bar\omega_q$. 

For the fSim construction of Sec.~\ref{subsec:fsim}, we work in the rotating frame defined by the unitary $U_R(t)=\exp\bigl[-i\tfrac{\Delta\omega_q\,t}{4}(\sigma_z^L-\sigma_z^R)\bigr]$, in which the inner doublet has vanishing average splitting and the polarized states sit at $\pm\bar\omega_q$. The entries proportional to $\sin2\theta_\text{so}$ and to $(1-\cos2\theta_\text{so})$ then drive off-resonant transitions detuned by $\bar\omega_q$ or $2\bar\omega_q$, respectively, and are discarded at leading order by the same RWA used to derive the gate constraints, provided their amplitudes remain small compared to $\bar\omega_q$. Under the approximation that $J_z=J_\perp$, we find the reasoning in Sec.~\ref{subsec:fsim} is therefore well justified even under SOC within the RWA.

For the ST encoding of Sec.~\ref{subsec:st}, the qubit subspace $\{\ket{S},\ket{T_0}\}$ is the inner doublet expressed in the singlet--triplet basis. The $\sin2\theta_\text{so}$ entries carry this subspace to $\ket{T_\pm}$, gapped by $\bar\omega_q$ from the operating point, with no matrix element within $\{\ket{S},\ket{T_0}\}$; the $(1-\cos2\theta_\text{so})$ entry is confined to the polarized-triplet block. The SOC therefore enters the ST qubit only through the renormalized inner-doublet parameters: the structure of $H_{\mathrm{ST}}$ in Eq.~(\ref{eq:HST}) is preserved, with the longitudinal axis $-(J/2)\tilde\sigma_z$ inheriting the substitution $J\to J_\perp(1+\cos2\theta_\text{so})/2$, while the transverse axis $(\Delta\omega_q/2)\tilde\sigma_x$ set by the per-dot splitting gradient is unmodified at any order in $\theta_\text{so}$.

\section{SU(2) gauge construction for smooth N\'eel fluctuations}
\label{app:neel}

In this appendix, we derive the leading coupling between the qubit and a smooth, weakly time-dependent fluctuation of the N\'eel vector. A complete treatment would include an inhomogeneous, dynamical N\'eel order $\mathbf{N}(\bm{r}, t)=N\bm{n}(\bm{r}, t)$ before the SW projections; such altermagnetic spin textures couple to itinerant electrons through emergent gauge fields~\cite{Maiani_2026_Conductivity, Schrade_2026_Altermagnetic}. However, since our goal is to identify the dominant low-energy decoherence channel, it is sufficient to start from the continuum Hamiltonian already derived in the main text, \cref{eq:bonding-continuum-final}.

In general, the confined electron samples the N\'eel texture at its average position, so the relevant object is simply $\bm n(\bm r(t),t)$, where the dot center $\bm r(t)$ may itself fluctuate due to charge noise acting on the harmonic confinement [see \cref{eq:edsr-displacement-components}]. By the chain rule, the total rate of change of the texture as seen by the electron is
\begin{equation}
\label{eq:neel-chain-rule}
\dot{\bm n}\equiv\frac{d}{dt}\bm n\big|_{\text{dot}}
=
\partial_t \bm n
+
\dot{r}_i\,\partial_i \bm n,
\end{equation}
which includes intrinsic texture dynamics and charge-noise-induced sampling of spatial gradients. The derivation below is most transparent in the co-moving frame of the dot.

We consider a smooth fluctuation about a collinear reference state polarized along $\hat{\bm z}$,
\begin{equation}
\label{eq:neel-noise-texture-decomposition}
\bm n(\bm r,t)
=
\hat{\bm z}
+
\delta \bm n_{\perp}(\bm r,t),
\qquad
\delta \bm n_{\perp}
=
\bigl(\delta n_x,\delta n_y,0\bigr),
\end{equation}
with $|\delta \bm n_{\perp}|\ll 1$. The fixed-length constraint enforces
\begin{equation}
\label{eq:neel-noise-fixed-length-condition}
\delta n_z
=
0
+
\mathcal{O}(\delta n^2).
\end{equation}

To align the instantaneous N\'eel vector with the spin quantization axis, we perform the local spin rotation
\begin{equation}
\label{eq:neel-local-rotation}
U(\bm r,t)
=
\exp\!\left[
-\frac{i}{2}
\bigl(
\delta n_x(\bm r,t)\sigma_y
-
\delta n_y(\bm r,t)\sigma_x
\bigr)
\right].
\end{equation}
It is convenient to introduce the associated SU(2) gauge connection
\begin{equation}
\label{eq:neel-gauge-connection}
A_\mu
\equiv
-\,i\,U^\dagger \partial_\mu U,
\qquad
\mu=t,x,y.
\end{equation}
To linear order in the fluctuation,
\begin{equation}
\label{eq:neel-gauge-connection-linear}
A_\mu
=
\frac{1}{2}
\Bigl[
(\partial_\mu \delta n_y)\sigma_x
-
(\partial_\mu \delta n_x)\sigma_y
\Bigr].
\end{equation}
The rotated Hamiltonian is then
\begin{equation}
\label{eq:neel-rotated-hamiltonian}
\widetilde H
=
U^\dagger H_0 U
-
i U^\dagger \partial_t U
=
H_0
+
A_t
+
\delta H_{\nabla}
+
\mathcal{O}(\delta n^2),
\end{equation}
where the spatial gauge fields enter through $U^\dagger k_i U=k_i+A_i$ and through the rotation of the spin matrices. Expanding to linear order gives
\begin{equation}
\label{eq:neel-gradient-correction}
\begin{split}
\delta H_{\nabla}
&=
\frac{1}{2m}
\Bigl(
\{k_x,A_x\}
+
\{k_y,A_y\}
\Bigr)
\\
&\quad
+
\eta
\Bigl(
\{k_x,A_x\}
-
\{k_y,A_y\}
\Bigr)\sigma_z
+
\alpha_R\,\delta R,
\end{split}
\end{equation}
with
\begin{equation}
\label{eq:neel-delta-r}
\begin{split}
\delta R
&=
\frac{1}{2}\{\delta n_y,k_x\}\sigma_z
-
\frac{1}{2}\{\delta n_x,k_y\}\sigma_z
\\
&\quad-
\frac{1}{2}
\bigl(
\partial_x \delta n_x
+
\partial_y \delta n_y
\bigr)\sigma_0.
\end{split}
\end{equation}

We now project the fluctuation Hamiltonian onto the qubit doublet. To leading order in the static spin--orbit admixture, it is sufficient to use the static basis $\ket{\uparrow},\ket{\downarrow}$ of \cref{eq:qubit-basis-def}, with qubit projector
\begin{equation}
\label{eq:neel-qubit-projector}
P_q
=
\ket{\uparrow}\bra{\uparrow}
+
\ket{\downarrow}\bra{\downarrow}.
\end{equation}
Inside this subspace we introduce Pauli matrices
\begin{equation}
\label{eq:neel-qubit-pauli}
\sigma_z
=
\ket{\uparrow}\bra{\uparrow}
-
\ket{\downarrow}\bra{\downarrow},
\qquad
\begin{aligned}
\sigma_+
&=
\ket{\uparrow}\bra{\downarrow},\\
\sigma_-
&=
\ket{\downarrow}\bra{\uparrow}.
\end{aligned}
\end{equation}
The projected fluctuation Hamiltonian is
\begin{equation}
\label{eq:neel-projected-hamiltonian}
\delta H_q
=
P_q\bigl(A_t+\delta H_{\nabla}\bigr)P_q.
\end{equation}

Since $A_t$ is purely off-diagonal in spin space, its diagonal matrix elements vanish,
\begin{equation}
\label{eq:neel-time-diagonal-vanish}
\bra{\uparrow}A_t\ket{\uparrow}=\bra{\downarrow}A_t\ket{\downarrow}=0,
\end{equation}
whereas the off-diagonal one is
\begin{equation}
\label{eq:neel-time-offdiagonal}
\bra{\uparrow}A_t\ket{\downarrow}
=
\frac{1}{2}
\bra{\phi_+}
\partial_t \delta n_y
+
i\,\partial_t \delta n_x
\ket{\phi_-}.
\end{equation}
For later use, define the orbital overlap
\begin{equation}
\label{eq:neel-orbital-overlap}
\Lambda
\equiv
\braket{\phi_+|\phi_-}
=
\prod_{i=x,y}
\sqrt{
\frac{2\ell_{i,+}\ell_{i,-}}{\ell_{i,+}^2+\ell_{i,-}^2}
}.
\end{equation}

Assuming that the N\'eel texture is smooth on the dot scale, with characteristic length $L_m$, we expand around the dot center,
\begin{equation}
\label{eq:neel-smooth-expansion}
\begin{aligned}
\partial_t \delta n_a(\bm r,t)
&=
\partial_t \delta n_a(\bm 0,t)
+
r_i\,\partial_i\partial_t \delta n_a(\bm 0,t)\\
&\quad
+
\mathcal{O}\left(\frac{\ell^2}{L_m^2}\right),
\qquad
a=x,y.
\end{aligned}
\end{equation}
Since $\phi_\pm(\bm r)$ are real and even in both $x$ and $y$, the terms linear in the coordinates integrate to zero. Therefore,
\begin{equation}
\label{eq:neel-time-offdiagonal-smooth}
\bra{\uparrow}A_t\ket{\downarrow}
=
\frac{\Lambda}{2}
\Bigl[
\partial_t \delta n_y(\bm 0,t)
+
i\,\partial_t \delta n_x(\bm 0,t)
\Bigr]
+
\mathcal{O}\left(\frac{\ell^2}{L_m^2}\right).
\end{equation}

For the spatial part, every qubit-active term in $\delta H_{\nabla}$ contains at least one spatial derivative of the texture and one factor of $k_i$ or an odd operator in the coordinates. The relevant matrix elements vanish by parity,
\begin{equation}
\label{eq:neel-parity-identities}
\bra{\phi_\pm}k_i\ket{\phi_\pm}=0,
\qquad
\bra{\phi_+}k_i\ket{\phi_-}=0,
\qquad
i=x,y.
\end{equation}
Hence the qubit-active part of $\delta H_{\nabla}$ starts only at order $\ell^2/L_m^2$. The only unsuppressed term is the scalar piece in $\delta R$, which produces a common-mode shift,
\begin{equation}
\label{eq:neel-common-mode-shift}
P_q\,\delta H_{\nabla}\,P_q
=
-\frac{\alpha_R}{2}
\bigl[
\nabla\cdot\delta\bm n_{\perp}(\bm 0,t)
\bigr]\sigma_0
+
\mathcal{O}\left(\frac{\ell^2}{L_m^2}\right).
\end{equation}

Collecting the results, the projected qubit Hamiltonian in the smooth-texture limit is
\begin{equation}
\label{eq:neel-final-projected-hamiltonian}
\begin{split}
\delta H_q
&=
-\frac{\alpha_R}{2}
\bigl[
\nabla\cdot\delta\bm n_{\perp}(\bm 0,t)
\bigr]\sigma_0
\\
&\quad
+
\frac{\Lambda}{2}
\Bigl[
\dot{n}_y(\bm 0,t)\,\sigma_x
-
\dot{n}_x(\bm 0,t)\,\sigma_y
\Bigr]
\\
&\quad
+
\mathcal{O}\left(\frac{\ell^2}{L_m^2},\delta n^2\right),
\end{split}
\end{equation}
where $\dot{n}_a\equiv dn_a/dt|_{\text{dot}}$ is the total time derivative of the transverse N\'eel component as seen by the confined electron [\cref{eq:neel-chain-rule}], encompassing intrinsic texture dynamics and charge-noise-driven sampling of spatial gradients on an equal footing. Because the qubit couples to $\dot{n}_a$ rather than to $n_a$ itself, low-frequency $1/f$ fluctuations that dominate in solid-state devices~\cite{Paladino_2014_1} are suppressed.

\Cref{eq:neel-final-projected-hamiltonian} shows that, for fixed-length fluctuations around a collinear background, the leading magnetic coupling to the qubit is purely transverse---there is no $\sigma_z$ term at linear order. Quasi-static N\'eel fluctuations therefore do not produce pure dephasing; the dominant magnetic decoherence channel is relaxation driven by the spectral weight of $\dot{\bm n}$ at the qubit frequency. Furthermore, the first-order charge-noise sweet spot persists as long as the texture is effectively uniform over the displacement range explored by the dot.

\subsection{\texorpdfstring{Material-level parametrization of $T_1$}{Material-level parametrization of T1}}
Here we detail the parametrization used to connect the $T_1$ scaling of \cref{eq:main-text-t1-scaling} to material properties. The $T_1$ rate is driven by the off-diagonal matrix element $\braket{\phi_+|n_a(\bm r,t)|\phi_-}$. For spatially inhomogeneous noise this generalizes the uniform-noise coupling $\Lambda\dot{n}_a(\bm 0)$ of \cref{eq:main-text-neel-noise} to a momentum integral weighted by the dot form factor
\begin{equation}
\label{eq:neel-form-factor-def}
F_d(\bm q)
=
\frac{1}{\Lambda}
\int d^2r\,\phi_+^*(\bm r)\phi_-(\bm r)\,e^{-i\bm q\cdot\bm r},
\end{equation}
normalized so that $F_d(\bm 0)=1$. The dot-averaged noise spectrum, distinguished from the local $\mathcal S_{n_a n_a}(\bm q,\omega)$ by its argument, is then
\begin{equation}
\label{eq:dot-averaged-neel-spectrum}
\mathcal S_{n_a n_a}(\omega)
=
\int
\frac{d^{D} q}{{(2\pi)}^{D}}
\left|F_d(\bm q)\right|^2
\mathcal S_{n_a n_a}(\bm q,\omega).
\end{equation}
The qubit eigenstates are squeezed Gaussians with spin-dependent lengths $\ell_{i,s_z}$ [\cref{eq:squeezedlengths}], so the cross-density $\phi_+^*(\bm r)\phi_-(\bm r)$ is itself a Gaussian with widths $\bar\ell_i^2 = \ell_{i,+}^2\ell_{i,-}^2/(\ell_{i,+}^2+\ell_{i,-}^2)$. To leading order in $m\eta$, $\bar\ell_{x,y}\approx\ell_{x,y}/\sqrt{2}$, giving
\begin{equation}
\label{eq:neel-dot-form-factor}
\left|F_d(\bm q)\right|^2
=
\exp\left[
-\frac{1}{2}
\left(
q_x^2\ell_x^2
+
q_y^2\ell_y^2
\right)
\right].
\end{equation}
We parametrize the retarded transverse susceptibility as
\begin{equation}
\label{eq:neel-retarded-susceptibility}
\chi^R_{aa}(\bm q,\omega)
=
\frac{\mathcal Z_a}{
\Omega_{a,\bm q}^2
-
\omega^2
-
i\gamma_{a,\bm q}\omega
},
\end{equation}
with the magnon dispersion
\begin{equation}
\label{eq:neel-mode-dispersion}
\Omega_{a,\bm q}
=
\sqrt{
\Delta_a^2
+
c_{a,x}^2q_x^2
+
c_{a,y}^2q_y^2
+
c_{a,z}^2q_z^2
}.
\end{equation}
Here $\Delta_a$ is the anisotropy or pinning gap of the transverse mode, $c_{a,i}$ are spin-wave velocities, $\gamma_{a,\bm q}$ is the damping, and $\mathcal Z_a$ is the spectral weight. Applying the fluctuation-dissipation relation,
\begin{equation}
\label{eq:neel-fluctuation-dissipation}
\mathcal S_{n_a n_a}(\bm q,\omega)
=
\frac{2}{1-e^{-\omega/(k_B T)}}
\operatorname{Im}\chi^R_{aa}(\bm q,\omega),
\end{equation}
and substituting into \cref{eq:main-text-t1-scaling} yields the material-level estimate
\begin{equation}
\label{eq:neel-material-t1}
\frac{1}{T_1}
=
\frac{\Lambda^2\omega_q^2}{4}
\sum_{a=x,y}
\int
\frac{d^{D} q}{(2\pi)^{D}}
\left|F_d(\bm q)\right|^2
\mathcal S_{n_a n_a}(\bm q,|\omega_q|),
\end{equation}
where the planar dot filters only in-plane momenta, so $D=2$ for a layered host and $D=3$ for a bulk one.

\section{Coulomb expansion for second-order charge noise}
\label{app:coulomb}

In this appendix we derive the second-order charge-noise sensitivity of the altermagnetic qubit and obtain the form factor $\mathcal{F}$ quoted in \cref{eq:formfactor} of the main text. We model a single charge fluctuator as a point-charge impurity at position $(x',y',z')$ relative to the QD center, with $r'=\sqrt{x'^2+y'^2+z'^2}$, where $z'$ is the vertical separation from the 2D electron gas. The Coulomb potential experienced by the confined electron at in-plane position $(x,y)$ is
\begin{equation}
    V_C(x,y) = \frac{e^2}{4\pi\epsilon\sqrt{(x-x')^2+(y-y')^2+z'^2}},
    \label{eq:VCfull}
\end{equation}
where $\epsilon$ is the host dielectric constant. Unprimed coordinates $(x,y)$ denote the electron position operators, consistent with the harmonic confinement in the main text, while primed coordinates $(x',y',z')$ label the impurity position.

Assuming the impurity is far from the dot center, $x,y\ll r'$, we Taylor-expand \cref{eq:VCfull} in the electron displacement around the origin to second order:
\begin{equation}
\begin{aligned}
    V_C(x,y) &\simeq V_C^{(0)} + V_C^{(1)} + V_C^{(2)}, \\[4pt]
    V_C^{(0)} &= \frac{e^2}{4\pi\epsilon r'}, \\[4pt]
    V_C^{(1)} &= \frac{e^2}{4\pi\epsilon r'^3}\bigl(x'\,x+y'\,y\bigr), \\[4pt]
    V_C^{(2)} &= \frac{e^2}{8\pi\epsilon r'^5}\bigl[(3x'^2-r'^2)\,x^2 + (3y'^2-r'^2)\,y^2 \\
    &\qquad\qquad\quad+ 6x'y'\,xy\bigr],
    \label{eq:Vexpand}
\end{aligned}
\end{equation}
with corrections of order $\mathcal{O}(x^n y^m/r'^{n+m+1})$ for $n+m>2$. Each successive order is suppressed by an additional factor of $\ell/r'$, where $\ell$ is a typical dot size. The first-order term can be rewritten as $V_C^{(1)} = e\delta F_x\,x + e\delta F_y\,y$, with effective electric fields
\begin{equation}
    \delta F_x = \frac{ex'}{4\pi\epsilon r'^3}, \qquad \delta F_y = \frac{ey'}{4\pi\epsilon r'^3},
    \label{eq:linfields}
\end{equation}
and the second-order term can similarly be written as $V_C^{(2)} = e\delta F_x^{(2)}\,x^2 + e\delta F_y^{(2)}\,y^2 + e\delta F_{xy}^{(2)}\,xy$, with
\begin{equation}
\begin{aligned}
    \delta F_x^{(2)} &= \frac{e(3x'^2-r'^2)}{8\pi\epsilon r'^5}, \qquad
    \delta F_y^{(2)} = \frac{e(3y'^2-r'^2)}{8\pi\epsilon r'^5}, \\
    \delta F_{xy}^{(2)} &= \frac{3ex'y'}{4\pi\epsilon r'^5}.
    \label{eq:quadfields}
\end{aligned}
\end{equation}

The first-order term $V_C^{(1)}$ is equivalent to a static, uniform in-plane electric field with components $\delta F_x$ and $\delta F_y$ given in \cref{eq:linfields}. For a harmonic potential, a uniform electric field rigidly displaces the minimum without changing the curvature:
\begin{equation}
    x\to x + \frac{e\delta F_x}{m\omega_x^2}, \qquad y\to y + \frac{e\delta F_y}{m\omega_y^2}.
\end{equation}
Since the harmonic lengths $\ell_{x,y}$ are set by the curvature and not the position of the minimum, the qubit splitting $\omega_q$ in \cref{eq:larmor} is unaffected. This is the \emph{first-order sweet spot}: the altermagnetic qubit is insensitive to charge-noise fluctuations at linear order by construction.

For a time-dependent fluctuator, the displacement oscillates and, through the Rashba spin--orbit coupling, can drive spin rotations. As shown in the EDSR derivation [\cref{eq:edsr-full-qubit-hamiltonian}], this coupling enters through the time derivative of the electric field, $\partial_t F(t)$. If the fluctuator oscillates at frequency $\omega_{\text{fl}}$, the induced spin-rotation amplitude scales as $\omega_{\text{fl}}F_i$, and is therefore strongly suppressed for the low-frequency $1/f$ noise that dominates in solid-state devices~\cite{Paladino_2014_1}.

The terms in $V_C^{(2)}$ that are diagonal in the harmonic basis ($x^2$ and $y^2$) directly renormalize the confinement spring constants:
\begin{equation}
    m\omega_x^2 \to m\omega_x^2 + 2e\delta F_x^{(2)}, \qquad
    m\omega_y^2 \to m\omega_y^2 + 2e\delta F_y^{(2)},
    \label{eq:springshift}
\end{equation}
with $\delta F_x^{(2)}$ and $\delta F_y^{(2)}$ given in \cref{eq:quadfields}. The cross term $\propto xy$ has vanishing expectation value in the harmonic ground state, $\langle 0|xy|0\rangle=0$, and therefore does not contribute at this order; it couples the ground state to the $|1_x,1_y\rangle$ excited state and produces corrections at higher order in perturbation theory.

The change in the spring constant translates into a shift of the inverse squared harmonic length. By expanding one obtains
\begin{equation}
    \delta\left(\frac{1}{\ell_i^2}\right) = em\ell_i^2\,\delta F_i^{(2)},
    \label{eq:deltalell}
\end{equation}
where $\delta F_x^{(2)} = e(3x'^2-r'^2)/(8\pi\epsilon r'^5)$ and similarly for $y$. From the qubit frequency in \cref{eq:larmor}, the resulting shift in the qubit splitting is
\begin{equation}
\begin{aligned}
    \delta\omega_q &= -\eta\left[\delta\left(\frac{1}{\ell_x^2}\right) - \delta\!\left(\frac{1}{\ell_y^2}\right)\right] \\
    &= -\eta \frac{me^2}{8\pi\epsilon} \frac{\ell_x^2(3x'^2{-}r'^2) - \ell_y^2(3y'^2{-}r'^2)}{r'^5}.
    \label{eq:deltaomegaSM}
\end{aligned}
\end{equation}
Introducing the effective Bohr radius of the host material, $a_B=4\pi\epsilon/(me^2)$, this simplifies to
\begin{equation}
    \delta\omega_q = -\frac{\eta}{a_0a_B}\;\mathcal{F}(x',y',z';\ell_x,\ell_y),
    \label{eq:deltaomegafinal}
\end{equation}
with the dimensionless form factor
\begin{equation}
    \mathcal{F} = a_0\,\frac{\ell_x^2(3x'^2-r'^2) - \ell_y^2(3y'^2-r'^2)}{r'^5},
    \label{eq:formfactorSM}
\end{equation}
recovering \cref{eq:deltaomega,eq:formfactor} of the main text. The result factorizes into a material-dependent prefactor $\propto \eta/(a_0a_B)$ and the purely geometric form factor $\mathcal{F}$, which encodes the $d$-wave anisotropic sensitivity of the qubit to the impurity position and dot shape.

The structure of $\mathcal{F}$ directly reflects the $d$-wave character of the altermagnetic spin splitting. For a circular dot ($\ell_x=\ell_y\equiv\ell$), the form factor reduces to
\begin{equation}
    \mathcal{F}\big|_{\ell_x=\ell_y} = \frac{3a_0\ell^2}{r'^5}(x'^2-y'^2),
\end{equation}
which vanishes along the diagonals $x'=\pm y'$. This means that a fluctuator placed exactly along the nodal lines produces no second-order dephasing, providing additional geometric protection that can be exploited in device design. Breaking the dot symmetry ($\ell_x\neq\ell_y$) reshapes the nodal structure, as shown in \cref{fig:4} of the main text, offering a tunable knob to minimize the sensitivity along the directions where charge traps are most likely to reside.

For an ensemble of fluctuators, each contributing independently with its own position $\mathbf{r}'_i$, the total qubit-frequency variance scales as
\begin{equation}
    \overline{(\delta\omega_q)^2} = \left(\frac{\eta}{a_0a_B}\right)^2 \sum_i \mathcal{F}_i^2,
\end{equation}
where the sum runs over all active fluctuators and the overline denotes averaging over their stochastic dynamics. The $1/r'^5$ decay of each $\mathcal{F}_i$ ensures that the dominant noise contribution comes from the nearest impurities, and the angular dependence inherited from the $d$-wave symmetry may lead to non-trivial correlations between the noise power spectral density and the crystallographic orientation of the device.

\section{Cavity quantum electrodynamics integration}
\label{app:cqed}

The single-electron qubit derived in \cref{sec:single-electron-qubit} can be embedded in a superconducting microwave resonator both for dispersive readout and for resonator-mediated long-range coupling. We sketch in this appendix the effective spin--photon Hamiltonian obtained by treating the cavity field as a quantized dipole drive on the dot, alongside the Rashba spin--orbit term already present in \cref{eq:continuum-hamiltonian}. The construction follows the second-order projection developed for spin qubits with spin--orbit coupling in microwave resonators~\cite{Bosco_2022_Fully, Michal_2023_Tunable, Sagaseta_2026_Switchable}.

\subsection{Effective spin--photon Hamiltonian}
\label{app:cqed-eff}

We consider a resonator mode of frequency $\omega_c$ whose vacuum electric field is polarized along $\hat x$ at the dot location, so that the dipole coupling to the confined electron reads
\begin{equation}
\label{eq:cqed-Hd}
H_d=-eE_{\mathrm{zpf}}\,x(a+a^\dagger),
\end{equation}
where $E_{\mathrm{zpf}}$ is the zero-point electric field amplitude at the dot location and $a,a^\dagger$ are the photon ladder operators. The Rashba spin--orbit term contained in \cref{eq:continuum-hamiltonian} acts on the dot as
\begin{equation}
\label{eq:cqed-HR}
H_R=\alpha_R(\sigma_y k_x-\sigma_x k_y),
\end{equation}
with $\alpha_R$ as defined in \cref{eq:bonding-continuum-coefficients}.

The unperturbed Hamiltonian is the dot Hamiltonian of \cref{eq:continuum-hamiltonian} with $\alpha_R\to 0$ together with the free cavity Hamiltonian $H_c=\omega_c a^\dagger a$,
\begin{equation}
\label{eq:cqed-H0}
H_0=\frac{\bm k^2}{2m}-\eta(k_x^2-k_y^2)\sigma_z+V(\bm r)+\omega_c a^\dagger a,
\end{equation}
and the perturbation is $V=H_d+H_R$, which we follow through to second order. The $d$-wave kinetic term $-\eta(k_x^2-k_y^2)\sigma_z$ in $H_0$ acts as a spin-dependent effective mass,
\begin{equation}
\label{eq:cqed-masses}
\frac{1}{M_{x,s_z}}=\frac{1}{m}-2s_z\eta,
\qquad
\frac{1}{M_{y,s_z}}=\frac{1}{m}+2s_z\eta,
\quad s_z=\pm,
\end{equation}
so that each spin sector of $H_0$ is an anisotropic harmonic oscillator with spin-resolved frequencies $\Omega_{x,s_z}=\omega_x\sqrt{1-2m\eta s_z}$ and $\Omega_{y,s_z}=\omega_y\sqrt{1+2m\eta s_z}$ [\cref{eq:spin-sector-frequencies}]. The qubit is encoded in the two orbital ground states $\{\ket{0,+},\ket{0,-}\}$, whose splitting is $\omega_q$ [\cref{eq:larmor}].

Within the qubit doublet, neither $H_d$ nor $H_R$ has matrix elements at first order: $H_d$ is spin-conserving while $H_R$ is odd in momentum and parity. The leading spin--photon coupling therefore appears at second order in $V$, obtained from the Schrieffer--Wolff transformation $S=S_d+S_R$ satisfying $[H_0,S]=V$ on the \emph{full} orbital spectrum of each spin sector; virtual transitions to the orbital first-excited states $\ket{1_x,\pm},\ket{1_y,\pm}$ of $H_0$ supply the intermediate states. The second-order effective Hamiltonian on the qubit subspace is then
\begin{equation}
\label{eq:cqed-SW2}
H^{(2)}_{\mathrm{eff}}=\tfrac{1}{2}\,P_q[S,H_d+H_R]P_q,
\end{equation}
with $P_q$ the projector onto $\{\ket{0,+},\ket{0,-}\}$ applied \emph{after} the SW expansion. The commutator splits into three contributions of the same order: $\tfrac12 P_q[S_d,H_d]P_q$ yields a spin-dependent polarizability term in $X^2$, the cross terms $\tfrac12 P_q([S_d,H_R]+[S_R,H_d])P_q$ yield the transverse couplings $X\sigma_x$ and $P\sigma_y$, and $\tfrac12 P_q[S_R,H_R]P_q$ renormalizes the qubit energy. For a centered separable dot with the cavity field along $\hat x$, only the $\alpha_R\sigma_y k_x$ branch of $H_R$ contributes to leading order; the $\sigma_x k_y$ branch is parity-suppressed by the harmonic ground state. Collecting the three contributions yields
\begin{equation}
\label{eq:cqed-Heff}
H_{\mathrm{eff}}
=
\frac{\omega_q}{2}\sigma_z
+
\tilde\omega_c a^\dagger a
+
g_X X\sigma_x
+
g_P P\sigma_y
+
\frac{\chi_{\mathrm{pol}}}{2}X^2\sigma_z,
\end{equation}
with $X=a+a^\dagger$, $P=i(a^\dagger-a)$, and the renormalized cavity frequency $\tilde\omega_c=\omega_c-(eE_{\mathrm{zpf}})^2\mathcal P_0$. The projection of $H_0$ onto the qubit doublet,
\begin{equation}
\label{eq:cqed-H0-qubit}
H_0\big|_{\mathrm{qubit}}=\frac{\omega_q}{2}\sigma_z+\omega_c a^\dagger a,
\end{equation}
is recovered as the unperturbed part of \cref{eq:cqed-Heff}. Eq.~\eqref{eq:cqed-Heff} coincides with the main-text effective Hamiltonian~\eqref{eq:Heff-main}.

Evaluating the commutators in \cref{eq:cqed-SW2} between the qubit ground states gives explicit expressions for the three coefficients. The transverse couplings are
\begin{equation}
\label{eq:cqed-gX}
g_X=\frac{\alpha_R\,eE_{\mathrm{zpf}}}{4}\left[\frac{1}{\Omega_{x,-}-\omega_q}-\frac{1}{\Omega_{x,+}+\omega_q}\right],
\end{equation}
\begin{equation}
\label{eq:cqed-gP}
g_P=-\frac{\alpha_R\,eE_{\mathrm{zpf}}}{4}\left[\frac{1}{\Omega_{x,-}-\omega_c}-\frac{1}{\Omega_{x,+}+\omega_c}\right],
\end{equation}
and reduce, for weak spin-dependent mass $\Omega_{x,+}\simeq\Omega_{x,-}\simeq\omega_x$, to the leading-order forms
\begin{equation}
\label{eq:cqed-gXgY-weak}
g_X\simeq\frac{\alpha_R\,eE_{\mathrm{zpf}}}{2}\frac{\omega_q}{\omega_x^2-\omega_q^2},
\qquad
g_P\simeq-\frac{\alpha_R\,eE_{\mathrm{zpf}}}{2}\frac{\omega_c}{\omega_x^2-\omega_c^2}.
\end{equation}
The $g_P$ coefficient has the structure of the standard co-moving-frame spin--photon coupling familiar from hole spin qubits~\cite{Bosco_2022_Fully}, while $g_X$ is controlled by the qubit splitting $\omega_q$ and vanishes at leading order for $\omega_q = 0$. The polarizability coefficient, coming from $H_d$ acting twice, reads
\begin{equation}
\label{eq:cqed-chidisp}
\chi_{\mathrm{pol}}=-(eE_{\mathrm{zpf}})^2\mathcal P_z,
\end{equation}
where, in terms of the spin-resolved dynamical $x$-polarizabilities
\begin{equation}
\label{eq:cqed-chis}
\mathcal P_{s_z}(\omega_c)=\frac{1}{M_{x,s_z}(\Omega_{x,s_z}^2-\omega_c^2)},
\qquad s_z=\pm,
\end{equation}
the spin-symmetric and spin-antisymmetric combinations are
\begin{equation}
\label{eq:cqed-Psym-Pasym}
\mathcal P_0=\frac{\mathcal P_++\mathcal P_-}{2},
\qquad
\mathcal P_z=\frac{\mathcal P_+-\mathcal P_-}{2},
\end{equation}
with $\mathcal P_0$ controlling the spin-independent cavity-frequency shift in $\tilde\omega_c$. For the strict harmonic potential of \cref{eq:continuum-hamiltonian} the spring constant $M_{x,s_z}\Omega_{x,s_z}^2=m\omega_x^2$ is spin-independent, so the static difference $\mathcal P_z(0)=0$: the polarizability shift is therefore an entirely dynamical effect, generated by the spin-dependent orbital level spacing. For weak $\eta$,
\begin{equation}
\label{eq:cqed-chiz-weak}
\mathcal P_z(\omega_c)\simeq\frac{2\eta\,\omega_c^2}{(\omega_x^2-\omega_c^2)^2},
\qquad
\chi_{\mathrm{pol}}\simeq-\frac{2(eE_{\mathrm{zpf}})^2\eta\,\omega_c^2}{(\omega_x^2-\omega_c^2)^2},
\end{equation}
in agreement with the leading-order expression \cref{eq:gX-gY-chipol-main} of the main text.

\subsection{Dispersive frame and total dispersive shift}
\label{app:cqed-disp}

For dispersive readout the cavity is operated in the regime $|\omega_q-\omega_c|\gg g_X,g_P$. The transverse part of \cref{eq:cqed-Heff} is then off-diagonal on the qubit subspace and can be eliminated by a second Schrieffer--Wolff transformation; combined with the rotating-wave approximation applied to the polarizability piece $X^2\sigma_z\to(2a^\dagger a+1)\sigma_z$, this brings the spin--photon Hamiltonian into the standard dispersive form.

Using $\sigma_x=\sigma_++\sigma_-$ and $\sigma_y=i(\sigma_--\sigma_+)$ together with $X=a+a^\dagger$, $P=i(a^\dagger-a)$, the transverse part of \cref{eq:cqed-Heff} rearranges into
\begin{equation}
\label{eq:cqed-Ht-decomposed}
\begin{aligned}
H_t\equiv g_X X\sigma_x+g_P P\sigma_y
&=
g_r(a\sigma_++a^\dagger\sigma_-)
\\
&\quad+
g_{\mathrm{cr}}(a^\dagger\sigma_++a\sigma_-),
\end{aligned}
\end{equation}
with rotating (Jaynes--Cummings) and counter-rotating (Bloch--Siegert) couplings
\begin{equation}
\label{eq:cqed-gr-gcr}
g_r=g_X-g_P,
\qquad
g_{\mathrm{cr}}=g_X+g_P.
\end{equation}
The generator that solves $[H_0,S]=H_t$ is
\begin{equation}
\label{eq:cqed-Slin}
S=\frac{g_r}{\omega_q-\omega_c}(a\sigma_+-a^\dagger\sigma_-)
+\frac{g_{\mathrm{cr}}}{\omega_q+\omega_c}(a^\dagger\sigma_+-a\sigma_-).
\end{equation}
The number-conserving part of $\tfrac12[S,H_t]$ then evaluates to
\begin{equation}
\label{eq:cqed-SHt-commutator}
\tfrac{1}{2}[S,H_t]\Bigr|_{\mathrm{disp}}
=\left(\frac{g_r^2}{\omega_q-\omega_c}+\frac{g_{\mathrm{cr}}^2}{\omega_q+\omega_c}\right)(a^\dagger a\sigma_z+\tfrac{1}{2}\sigma_z),
\end{equation}
contributing a dispersive shift
\begin{equation}
\label{eq:cqed-chilin}
\chi_{\mathrm{lin}}
=\frac{g_r^2}{\omega_q-\omega_c}+\frac{g_{\mathrm{cr}}^2}{\omega_q+\omega_c}.
\end{equation}
Combining $\chi_{\mathrm{lin}}$ with the RWA-reduced polarizability piece $\chi_{\mathrm{pol}}a^\dagger a\sigma_z$, and absorbing the total qubit Lamb shift $(\chi_{\mathrm{pol}}+\chi_{\mathrm{lin}})\sigma_z/2$ into the renormalized splitting $\omega_q^\text{ren}=\omega_q+\chi_{\mathrm{pol}}+\chi_{\mathrm{lin}}$, yields the dispersive-frame Hamiltonian
\begin{equation}
\label{eq:cqed-Hdisp}
\begin{aligned}
H_{\mathrm{disp}}
&=\frac{\omega_q^\text{ren}}{2}\sigma_z
+\tilde\omega_c a^\dagger a
+\chi_{\mathrm{tot}}a^\dagger a\sigma_z,
\\
\chi_{\mathrm{tot}}&=\chi_{\mathrm{pol}}+\chi_{\mathrm{lin}},
\end{aligned}
\end{equation}
which coincides with the main-text \cref{eq:Hdisp-main} where we have taken $\omega_q^\text{ren}\rightarrow \omega_q$ to alleviate the notation.

The two contributions to $\chi_{\mathrm{tot}}$ have distinct origins in the microscopic parameters. The polarizability piece scales as $\chi_{\mathrm{pol}}\propto(eE_{\mathrm{zpf}})^2\eta$ at leading order in $\eta$ [\cref{eq:cqed-chiz-weak}]. The linear-coupling piece scales as $\chi_{\mathrm{lin}}\propto\alpha_R^2(eE_{\mathrm{zpf}})^2$, since $g_X,g_P\propto\alpha_R\,eE_{\mathrm{zpf}}$ [\cref{eq:cqed-gX,eq:cqed-gP}]; its $\eta$ dependence enters through $\omega_q$ and through the splitting $\Omega_{x,+}-\Omega_{x,-}$. The overall scale $(eE_{\mathrm{zpf}})^2$ factors out of $\chi_{\mathrm{tot}}$, so the genuine competition parameter between the two mechanisms is $\alpha_R^2$ relative to $\eta$, with $\omega_c$ controlling the detuning structure. The two pieces can have the same or opposite sign depending on the dot geometry, yielding a cancellation locus $\chi_{\mathrm{tot}}=0$ in the $(\ell_x,\ell_y)$ plane distinct from the spin-inversion locus $\omega_q=0$.

In the transverse-coupling form \cref{eq:cqed-Heff}, $g_X X\sigma_x$ and $g_P P\sigma_y$ mediate cavity-bus interactions between physically separated dots~\cite{Blais_2021_Circuit,Dijkema_2024_Cavity}; in the dispersive form \cref{eq:cqed-Hdisp}, $\chi_{\mathrm{tot}}a^\dagger a\sigma_z$ supplies the qubit-state-dependent cavity shift used for circuit-QED readout in \cref{subsec:cqed-readout}. An extrinsic linear longitudinal $X\sigma_z$ or $P\sigma_z$ coupling, absent here, can be engineered by displacing the dot off the resonator-field maximum or by linearizing the polarizability term around a coherent-drive amplitude $\langle X\rangle=X_{\mathrm{cl}}$, in which case the polarizability piece produces $-(eE_{\mathrm{zpf}})^2\mathcal P_z X_{\mathrm{cl}}\,\delta X\,\sigma_z$.

\bibliography{amspinqb}

\end{document}